\documentclass[aps,prd,twocolumn,amsmath,amssymb,nofootinbib,longbibliography]{revtex4-2}
\usepackage{amsfonts,amsthm,mathrsfs}
\usepackage{graphicx}
\usepackage[usenames,dvipsnames,svgnames]{xcolor}
\usepackage{times}
\usepackage{inputenc}
\usepackage[normalem]{ulem}
\usepackage{url}
\usepackage{natbib}
\usepackage{subfigure}
\usepackage{booktabs}
\usepackage{cancel}
\usepackage[colorlinks=true,citecolor=blue,urlcolor=blue,linkcolor=blue]{hyperref}

\def\That{{\hat{T}}}
\def\lambdahat{{\hat{\lambda}}}
\def\qbar{{\bar{q}}}

\begin{document}
\preprint{}
\title{
Finite-Density Dynamics of Chemically Equilibrating QGP in Conformal Gubser Flow and Hard Thermal Photon Production}

\author{Lakshmi J. Naik}
\email{jn\_lakshmi@cb.students.amrita.edu}

\author{V. Sreekanth}
\email{v\_sreekanth@cb.amrita.edu}

\affiliation{Department of Physics, Amrita School of Physical Sciences Coimbatore, Amrita Vishwa Vidyapeetham, India}

\date{\today}
\begin{abstract}
We study the chemical equilibration of a hot and dense quark-gluon plasma (QGP) at finite baryon density produced in relativistic heavy-ion collisions within conformal Gubser flow. Chemical non-equilibrium is incorporated through fugacity parameters in the parton phase-space distribution functions, whose evolution is governed by master rate equations coupled to the hydrodynamic expansion with transverse flow. We analyse the interplay between chemical equilibration and finite-density dynamics, and investigate its impact on hard thermal photon production. We observe that both finite density and transverse expansion delay chemical equilibration, leading to a chemically undersaturated medium with quarks lagging behind gluons. While the overall thermal photon yield from the expanding system is suppressed in the non-equilibrium scenario, we find an enhanced early-time contribution to high $p_T$ photon production. By analyzing the instantaneous photon emission in presence of chemical non-equilibrium, we demonstrate that the rates exhibit a distinct temporal structure arising from the interplay of rapid cooling and evolving fugacities. These features may provide potential observable signatures of chemical equilibration dynamics in the QGP.

\end{abstract}
\maketitle

\section{Introduction}

Investigations of hot and dense QCD matter created in ultra-relativistic heavy-ion collision experiments at RHIC and the LHC have resulted in the discovery of the quark–gluon plasma (QGP) - a deconfined state of quarks and gluons exhibiting remarkable collective behavior and low viscosity~\cite{Hirano:2005wx,Voloshin:2008dg}. Recent beam-energy–scan programs at RHIC and future experiments at FAIR and NICA have renewed interest in understanding the dynamics of QGP at finite density~\cite{CBM:2016kpk,STAR:2017sal,Senger:2021cfo}. 
Relativistic hydrodynamics has emerged as a powerful framework to describe the evolution of this strongly interacting medium, successfully explaining a wide range of collective flow observables and providing insights into the bulk dynamics of the QGP (See Ref.~\cite{Romatschke:2017ejr} and references therein). Most of the hydrodynamical modeling and studies of the expanding QGP matter are based on the assumption that system formed is in thermal and chemical equilibrium~\cite{Teaney:2000cw,Muronga:2003ta,Romatschke:2007mq,Song:2007ux,Gale:2013da,Romatschke:2017ejr}, although this is not strongly supported by arguments based on the first principles. On the contrary, early gluon dominance and quark undersaturation indicate different timescales for thermal and chemical equilibration.  
\par 
Equilibration of hot QCD matter created in heavy-ion collisions has been looked at with great interest over the time. 
Both perturbative and kinetic theory analysis of QCD have indicated towards different timescales for thermalization and chemical equilibration, even though there is ambiguity regarding the exact nature and timescales involved~\cite{Shuryak:1992wc,Geiger:1992si,Biro:1993qt,Levai:1994dx,Wong:1996va,Baier:2000sb,Berges:2013eia,Epelbaum:2013ekf,Kurkela:2015qoa,Keegan:2016cpi,Kurkela:2018vqr,Kurkela:2018oqw,Kurkela:2018xxd,Du:2020zqg,Cabodevila:2023htm}. 
However, these investigations only reinforce the importance of study of chemical equilibration dynamics in QGP evolution and subsequent analysis of observables from the fireball. 
Perturbative analysis of chemical non-equilibrium generally has held the picture that after the rapid thermalisation, chemical equilibration occurs - first for the gluons owing to the high gluon-gluon cross-sections implying a slower evolution of quark (antiquark) equilibration in QGP.
In most of such studies, non-equilibrium effects are introduced via fugacity parameters in parton distribution functions. Master rate equations are then constructed by identifying the most prominent parton reactions driving the equilibration. By calculating the involved rates of the reactions, rate equations are solved together with the underlying hydrodynamical equations with the appropriate heavy-ion collision geometry~\cite{Biro:1993qt,Levai:1994dx}. Generally, one form or other of the J\"uttner distribution function is used with fugacities to describe the system dynamics. Using the one-dimensional longitudinal expansion, chemical equilibration of the QGP matter was studied in Refs.~\cite{Biro:1993qt,Levai:1994dx}. Chemical equilibration of the viscous QGP was investigated using causal second-order hydrodynamics~\cite{Bhatt:2009zg} in such a set-up.
Entropy generation under chemical equilibration was explored in Refs.~\cite{Elliott:1999uz,El:2010mt,Vovchenko:2015yia}. The effect of finite baryon density on chemical equilibration was introduced and analysed in Refs.~\cite{Elliott:1999uz,Dutta:1999hj}, as well. There has been attempts to include the effect of radial expansion in chemical non-equilibrium scenario by superimposing phenomenological flow profiles, typically linear or self-similar radial velocity fields~\cite{Kampfer:1995mg,Srivastava:1996qd,Elliott:1999uz,Gelis:2004ep,Vovchenko:2016ijt}. While these models captured qualitative effects of transverse flow they lacked a symmetry-based hydrodynamic foundation. 
\par 
Analytical modeling of the expansion of the fireball of matter produced in heavy-ion collisions using relativistic hydrodynamics has a long history~\cite{Landau:1953gs,Bjorken:1982qr,Csorgo:2006ax,Bialas:2007iu,Gubser:2010ui,Gubser:2010ze,Shi:2022iyb}. In the very successful one-dimensional model of Bj\"orken, inspired by the phenomenological results, a boost-invariant expansion of the fireball along the collisional axis is assumed~\cite{Bjorken:1982qr}. The Gubser prescription provides analytical solutions to underlying hydrodynamics by including the radial expansion of the fireball while maintaining a boost-invariant expansion in longitudinal direction~\cite{Gubser:2010ui,Gubser:2010ze}. Gubser geometry has recently emerged as a powerful semi-analytic framework to study diverse aspects of heavy-ion collision dynamics, and has been employed to obtain hydrodynamic solutions of the expanding QGP medium across a variety of scenarios, including viscous, anisotropic, magnetic field, spin, and attractor dynamics~\cite{Marrochio:2013wla,Denicol:2014tha,Denicol:2014xca,Nopoush:2014qba,Martinez:2017ibh,Chattopadhyay:2018apf,Shokri:2018qcu,Wang:2021wqq,Behtash:2017wqg,Denicol:2018pak,Ingles:2025yrv,Singh:2024emy}. Coming to the various observables from QGP, within Gubser prescription, the flow harmonics were analysed 
in presence of viscosity~\cite{Hatta:2014upa}, magnetic field~\cite{Gursoy:2014aka} and finite baryon density~\cite{Hatta:2015era}. Recently, quarkonia suppression and charmonium production were studied with the help of Gubser solutions in presence of shear viscosity in Refs.~\cite{Bagchi:2023vfv,Singh:2025xrd}, respectively. Thermal photon spectra was analysed using Gubser flow within ideal hydrodynamics in Ref.~\cite{Paquet:2023bdx}. Thermal dilepton emission from a viscous QGP medium was studied, by the present authors, using second-order Israel-Stewart causal relativistic hydrodynamical Gubser solutions by varying the parameter which characterizes the transverse system size and found that the effective temperature for a smaller system to be higher~\cite{Naik:2025pjt}. Dilepton flow parameters in presence of dissipative effects were also analysed within the same geometry recently~\cite{Dwibedi:2025xho}. 
It will be interesting to study the effect of chemical equilibration on the QGP evolution and observable signals such as hard thermal photons, within the framework of Gubser flow.
\par 
Thermal photons and dileptons are considered as penetrating probes of the QGP, since they interact only electromagnetically, once produced they escape the medium without rescattering, thus carrying direct information about the conditions at the point of their emission~\cite{Feinberg:1976ua,Shuryak:1978ij,Kapusta:1991qp,Peitzmann:2001mz,Rapp:2016xzw,David:2019wpt,Geurts:2022xmk}. Because they are emitted throughout the fireball evolution, their study is of extreme importance to better our understanding of the different stages and properties of the hot and or dense evolving medium~\cite{Bhatt:2011kx,Linnyk:2013hta,Chandra:2015rdz,Vujanovic:2016anq,Naik:2020jfc,Gotz:2021dco,Naik:2022pyk,Garcia-Montero:2023lrd,Churchill:2023vpt}. 
The most prominent sources of hard thermal photon production in QGP are Compton scattering, quark–antiquark annihilation, bremsstrahlung, annihilation with scattering~\cite{Kapusta:1991qp,Aurenche:1998nw,Arnold:2001ms,Steffen:2001pv}. Unlike hadronic observables, photon spectra are sensitive to the entire space-time history of the fireball, including its temperature and density profiles, collective flow, effect of dissipation, and many other non-equilibrium phenomena~\cite{Traxler:1994hy,Schenke:2006yp,Bhatt:2010cy,Shen:2013cca,Bhattacharya:2015ada,Hauksson:2017udm,Naik:2021yph,Xiong:2025koa}. Photon production has been looked at in the context of chemical equilibration as well in number of works~\cite{Kampfer:1994rr,Strickland:1994rf,Traxler:1995kx,Dutta:1999dy,Mustafa:2000sg,Dutta:2001ii,Gelis:2004ep,Long:2005cn,Bhatt:2009zg,Monnai:2014kqa,Srivastava:2016hwr,Gordeev:2025vog}, 
affirming the sensitivity of thermal photons to the early, chemically evolving stages of the medium.
\par 
In this work, we investigate the chemical equilibration of QGP matter at finite baryon density within the framework of conformal Gubser flow. Embedding chemical non-equilibrium and finite-density dynamics in this geometry enables a self-consistent treatment of expansion, cooling, and transverse flow within a controlled semi-analytic setup. Chemical non-equilibrium is incorporated through partonic fugacities whose evolution is consistently coupled to the hydrodynamic expansion and conservation laws. 
Within this framework, we analyse the interplay between finite density, chemical equilibration, and transverse expansion, and study its impact on hard thermal photon production by evaluating the non-equilibrium photon production rates for the relevant processes. In particular, we demonstrate that chemical non-equilibrium enhances the early-time contribution to photon emission, which in turn leads to a relative enhancement of the high-$p_T$ photon yield. This feature provides a potential observable signature of chemical equilibration dynamics in the QGP.
\par 
The paper is structured as follows. In Section~\ref{Sec:Gubser_flow}, we review the Gubser solutions of relativistic hydrodynamics. We study the chemical equilibration in QGP at finite chemical potential within Gubser flow in~\ref{Sec:ChemEq}. The hard thermal photon production under chemical non-equilibrium is presented in Section~\ref{Sec:photon}. Section~\ref{Sec:results} is devoted to the results and its discussion. We summarize our results in Section~\ref{Sec:conclusion}. 
\par 
\textit{Notation and conventions}: Throughout the manuscript we set $c=\hbar=1$ and the metric convention of $(1,-1,-1,-1)$ is followed.

\section{Gubser Flow} \label{Sec:Gubser_flow}

In this section, we review the aspects of Gubser model developed in Refs.~\cite{Gubser:2010ze,Gubser:2010ui}. 
At ultra-relativistic energies, the heavy-ion collision scenario can be described most conveniently using the
Milne coordinates $x^\mu = (\tau, r, \phi, \eta_s)$, with the metric measure $ds^2 = d\tau^2 -(dr^2 + r^2 d\phi^2) - \tau^2 d\eta_s^2$. Here, $\tau = \sqrt{t^2 - z^2}$ and $\eta_s = \tanh^{-1}(z/t)$ are the proper time and space-time rapidity respectively, $r=\sqrt{x^2 + y^2}$ is the radial distance from the center of fireball and $\phi = \tan^{-1} (y/x)$ is the azimuthal angle in transverse plane. The Bj\"orken's one-dimensional (1-D) scaling solution~\cite{Bjorken:1982qr} expresses the QGP dynamics in simplified form. Gubser~\cite{Gubser:2010ze} developed a generalization for the 1-D Bj\"orken flow by considering radial expansion, which is valid only for conformal fluids. Similar to the Bj\"orken model, the system within this geometry has longitudinal boost-invariance along the $\eta_s$ direction and invariance under reflections along the $\eta_s$ axis. Unlike the Bj\"orken flow, the system is not homogeneous in the transverse plane, but has conformal symmetry $SO(3)$.  

Gubser flow can be described using de Sitter coordinates by performing Weyl rescaling of the metric measure in Milne coordinates $i.e.,$ $ds^2 \rightarrow ds^2/\tau^2 \equiv d\hat{s}^2$, followed by a coordinate transformation which replaces $\tau$ and $r$ with the {\it Gubser coordinates} $\rho$ and $\theta$: 
\begin{eqnarray}
    \sinh\rho &\equiv& -\frac{1-(q\tau)^2+(qr)^2}{2q\tau}, \label{Eq:rho}\\
    \tan\theta &\equiv& \frac{2qr}{1+(q\tau)^2-(qr)^2};\label{Eq:theta}
\end{eqnarray}
where, $q$ denotes an arbitrary inverse length scale which sets the transverse size of the system. Note that, we obtain the Bj\"orken flow solutions within the limit $q\rightarrow 0$. The Weyl rescaled metric element in the new coordinates $(\rho, \theta, \phi, \eta_s)$ is now given by
\begin{eqnarray}
  d\hat{s}^2 &=& d\rho^2 - \cosh^2 \rho\,(d\theta^2 +  \sin^2\theta\,d\phi^2) - d\eta_s^2 . 
 \label{Eq:metric_gubser}
\end{eqnarray}
Gubser flow is static in the de Sitter coordinates $i.e.,$ the fluid 4-velocity is 
\begin{equation}\label{Eq:4vel}
 \hat{u}^\mu = (1,0,0,0).   
\end{equation}
In Milne coordinates, this corresponds to the non-vanishing components
\begin{eqnarray*}
 u_\tau  &=& \frac{\partial \rho}{\partial \tau} \hat{u}_\rho = \frac{1}{\sqrt{1-v_{r}(\tau,r)^2}},  \\
 u_r &=& \frac{\partial \rho}{\partial r} \hat{u}_\rho = \frac{-v_{r}(\tau,r)}{\sqrt{1-v_{r}(\tau,r)^2}}; \label{flow_milne}
\end{eqnarray*} 
where the transverse velocity is
\begin{eqnarray}
 v_{r}(\tau,r) &=& \frac{2q^2\tau r}{1+(q\tau)^2 + (qr)^2} \cdot\label{Eq:vel-milne}
\end{eqnarray}

 Also, Weyl rescaling makes all the macroscopic variables dimensionless and renders the fluid homogeneous with all the fields depending only on the de Sitter time coordinate $\rho$.  
In the following, all the quantities depending on Gubser coordinates are denoted by a hat. 
For the given metric, $i.e.$, Eq.~\eqref{Eq:metric_gubser}, we note that the non-vanishing Christoffel symbols of second kind are
\begin{eqnarray*}
\hat{\Gamma}_{\theta\theta}^\rho &=& 
 \cosh\rho \sinh\rho, \qquad   
 \hat{\Gamma}_{\phi\phi}^\rho = \sin^2\theta\cosh\rho \sinh\rho, \\
 \hat{\Gamma}_{\rho\theta}^\theta &=&\hat{\Gamma}_{\rho\phi}^\phi = \tanh\rho,\quad\hspace{0.25cm}
\hat{\Gamma}_{\phi\phi}^\theta = -\sin\theta\cos\theta,\\ 
\hat{\Gamma}_{\phi\theta}^\phi &=& \hat{\Gamma}_{\theta\phi}^\phi = \cot\theta. \end{eqnarray*}
Now, the covariant derivative is given by $\hat{D}_\mu A^\nu= \hat{\partial}_\mu A^\nu + \hat{\Gamma^\nu}_{\mu\rho} A^\rho$ and we get the scalar expansion rate for the Gubser flow as 
\begin{equation}
   \hat{\Theta} = \hat{D}_\mu\hat{u}^\mu=2\tanh\rho. 
\end{equation}

The conformal Gubser flow provides a semi-realistic picture of QGP evolution and can be considered to be well suited to study the effect of finite density dynamics on chemical equilibration. 
Next, we derive the evolution equations describing chemical non-equilibrium in QGP employing the Gubser flow at finite baryon density. 

\section{Chemical Equilibration in QGP} \label{Sec:ChemEq}

\subsection{Master Equations}

As mentioned before, the matter created in heavy-ion collisions is not chemically equilibrated and the number densities of quark (antiquark) and gluons also evolve during the course of expansion of QGP. We assume that the elastic collisions between the partons are sufficiently rapid so that the matter is in thermal equilibrium and hence relativistic hydrodynamics can be employed to study the dynamics of energy density. We invoke master equations to study the equilibration of parton densities. 

In order to study chemical non-equilibrium in QGP, we assume thermal phase space distributions of quarks, antiquarks and gluons of the form 
\begin{eqnarray}\label{Eq:dist}
   f_{q, \bar{q}} = \frac{\lambda_{q, \bar{q}} e^{\pm x_q}}{\lambda_{q, \bar{q}} e^{\pm x_q} + e^{\beta (u \cdot p)}}, \quad\quad   f_g = \frac{\lambda_g }{e^{\beta (u \cdot p)} - \lambda_g};
\end{eqnarray}
where, ${\lambda}_{q, \bar{q}}$ and $\lambda_g$ represent the non-equilibrium fugacities of the partons, $u^\mu$ is the four-velocity of the fluid element and $\beta = 1/T$. Here, $x_q = \mu_q/T$ with $\mu_q$ being the quark (antiquark) chemical potential. Note that the parton fugacities account for the deviation from chemical equilibrium. For a fully equilibrated QGP, the fugacity parameters tend to unity $i.e.,\,\lambda_{i} \rightarrow 1$ ($i\equiv (q, \bar{q},g)$) and in an equilibrating medium, they lie within the range $0 \leq \lambda_{i} \leq 1$. 

Redefining $e^{\pm x_q} \lambda_{q,\bar{q}} = \lambda_{Q(\bar{Q})}$, one can rewrite the quark distribution functions as 
\begin{eqnarray}
    f_{q,\bar{q}} = \frac{\lambda_{Q,\bar{Q}}}{\lambda_{Q,\bar{Q}} + e^{\beta (u \cdot p)}}.
\end{eqnarray}
We also assume $\lambda_q = \lambda_{\bar{q}}$, so that when the quark chemical potential vanishes ($\mu_q \rightarrow 0$), $\lambda_Q = \lambda_{\bar{Q}} = \lambda_q$. 

In order study chemical equilibration in QGP, we consider the following four processes which are the dominant reaction mechanisms that results in the equilibration of each parton flavor~\cite{Biro:1993qt}:
\begin{eqnarray}
    gg \leftrightarrow ggg, \quad\quad
    gg \leftrightarrow q\bar{q}.
\end{eqnarray}
Considering the above processes, the evolution of the quark, antiquark and gluon densities ($n_{i}$) are governed by the master equations~\cite{Biro:1993qt,Bhatt:2009zg}
\begin{eqnarray} \label{Eq:Rate-Eqs-full}
    {\partial}_\mu ({n}_g {u}^\mu) &=& {n}_g ({R}_{2\rightarrow 3} - {R}_{3\rightarrow 2}) - ({n}_g {R}_{g\rightarrow q} - {n}_q{R}_{q\rightarrow g}),\nonumber \\
    {\partial}_\mu ({n}_q {u}^\mu) &=& {\partial}_\mu ({n}_{\bar{q}} {u}^\mu) =  ({n}_g {R}_{g\rightarrow q} - {n}_q {R}_{q\rightarrow g}).
\end{eqnarray}
The quantities ${R}_{2\rightarrow 3}$ and $R_{3\rightarrow 2}$ are the rates for the reaction $gg \rightarrow ggg$ and its reverse. Similarly, the rates for $gg \rightarrow q\bar{q}$ and its reverse process are given by $R_{g\rightarrow q}$ and $R_{q\rightarrow g}$ respectively. 
To simplify the above equations, we adopt factorized phase-space distributions given by~\cite{Dutta:1999dy}
\begin{eqnarray} \label{Eq:fac-dist}
     f_{q,\bar{q}} &=& \frac{\lambda_{Q,\bar{Q}}}{1 + e^{\beta (u \cdot p)}} \equiv \lambda_{Q,\bar{Q}} f_{q,\bar{q}}^\textrm{eq}, \nonumber \\ 
  f_g &=& \frac{\lambda_g }{e^{\beta (u \cdot p)} - 1} \equiv \lambda_g f_g^\textrm{eq};
\end{eqnarray}
where, $f_{q,\bar{q}}^\textrm{eq}$ and $f_g^\textrm{eq}$ are the Fermi-Dirac and Bose-Einstein distribution functions. 
Within this approximation, Eqs.~\eqref{Eq:Rate-Eqs-full} can be now rewritten as~\cite{Bhatt:2009zg}
\begin{eqnarray}
    \partial_\mu (n_q u^\mu) &=& \partial_\mu (n_{\bar{q}} u^\mu) =R_2 n_g \left( 1 - \frac{n_q n_{\bar{q}} \tilde{n}_g^2}{\tilde{n}_q^2 n_g^2}\right), \label{Eq:Rate-Eq1-fact} \\
     \partial_\mu (n_g u^\mu) &=& R_3 n_g \left(1-\frac{n_g}{\tilde{n}_g}\right) \nonumber\\
     &&- R_2 n_g \left( 1 - \frac{n_q n_{\bar{q}} \tilde{n}_g^2}{\tilde{n}_q^2 n_g^2}\right); \label{Eq:Rate-Eq2-fact}
\end{eqnarray}
where, $\tilde{n}_k$ denote number density of the parton $k$ in a fully equilibrated QGP. We note that adopting the factorized distribution functions resulted the RHS of the rate equations (Eqs.~\eqref{Eq:Rate-Eq1-fact} and \eqref{Eq:Rate-Eq2-fact}) to have the same form as that obtained in Ref.~\cite{Biro:1993qt}. 
Now, the number densities of the parton species in the rest frame of the fluid are calculated as
\begin{equation}
    n_{i} = \frac{\gamma_{i}}{(2\pi)^3} \int f_{i}\,d^3 p,
\end{equation}
where, the degeneracy factors of the partons are taken to be $\gamma_{q, \bar{q}} = N_s\times N_c \times N_f$ and $\gamma_g = 2(N_c^2 -1)$. We take $N_s =2$, $N_f = 2.5$ and $N_c =3$ in our analysis. 
By adopting the factorized distribution, we obtain the quark, antiquark and gluon number densities as 
\begin{eqnarray}
    n_{q,\bar{q}} &=& \frac{3\gamma_{q, \bar{q}}} {4\pi^2} \zeta(3) {T}^3 {\lambda}_{Q,\bar{Q}} \equiv b_1 T^3 {\lambda}_{Q,\bar{Q}} , \\
    n_g &=& \frac{\gamma_g}{\pi^2} \zeta(3) T^3 {\lambda}_g \equiv a_1  T^3 {\lambda}_g,
\end{eqnarray}
respectively. Here, $\zeta$ is the Riemann zeta function defined as
\begin{equation}
    \zeta(n) = \sum_{k=1}^\infty \frac{1}{k^n}\cdot 
\end{equation}
Further, in a fully equilibrated QGP, the number density expressions become 
\begin{eqnarray}
   \tilde{n}_{q/\bar{q}} = b_1 T^3 e^{\pm x_q}, \quad\quad \tilde{n}_g = a_1 T^3.
\end{eqnarray}
\par
In the rate equations, Eqs.~\eqref{Eq:Rate-Eq1-fact} and \eqref{Eq:Rate-Eq2-fact}, we have $R_2 = \sigma_2 n_g/2$ and $R_3 = \sigma_3 n_g/2$ with $\sigma_3$ and $\sigma_2$ being the thermally averaged velocity-weighted scattering cross-sections given by
\begin{eqnarray}
\sigma_2 = \langle\sigma (gg \leftrightarrow q\bar{q}) \rangle, \quad \sigma_3 = \langle\sigma (gg \leftrightarrow ggg) \rangle.
\end{eqnarray}
Now, the rates $R_2$ and $R_3$ in the presence of chemical potential are given by~\cite{Levai:1994dx,Dutta:1999hj} 
\begin{eqnarray}
    R_2 &\approx& 0.24 N_f \alpha_s^2 \lambda_g T \ln \left( \frac{1.65}{\alpha_s\Lambda}\right),  \label{Eq:rate-R2}\\
    R_3 &=& \frac{32}{3a_1} \frac{\alpha_s}{\lambda_g} \Bigg[ \lambda_g + \lambda_q\frac{N_f}{6}\cosh x_q\Bigg]^2 \nonumber \\
    &&\times \Bigg[1 + \frac{2}{9}\frac{m_D^2}{T^2}\Bigg]^2 I(\lambda_g, \lambda_q, x_q); \label{Eq:rate-R3}
\end{eqnarray}
where, $\Lambda = \lambda_g + (\lambda_q/2)\cosh x_q$ and
\begin{eqnarray}
I(\lambda_g, \lambda_q, x_q) &=& \int_1^{\sqrt{s_{\tiny M}}\lambda_f}  dx \int_0^{s_{\tiny M}/(4m_D^2)} dz \frac{z}{(1+z)^2}  \nonumber \\
&&\times \Bigg[\frac{\cosh^{-1}\sqrt{x}}{x\sqrt{[x + (1+z)x_D]^2-4xzx_D}}  \nonumber \\
&&+\frac{1}{s_{\tiny M}\lambda_f^2} \frac{\cosh^{-1}\sqrt{x}}{\sqrt{[1+x(1+z)y_D]^2 - 4xzy_D}}\Bigg], \nonumber
\end{eqnarray}
with $x_D = m_D^2\lambda_f^2$, $y_D = m_D^2/s_{\tiny M}$, and $s_{\tiny M}=18 T^2$. Further, the expressions for the Debye screening mass, $m_D$, is 
\begin{eqnarray}
    m_D^2 = 4\pi \alpha_s T^2\Bigg[ \lambda_g + \lambda_q\frac{N_f}{6}\cosh x_q\Bigg]; \label{Eq:debye}
\end{eqnarray}
and the mean free path for elastic scattering is
\begin{eqnarray}
 \lambda_f^{-1}   &=& \pi \alpha_s^2 a_1 T \lambda_g \Bigg[ \frac{9}{2}\frac{T^2}{m_D^2} \Bigg]\Bigg[ 1 + \frac{2}{9}\frac{m_D^2}{T^2}\Bigg]^{-1}.
\end{eqnarray}
In all these expressions, $\alpha_s$ denotes 
the strong coupling constant and we take $\alpha_s = 0.3$.
\par 
Further, we note the expressions for quark and gluon energy densities obtained using the factorized distribution functions
\begin{eqnarray}
    \varepsilon_{q,\bar{q}} &=& b_2 T^4 \lambda_{Q, \bar{Q}}, \\
    \varepsilon_g &=& a_2 T^4 \lambda_g;
\end{eqnarray}
where, $a_2 = \pi^2 \gamma_g/30$, $b_2 = 7\pi^2\gamma_{q,\bar{q}}/240$, and we note that $\varepsilon_i=3P_i$ in all the cases. 
%
\subsection{Hydrodynamic evolution and chemical equilibration under Gubser flow}

Now, we derive the evolution equations for the number densities of quarks, antiquarks and gluons by invoking the Gubser symmetry. As discussed in Section~\ref{Sec:Gubser_flow}, Weyl rescaling of the metric make the fluid homogeneous with all the fields depending only on the Gubser time coordinate $\rho$. In
this scenario, we can rewrite 
Eqs.~\eqref{Eq:Rate-Eq1-fact} and \eqref{Eq:Rate-Eq2-fact} as
\begin{eqnarray}
     \hat{D}_\mu (\hat{n}_q \hat{u}^\mu) &=& \hat{R}_2 \hat{n}_g \left(1- \frac{\hat{n}_q \hat{n}_{\bar{q}} \tilde{\hat{n}}_g^2}{\tilde{\hat{n}}_q \tilde{\hat{n}}_{\bar{q}} \hat{n}_g^2} \right) \\
   \hat{D}_\mu (\hat{n}_g \hat{u}^\mu) &=& \hat{R}_3 \hat{n}_g \left(1- \frac{\hat{n}_g}{\tilde{\hat{n}}_g} \right) \nonumber \\
   &&- \hat{R}_2\hat{n}_g \left(1- \frac{\hat{n}_q \hat{n}_{\bar{q}} \tilde{\hat{n}}_g^2}{\tilde{\hat{n}}_q \tilde{\hat{n}}_{\bar{q}} \hat{n}_g^2} \right);
\end{eqnarray}
where, $\hat{D}_\mu$ is the covariant derivative. As mentioned before, we represent all the quantities depending on $\rho$ with a {\it hat}. Using the definitions of parton densities $\hat{n}_{q,\bar{q}} = b_1 \That^3 \lambdahat_{Q,\bar{Q}}$, $\hat{n}_g =a_1 \That^3 \lambdahat_g$ and the covariant derivative $\hat{D}_\mu$,  
we obtain the evolution equations for quark and gluon equilibration as follows:
\begin{eqnarray}
   \frac{3 \That'}{\That} + \frac{\lambdahat_Q'}{\lambdahat_Q} + 2\tanh \rho &=& \hat{R}_2 \frac{a_1}{b_1} \left(\frac{\lambdahat_g}{\lambdahat_Q} - \frac{\lambdahat_{\bar{Q}}}{\lambdahat_g} \right), \label{Eq:lq-evo}\\
    \frac{3 \That'}{\That} + \frac{\lambdahat_g'}{\lambdahat_g} + 2\tanh \rho &=& \hat{R}_3 (1 - \lambdahat_g) \nonumber\\
    &&- \hat{R}_2 \left(1 - \frac{\lambdahat_Q \lambdahat_{\bar{Q}}}{\lambdahat_g^2} \right); \label{Eq:lg-evo}
\end{eqnarray}
where, the {\it prime} denotes derivative over $\rho$. The rates $\hat{R}_2$ and $\hat{R}_3$ are functions of the coordinate $\rho$. We note that, in the absence of radial expansion, the above equations reduce to the ones obtained for vanishing viscosity and chemical potential within Bj\"orken flow~\cite{Bhatt:2009zg}. Further, from the baryon number ($\hat{n}_B = \hat{n}_q - \hat{n}_{\bar{q}}$) conservation 
equation, $\hat{D}_\mu(\hat{n}_B \hat{u}^\mu)
=0$, we obtain
\begin{eqnarray}
     \frac{3 \That'}{\That} + \frac{\lambdahat_Q' - \lambdahat'_{\bar{Q}}}{\lambdahat_Q - \lambdahat_{\bar{Q}}} + 2\tanh\rho &=& 0. \label{Eq:lqbar-evo}
\end{eqnarray}
\par
The dynamical equation governing the evolution of temperature of the QGP within Gubser flow can be obtained by the conservation of energy momentum tensor, $\hat{u}_\nu \hat{D}_\mu \hat{T}^{\mu\nu} = 0$, where $\hat{T}^{\mu\nu} = (\hat{\varepsilon} + \hat{P})\hat{u}^\mu \hat{u}^\nu - \hat{P}\hat{g}^{\mu\nu}$. Here, $\hat{\varepsilon}$ and $\hat{P}$ are the energy density and pressure respectively, in the Gubser coordinates 
and the metric is given by Eq.~\eqref{Eq:metric_gubser}. This leads to  
\begin{eqnarray}
    \hat{u}^\mu \hat{\partial}_\mu \hat{\varepsilon} + (\hat{\varepsilon} + \hat{P}) \hat{D}_\mu \hat{u}^\mu = 0, \label{Eq:energy_cons}
\end{eqnarray}
where, $\hat{\partial}_\mu$ denotes the partial derivative. 
Noting $\hat{u}^\mu \hat{\partial}_\mu \equiv d/d\rho$ and $\hat{D}_\mu \hat{u}^\mu =2\tanh\rho$, we obtain
\begin{eqnarray} 
     \frac{d \hat{\varepsilon}}{d\rho} + \frac{8}{3} \hat{\varepsilon} \tanh\rho  = 0.\label{Eq:eps-evo-gubser} 
\end{eqnarray}
The ultra-relativistic EoS for a massless quark-gluon system under chemical non-equilibrium, to close the above dynamical evolution equation, is given by ($\hat{\varepsilon} = \hat{\varepsilon}_g + \hat{\varepsilon}_q + \hat{\varepsilon}_{\bar{q}} = 3\hat{P}$):
\begin{eqnarray} \label{Eq:tot-eps}
    \hat{\varepsilon} &=& \Big[a_2 \lambdahat_g + b_2 (\lambdahat_Q + \lambdahat_{\bar{Q}})\Big]\That^4. 
\end{eqnarray}
Adopting the above EoS, Eq.~\eqref{Eq:eps-evo-gubser} can be rewritten as
\begin{eqnarray} 
    \frac{4 \That'}{\That} + \frac{a_2 \lambdahat'_g + b_2(\lambdahat'_Q + \lambdahat'_{\bar{Q}})}{a_2\lambdahat_g + b_2(\lambdahat_Q + \lambdahat_{\bar{Q}})} + \frac{8}{3}\tanh \rho = 0.  \label{Eq:T-evo}
\end{eqnarray}

Equations \eqref{Eq:lq-evo}, \eqref{Eq:lg-evo}, \eqref{Eq:lqbar-evo}, and \eqref{Eq:T-evo} must be solved simultaneously by providing the initial conditions: $\hat{T}(\rho_0) = T(\tau_0, r_0)\tau_0$ and $\lambdahat_{q,\qbar,g}(\rho_0) = \lambda_{q,\qbar,g}(\tau_0, r_0)$, where $\tau_0$ and $r_0$ are the initial values of proper time and radial coordinate respectively; and $\rho_0$ is the initial de Sitter time. Also, the initial value of chemical potential is fixed as $\hat{x}_q (\rho_0)= x_q(\tau_0, r_0)$. The above quantities, once obtained, can be expressed in Milne coordinates through the transformation equations:
\begin{eqnarray}
    T(\tau, r) &=& \That(\rho)/\tau,\nonumber \\
    \lambda_{q, \bar{q}, g}(\tau, r) &=& \lambdahat_{q, \bar{q}, g}(\rho),  \\
   \mu_q(\tau, r) &=& \hat{\mu}_q (\rho)/\tau. \nonumber\label{Eq:transformation}
\end{eqnarray}

We note that, in a fully chemically equilibrated medium ($\lambda_{q,\bar{q}, g} \rightarrow 1$), the equation for temperature and chemical potential decouple from each other. From Eq.~\eqref{Eq:lg-evo}, the evolution equation for temperature can be obtained as
\begin{equation}
    \frac{3\That'}{\That} + 2\tanh\rho = 0;
\end{equation}
which on solving, gives the temperature profile for the equilibrated matter as
\begin{eqnarray}
    \That_\textrm{eq} = \That_0 \left( \frac{\cosh \rho_0}{\cosh \rho}\right)^{2/3}.  \label{Eq:T-eq}
\end{eqnarray}
Using the above solution in Eq.~\eqref{Eq:lq-evo}, we obtain the evolution equation for $\mu_q$ as
\begin{eqnarray} 
    \hat{\mu}_q' = -\frac{2}{3}\hat{\mu}_q \tanh\rho;
\end{eqnarray}
and the expression for chemical potential in equilibrium is found as
\begin{equation}
    \hat{\mu}_\textrm{eq} = \hat{\mu}_q^0 \left( \frac{\cosh \rho_0}{\cosh \rho}\right)^{2/3}.  \label{Eq:mu-eq}
\end{equation}
It must be noted that solving Eq.~\eqref{Eq:lqbar-evo} or Eq.~\eqref{Eq:T-evo} in the limit $\lambda_i \rightarrow 1$ results in the same solution given by Eq.~\eqref{Eq:mu-eq}. Further, it can be seen that the presence of chemical potential does not affect the evolution of temperature when the matter is in chemical equilibrium and the obtained temperature profile is same as that in Ref.~\cite{Gubser:2010ze}. 
\par 
Further, we observe that the 
evolution equations for $T$, $x_q$, $\lambda_q$ and $\lambda_g$ corresponding to the one-dimensional Bj\"orken geometry under chemical non-equilibrium are given as (in Milne coordinates)~\cite{Dutta:1999hj,Bhatt:2009zg}
\begin{eqnarray} 
\begin{split}
\frac{4 \dot{T}}{T} + \frac{a_2 \dot{\lambda}_g + b_2(\dot{\lambda}_Q + \dot{\lambda}_{\bar{Q}})}{a_2\lambda_g + b_2(\lambda_Q + \lambda_{\bar{Q}})}+ \frac{4}{3\tau}  &= 0,  \\
    \frac{3 \dot{T}}{T} + \frac{\dot{\lambda}_Q - \dot{\lambda}_{\bar{Q}}}{\lambda_Q - \lambda_{\bar{Q}}} + \frac{1}{\tau} &= 0,  \\  
\frac{3 \dot{T}}{T} + \frac{\dot{\lambda}_Q}{\lambda_Q} + \frac{1}{\tau} &= R_2 \frac{a_1}{b_1}\left(\frac{\lambda_g}{\lambda_Q} - \frac{\lambda_{\bar{Q}}}{\lambda_g} \right),\\
    \frac{3 \dot{T}}{T} + \frac{\dot{\lambda}_g}{\lambda_g} + \frac{1}{\tau} &= R_3 (1 - \lambda_g)  \\
     &\,\,\,\,\,\,\,\,- R_2 \left(1 - \frac{\lambda_Q \lambda_{\bar{Q}}}{\lambda_g^2} \right); \label{Eq:evo-bjorken}
   \end{split}
\end{eqnarray}
where, $dot$ denotes derivate over the proper time $\tau$. 
\subsection{Entropy Generation}

Under chemical non-equilibrium, the entropy of the system increases until the matter equilibrates and becomes conserved once a local thermodynamic equilibrium is achieved. The fundamental relation in thermodynamics is given by
\begin{eqnarray}
    (\hat{\varepsilon} + \hat{P}) = \That \hat{s} + \sum_i \hat{\mu}_i \hat{n}_i, \label{Eq:thermodyn}
\end{eqnarray}
where, $\hat{s}$ is the entropy density and $i=(q,\bar{q},g)$ and from the first law of thermodynamics, we can write
\begin{eqnarray}
  \hat{u}^\mu \hat{\partial}_\mu\hat{\varepsilon} = \That \hat{u}^\mu \hat{\partial}_\mu \hat{s} + \sum_i \hat{\mu}_i \hat{u}^\mu \hat{\partial}_\mu \hat{n}_i.  \label{Eq:thermodyn-1}
\end{eqnarray}
Substituting the above relations in Eq.~\eqref{Eq:energy_cons}, we obtain 
\begin{eqnarray}
    \hat{D}_\mu (\hat{s}\hat{u}^\mu) = - \sum_i \left[\frac{\hat{\mu}_i}{\hat{T}} + \ln \hat{\lambda}_i\right] \hat{D}_\mu (\hat{n}_i \hat{u}^\mu).
\end{eqnarray}
The analysis of the above equation during the evolution gives the estimate of entropy generation due to the presence of the chemical non-equilibrium in the finite density system. We note that for a chemically equilibrated QGP, the conservation law 
\(\hat{D}_\mu (\hat{n}_i \hat{u}^\mu) = 0\) holds, implying the conservation of entropy, 
\(\hat{D}_\mu (\hat{s} \hat{u}^\mu) = 0\), 
with the corresponding solution
\begin{eqnarray}
 \hat{s}\cosh^2\rho = \mathrm{constant}
\end{eqnarray}
under ideal Gubser flow. 

Now, in order to study the effect of chemical non-equilibrium on signals coming from the finite density QGP, we look into the thermal photon production.

\section{Hard thermal photon production} \label{Sec:photon}

\begin{figure*} 
  \centering
  \subfigure
  []{\includegraphics[width=0.49\textwidth]{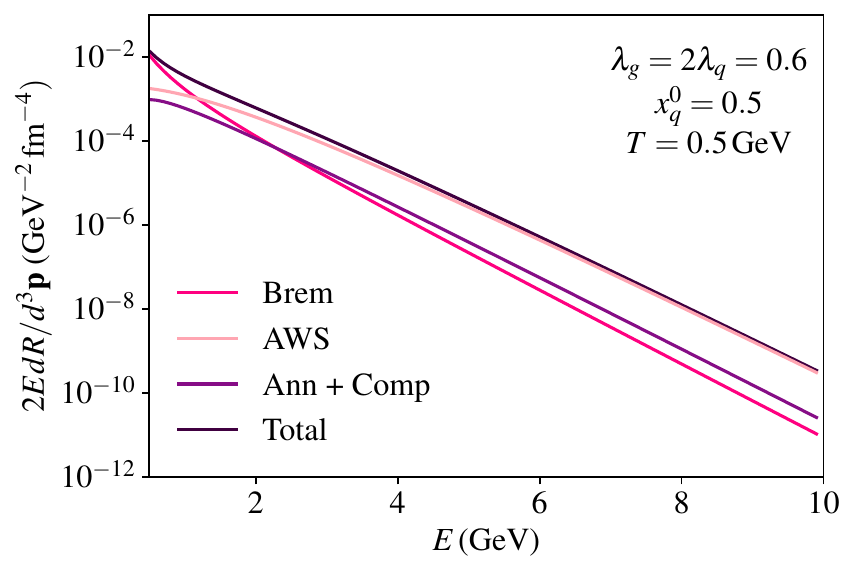}\label{fig:rate-non-eq}}
  \subfigure[]{\includegraphics[width=0.49\textwidth]{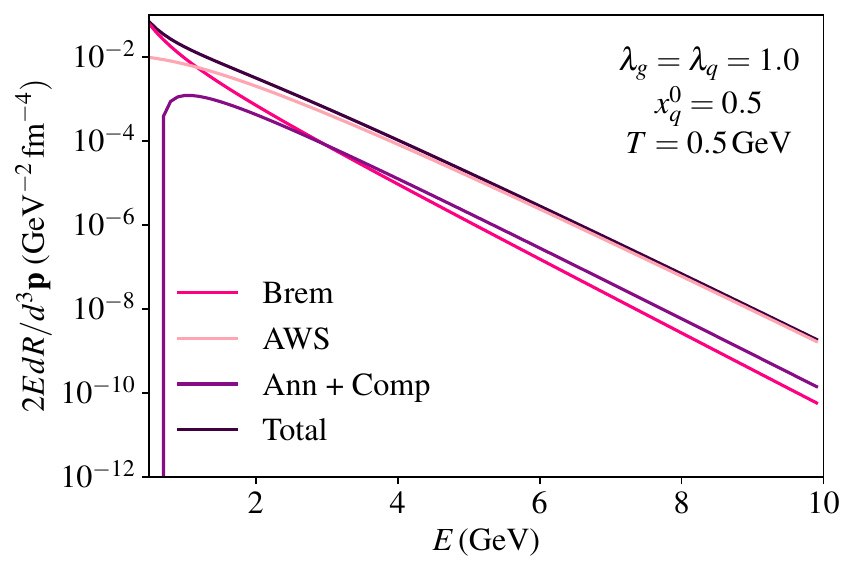}\label{fig:rate-eq} 
  } 
  \caption{Hard thermal photon production rates from chemically {\bf (a)} non-equilibrated ($\lambda_g=2\lambda_q=0.6$) and {\bf (b)} equilibrated ($\lambda_g=\lambda_q=1$) QGP at finite baryon density ($x_q=\mu_q/T=0.5$), for a fixed temperature $T=0.5$ GeV as a function of energy $E$.}
  \label{fig:photon-rate}
  \end{figure*}

In this section, we discuss the thermal photon production from heavy-ion collisions at finite chemical potential. We consider only the hard photon contribution in our analysis. The major processes involved in the hard thermal photon production from one-loop level are the annihilation (Ann) $q\bar{q} \rightarrow g \gamma$ and the Compton scattering (Comp)  $q (\bar{q})g \rightarrow q (\bar{q})\gamma$. The expression for photon production rate for these processes is given by~\cite{Kapusta:1991qp}
\begin{eqnarray}
    2E\frac{dR}{d^3 {\bf p}} &=& \frac{\mathcal{N}}{(2\pi)^8} \int \frac{d^3{\bf p}_1}{2E_1} \frac{d^3{\bf p}_2}{2E_2} \frac{d^3{\bf p}_3}{2E_3} \nonumber \\
    && \times f_1(E_1) f_2(E_2) (1 \pm f_3(E_3)) \nonumber \\
    && \times \delta(p_1 + p_2 - p_3 - p) \sum |\mathcal{M}|^2; \label{Eq:ph-rate-1}
\end{eqnarray}
where, $f_{1,2,3}$ represent the distribution functions corresponding to each process. Here, $p$ is the four-momentum of the photon, $p_{1,2,3}$ are the four-momenta of the partons, $\mathcal{N}$ is the degeneracy factor and $|\mathcal{M}|^2$ are the square of the matrix elements depending on the processes. For the annihilation process, we have~\cite{Kapusta:1991qp,Traxler:1994hy}, 
\begin{eqnarray}
 \sum |\mathcal{M}|_\textrm{Ann}^2 = \frac{2^9\times 5}{9}\pi^2\alpha \alpha_s \frac{u^2 + t^2}{ut},
\end{eqnarray}
and $\mathcal{N}=20$ considering the $u$ and $d$ quarks; whereas, for the Compton process~\cite{Kapusta:1991qp,Traxler:1994hy},
\begin{eqnarray}
 \sum |\mathcal{M}|_\textrm{Comp}^2 = -\frac{2^9\times 5}{9}\pi^2\alpha \alpha_s \frac{s^2 + t^2}{st},
\end{eqnarray}
and $\mathcal{N}=320/3$. Here, $u$, $t$ and $s$ are the Mandelstam variables. We determine the photon production rate by employing the factorized distribution functions Eqs.~\eqref{Eq:fac-dist} in Eq.~\eqref{Eq:ph-rate-1}. Using this approximation, the product of distribution functions appearing in Eq.~\eqref{Eq:ph-rate-1} can be written as
\begin{eqnarray}
    f_1 f_2(1 \pm f_3) &=& \lambda_1 \lambda_2 \lambda_3 f_1^\textrm{eq}  f_2^\textrm{eq} (1 \pm  f_3^\textrm{eq}) \nonumber\\
    &&+ \lambda_1 \lambda_2 (1-\lambda_3)  f_1^\textrm{eq} f_2^\textrm{eq}.
\end{eqnarray}
Adopting the above expression, the one-loop photon production rate can be written as the sum of~\cite{Dutta:1999dy}
\begin{eqnarray}
    \left(2E\frac{dR}{d^3 {\bf p}}\right)_\textrm{Ann} &=& \frac{5\alpha \alpha_s}{9\pi^2} T^2 e^{-E/T}\Bigg[\lambda_Q \lambda_{\bar{Q}}\lambda_g \nonumber\\ 
    && \times \Bigg\lbrace \frac{2}{3} \ln\left( \frac{4ET}{k_c^2}\right) - 1.43 \Bigg\rbrace \nonumber\\ 
    &&+ \frac{2}{\pi^2} \lambda_Q \lambda_{\bar{Q}}(1-\lambda_g) \nonumber\\ &&\times\Bigg\lbrace -2-2\gamma + 2\ln \left( \frac{4ET}{k_c^2}\right) \Bigg\rbrace \Bigg] \\ 
    \textrm{and}\quad\quad\quad\quad&& \nonumber\\
    \left(2E\frac{dR}{d^3 {\bf p}}\right)_\textrm{Comp} &=& \frac{5\alpha \alpha_s}{9\pi^2} T^2 e^{-E/T}\Bigg[ (\lambda_Q^2 \lambda_g + \lambda_{\bar{Q}}^2\lambda_g) \nonumber\\ 
    && \times \Bigg\lbrace \frac{1}{6} \ln \left( \frac{4ET}{k_c^2}\right) + 0.0075\Bigg\rbrace \nonumber \\
    && + \frac{1}{\pi^2} \left[ \lambda_Q \lambda_g(1-\lambda_Q) + 
    \lambda_{\bar{Q}}\lambda_g(1 - \lambda_{\bar{Q}})\right] \nonumber\\
    &&\times \Bigg\lbrace 1- 2\gamma + 2\ln \left( \frac{4ET}{k_c^2}\right)
    \Bigg\rbrace\Bigg]; 
\end{eqnarray}
where, $k_c^2 = 2m_q^2$ is the cutoff parameter with $m_q^2$ being the thermal mass of the quark given in terms of Debye screening mass (Eq.~\eqref{Eq:debye}) as $m_q^2=m_D^2/9$. 

We note that with $x_q =0$, the above non-equilibrium rate expressions match with the results of Ref.~\cite{Traxler:1995kx}. Also for an equilibrated QGP with $x_q =0$, the above results converge to that of Refs.~\cite{Kapusta:1991qp,Traxler:1994hy}.

Next, we determine the photon production rate from two-loop processes such as bremsstrahlung (Brem) and annihilation with scattering (AWS) in presence of chemical non-equilibrium with particle fugacities defined in our analysis. The photon production rate for the bremsstrahlung process is given as~\cite{Dutta:2001ii}
\begin{eqnarray}
    \left(2E \frac{dR}{d^3 {\bf p}} \right)_\textrm{Brem} &=& \frac{2 N_c C_F \alpha \alpha_s}{\pi^5} \left(\sum_f e_f^2\right) \frac{T}{E^2} (J_T - J_L)  \nonumber \\
    &&\times f_g(E)\int_0^\infty dp (p^2 + (p+E)^2) \Big[ f_q(p)  \nonumber \\
    &&- f_q(p+E)+ f_{\bar{q}}(p) - f_{\bar{q}}(p+E) \Big],
\end{eqnarray}
where, $C_F = (N_c^2 - 1)/2N_c$ and $e_f$ is the electric charge of quark with flavor $f$. The factor $J_T-J_L$ is evaluated as follows~\cite{Aurenche:2002pd}: 
\begin{eqnarray}
    J_T-J_L &=&  \pi F\left( \frac{4 M_{\infty}^2}{3m_g^2}\right),
\end{eqnarray}
with 
\begin{eqnarray*}
F(z) &=& \int_0^1 dw \, \frac{\tanh^{-1}w}{(z-1)w^2 + 1} \cdot
\end{eqnarray*}
Here, $M_\infty^2 = 2m_q^2$ is the asymmetric fermion thermal mass and $m_g^2 = m_D^2/3$ denotes the thermal mass of gluon. The above rate expression can be further simplified by evaluating the momentum integrals with the expressions for non-equilibrium distribution functions considered. Thus we obtain, 
\begin{eqnarray}
    \left(2E \frac{dR}{d^3 {\bf p}} \right)_\textrm{Brem} &=& \frac{4 N_c C_F \alpha \alpha_s}{\pi^5} \left(\sum e_f^2\right) \frac{T^2}{E^2} (J_T - J_L) \lambda_q \nonumber \\
     && \times f_g(E)
    \cosh\left( \frac{\mu_q}{T}\right)\Bigg[ E^2\ln \left( \frac{2}{1+e^{-E/T}}\right) \nonumber\\
    &&+T^2 \Bigg\lbrace 3 \zeta(3)+4 \textrm{PolyLog}[3, -e^{-E/T}] \Bigg\rbrace  \nonumber\\
    &&+ ET\Bigg\lbrace \frac{\pi^2}{6} 
    + 2\textrm{PolyLog}[2,-e^{-E/T}]  \Bigg\rbrace \Bigg]; \nonumber\\
\end{eqnarray}
where, we denote the polylogarithm of order $n$ and argument $z$ as
\begin{equation}
    \textrm{PolyLog}[n, z] = \sum_{k = 1}^\infty \frac{z^k}{k^n} \cdot 
\end{equation}
We note that for a chemically equilibrated QGP with $\mu_q =0$, the above rate reduces to the result of Ref.~\cite{Aurenche:1998nw}.
\par
Similarly, for the AWS process, with the rate given by~\cite{Dutta:2001ii}
\begin{eqnarray}\label{Eq:AWS-f}
 \left(2E \frac{dR}{d^3 {\bf p}} \right)_\textrm{AWS} &=& \frac{2 N_c C_F \alpha \alpha_s}{\pi^5} \left(\sum e_f^2\right) \frac{T}{E^2} (J_T - J_L)  \nonumber \\
    &&\times f_g(E) \int_0^E dp (p^2 + (E-p)^2)  \nonumber \\
    &&\times \Big[ f_{q}(-p) - f_q(E-p) \Big],  
\end{eqnarray}
we obtain the following expression with the factorized parton distribution functions,
\begin{eqnarray}
 \left(2E \frac{dR}{d^3 {\bf p}} \right)_\textrm{AWS}
     &=& \frac{2 N_c C_F \alpha \alpha_s}{\pi^5} \left(\sum e_f^2\right) \lambda_q T^2 (J_T - J_L)  \nonumber \\
    && \times f_g(E) e^{\mu_q/T} \Bigg[ \frac{2T}{E} \Bigg\lbrace\textrm{PolyLog}\left[2, -e^{E/T} \right]  \nonumber\\
    && - \textrm{PolyLog}\left[2, -e^{-E/T} \right] \Bigg\rbrace - \frac{E}{T} \nonumber \\ && - \frac{6T^2}{E^2} \zeta(3)+ 3 \ln  \left(\frac{1 + e^{E/T}}{2}\right) \nonumber\\
    &&+ \frac{4T^2}{E^2} \Bigg\lbrace\textrm{PolyLog}\left[3, -e^{E/T} \right] \nonumber\\
    &&  - \textrm{PolyLog}\left[3, -e^{-E/T} \right] \Bigg\rbrace\Bigg].
\end{eqnarray}
In this work, we use the non-equilibrium factorized distribution in the presence of finite density to determine the AWS rate; whereas, in Ref.~\cite{Aurenche:1998nw}, the authors derive the equilibrium rate at vanishing chemical potential within the limit $E>>T$. We note that, the aforementioned equilibrium expression is retrieved under the same assumptions, $\mu_q=0$ and $\lambda_i=1$, using Eq.~\eqref{Eq:AWS-f}.

\begin{figure*} 
   \centering
     \subfigure
     []{
    \includegraphics[width=8.7cm,height=5.8cm]{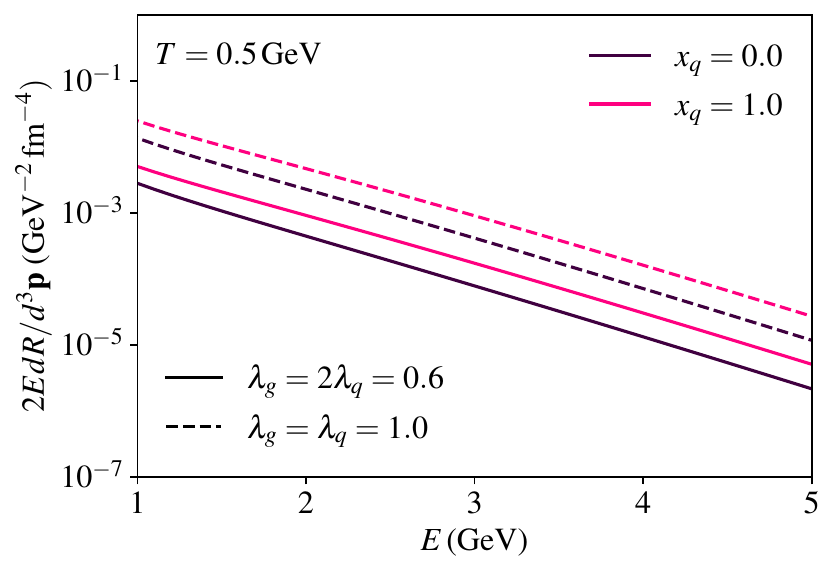}\label{fig:rate-total}} \quad 
    \subfigure[]{\includegraphics[width=8.7cm,height=5.8cm]{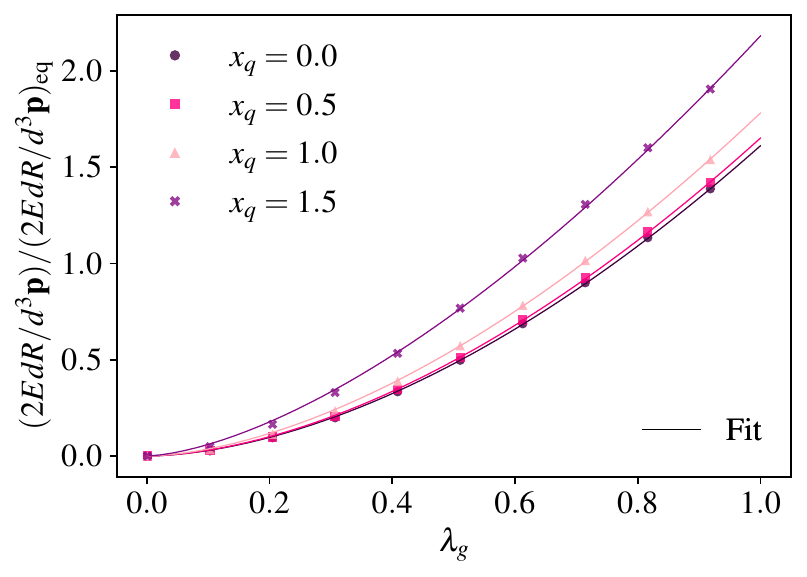}\label{fig:rate-ratio-lg}}
  \caption{{\bf (a)} Total contribution to hard thermal photon rate from QGP, for a fixed temperature $T=0.5$ GeV as a function of energy $E$, varying the quark chemical potential $x_q$ values. The solid lines denote rates from non-equilibrium QGP and dashed lines represent the equilibrium cases. {\bf (b)} Scaling of normalised total photon rate with gluon fugacity $\lambda_g$, for different value of $x_q$.  Lines show the derived fits $a \lambda_g^{n}$ each $x_q$ values considered (Table~\ref{tab:fit_coeffs}). The quark fugacity is fixed to be $\lambda_q =\lambda_g/2$, whereas temperature $T=0.5$ GeV and energy $E=1.5$ GeV.}
  \end{figure*}
\par
In Fig.~\ref{fig:photon-rate}, 
we depict the above derived non-equilibrium hard thermal photon production rates at finite density as a function of
energy ($E$), for the fixed temperature $T=0.5$ GeV and quark chemical potential $x_q=\mu_q/T=0.5$ GeV. The gluon and quark fugacity values are set as $\lambda_g=2\lambda_q=0.6$, since gluon population is expected to be higher compared to quarks~\cite{Bhatt:2009zg}. 
For reference, we also plot the equilibrium ($\lambda_g=\lambda_q=1$) photon rates as well. One can see that presence of chemical non-equilibrium suppresses the photon production rate for every process under consideration for any value of $E$. For the contribution from one-loop processes, {\it i.e.}; annihilation (Ann) and Compton scattering (Comp), suppression is about $50\%$ ($78\%$) at $E=1(3.5)$ GeV. For the two-loop processes bremsstrahlung (Brem) and annihilation with scattering (AWS) reduction in the rate is $\sim 82\%$ for all $E$. Subsequently, the total hard thermal photon production rate is suppressed in presence of non-equilibrium with $\sim 79.5\% \,(81.3\%)$ at $E=1(3.5)$ GeV. 
In Fig.~\ref{fig:rate-total}, we display the total photon production rate as before, for $x_q=1.0$ and $x_q=0$. We also plot the corresponding equilibrium case with dashed lines. Compared to the $x_q=0.5$ case considered before, we observe that suppression is $79.6\%$ ($81\%$) at $E=1(3.5)$ GeV. We note that our results are qualitatively similar to the trend obtained for photon rates (1-loop) from QGP at finite chemical potential for equilibrium~\cite{Traxler:1994hy} and non-equilibrium~\cite{Dutta:1999dy} scenarios. 
\par
Next, in Fig.~\ref{fig:rate-ratio-lg}, we depict the scaling of total photon rate with $\lambda_g$, by normalizing the rate with that obtained at equilibrium, for different $x_q$. We fix $T=0.5$ GeV, $E=1.5$ GeV and $\lambda_q = \lambda_g/2$ for this analysis. We see that the rate increases with $\lambda_g$ for all values of $x_q$ considered. This increase is observed to be steeper for $x_q = 1.5$. The rates can be fitted to a simple function $a \lambda_g^{n}$, where the fit parameters $a$ and $n$ corresponding to each $x_q$ have been tabulated in Table~\ref{tab:fit_coeffs}. We find that rates corresponding to finite values of $x_q$ scales with $\lambda_g$ with powers $n<2$. For example, with $x_q = 0$ (1.0), the rate scales as $\lambda_g^{1.75}\,(\lambda_g^{1.78})$. 
\begin{table}[h]
\centering
\begin{tabular*}{5cm}{@{\extracolsep{\fill}} ccc}
\toprule
$x_q$ & $a$ & $n$  \\
\midrule
0.0 & 1.61 &  1.75  \\
0.5 & 1.65 &  1.74  \\
1.0 & 1.78 &  1.69  \\
1.5 & 2.18 &  1.56  \\
\bottomrule
\end{tabular*}
\caption{Fit (Rate $=a \lambda_g^{n}$) coefficients $a$ and $n$ corresponding to each value of $x_q$.}
\label{tab:fit_coeffs}
\end{table}

\par 
Thus, we see that the chemical non-equilibrium has significant effect on the production rates of photons and it will be interesting to see how this affects the spectra from the expanding fireball within Gubser flow. 
Now, the thermal photon spectra from expanding QGP is obtained by convoluting the total emission rate with the space-time history of collisions. The 4-momentum of the photon is parameterized as $p^\mu = \Big(p_T \cosh(y-\eta_s), p_T \cos(\phi_p - \phi), p_T \sin(\phi_p - \phi)/r, p_T \sinh(y-\eta_s)/\tau \Big)$. Noting the definition of fluid 4-velocity in Gubser flow (Eq.~\eqref{Eq:vel-milne}), the photon energy in a general frame can be obtained as
\begin{eqnarray}
    u\cdot p &=& u_\tau p_T \cosh(y-\eta_s) - u_r p_T \cos(\phi_p - \phi). \label{Eq:udotp}
\end{eqnarray}
In order to terminate the hydrodynamical evolution, we need to define the freeze-out hypersurface describing the end of QGP phase. There are several approaches to fix this surface and we proceed with fixing the energy density value: 
\begin{equation}  \label{Eq:freeze_out}
    \varepsilon (T, \mu_q, \lambda_i) = \varepsilon_f. 
\end{equation}
In this study, we fix $\varepsilon_f=1$ GeV/fm$^{3}$ and we note that for a completely
equilibrated plasma at zero baryon density, this value corresponds to a freeze-out temperature of 0.153 GeV. 
Now, the total photon emission spectra within Gubser flow can be obtained as
\begin{align}
    \frac{dN}{p_Tdp_T dy} =& \frac{1}{2} \int_0^{2\pi} d\phi_p \int_{\tau_0}^{\infty} d\tau\,\tau \int_0^\infty dr\,r \int_0^{2\pi} d\phi \nonumber\\
    &\times\int_{-\infty}^\infty d\eta_s \sum_i \left(2E \frac{dR}{d^3 {\bf p}}\right)_i \Theta ({\small \varepsilon >\varepsilon_f}); \label{Eq:ph-yield} \\
    =& \int_{\tau_0}^{\infty} d\tau\,\tau\,\mathscr{R}(p_T, \tau)
\end{align}
where, the Heaviside step function $\Theta$ restricts the evolution till the freezeout criteria defined by $\varepsilon_f$.

\begin{figure*}
  \centering
  \subfigure
  []{\includegraphics[width=8.5cm,height=5.8cm]{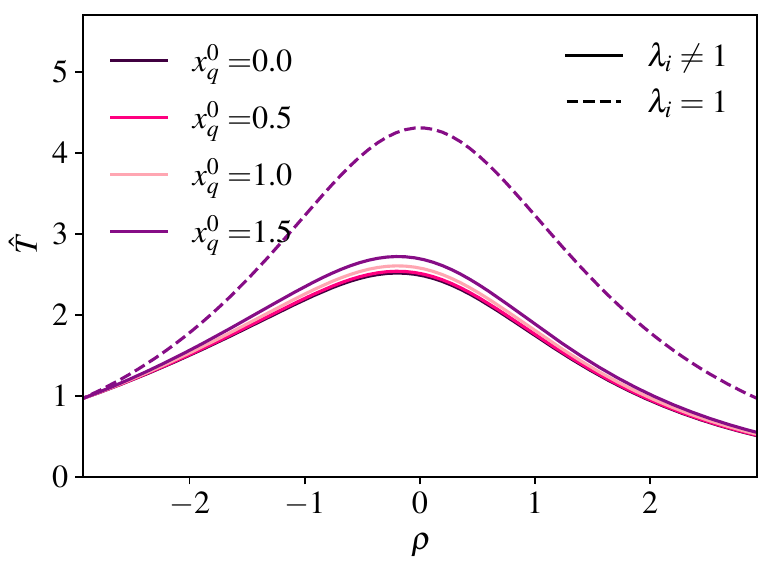}\label{fig:T-rho} 
  }  \quad
  \subfigure[]{\includegraphics[width=8.5cm,height=5.8cm]{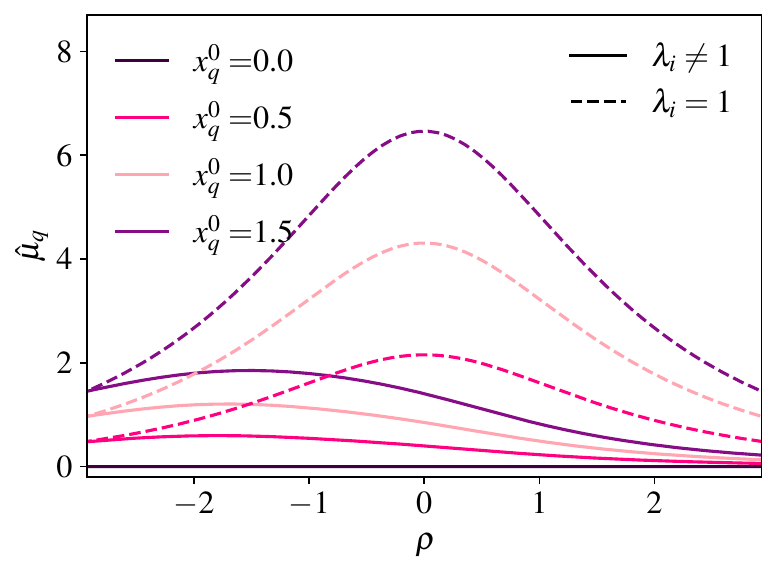}\label{fig:mu-rho}}
  \caption{Evolution of {\bf (a)} temperature and {\bf (b)} chemical potential of the chemically equilibrating QGP under Gubser flow, by varying the initial quark chemical potential $x_q^0$. Evolution of the quantities from an equilibrated QGP are denoted by dashed lines.}
  \label{fig:evo-rho-1}
  \end{figure*}

\begin{figure*} 
  \centering
  \subfigure
  []{\includegraphics[width=8.5cm,height=5.8cm]{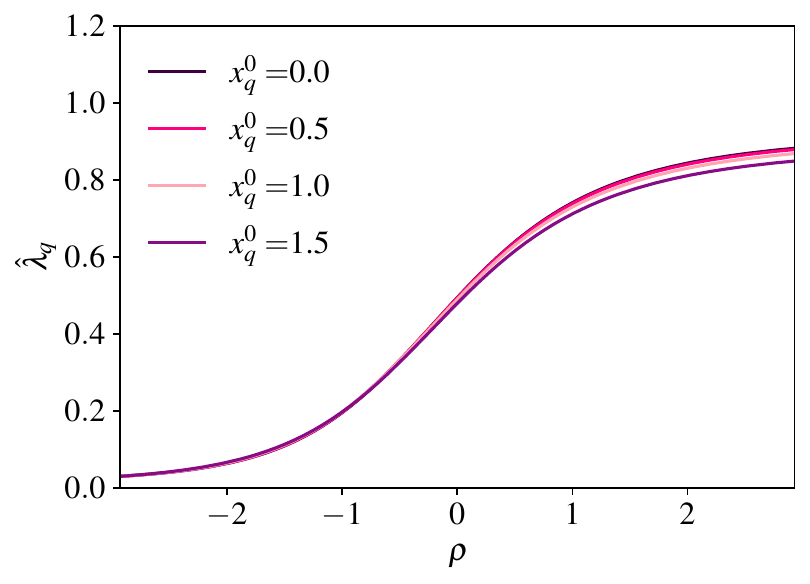}\label{fig:lq-rho}} \quad
  \subfigure[]{\includegraphics[width=8.5cm,height=5.8cm]{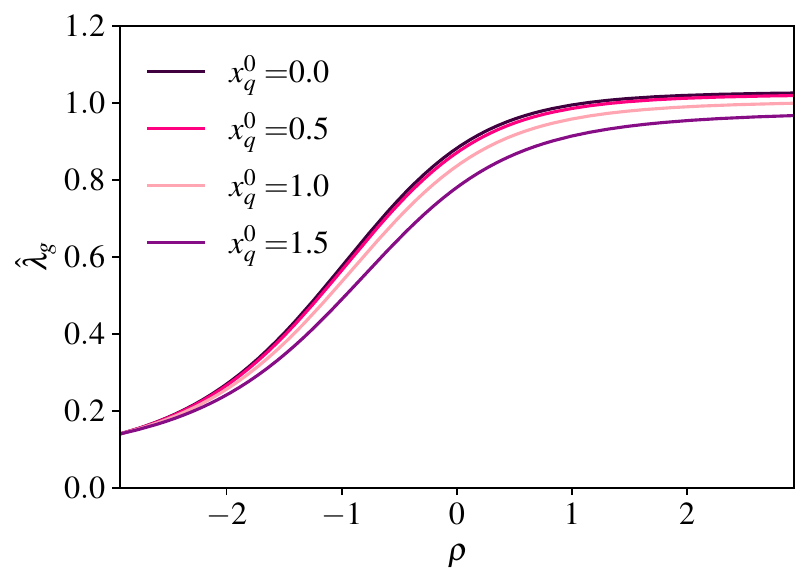}\label{fig:lg-rho} 
  } 
  \caption{Evolution of {\bf (a)} quark and {\bf (b)} gluon fugacities of the chemically equilibrating QGP under Gubser flow, by varying the initial quark chemical potential.}
  \label{fig:evo-rho-2}
  \end{figure*}

\section{Results and discussion} \label{Sec:results}

 \begin{figure} 
  \centering
    \includegraphics[width=8.5cm,height=5.8cm]{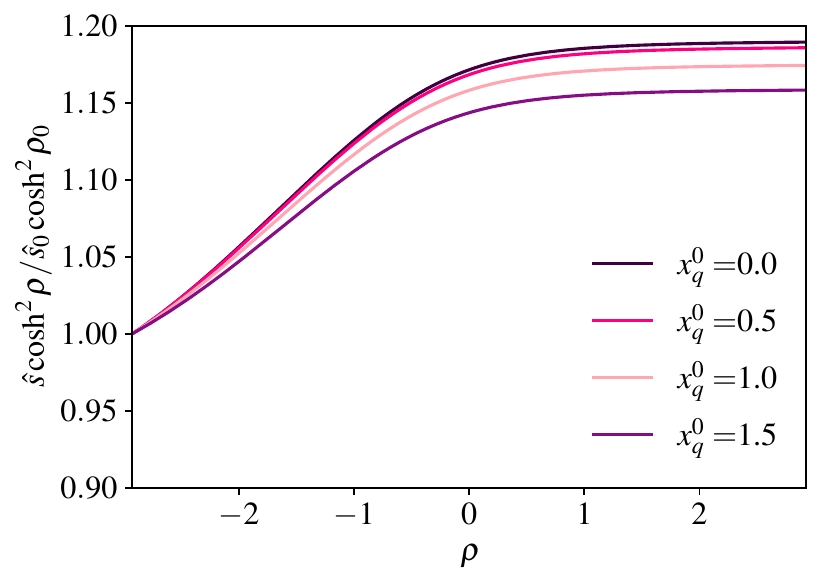} 
  \caption{Evolution of the entropy density as a function of Gubser coordinate $\rho$, for different initial quark chemical potential values.}
  \label{fig:s-rho}
  \end{figure}

We numerically solve Eqs.~\eqref{Eq:lq-evo}, \eqref{Eq:lg-evo}, \eqref{Eq:lqbar-evo}, and \eqref{Eq:T-evo} to study the chemical equilibration in QGP medium in the presence of finite chemical potential. We consider the following initial conditions~\cite{Biro:1993qt}: the initial proper time is taken as $\tau_0 = 0.23$ fm and at the center of the fireball ($r_0=0$ fm), the values of initial temperature and parton fugacities are taken as $T(\tau_0, r_0) \equiv T_0 = 0.83$ GeV, $\lambda_q (\tau_0, r_0) \equiv \lambda_q^0 = 0.03$, and $\lambda_g(\tau_0, r_0) \equiv \lambda_g^0 = 0.14$, respectively. These values are motivated from HIJING calculations~\cite{Wang:1991hta,Traxler:1995kx}. The chosen initial fugacity values correspond to an extreme chemically undersaturated scenario intended to isolate the effects of chemical equilibration dynamics. The initial de Sitter time, $\rho_0 = \rho(\tau_0, r_0)$ is calculated from the transformation given by Eq.~\eqref{Eq:rho} and the Gubser parameter is fixed to be $q=1/4.3$ fm$^{-1}$~\cite{Gubser:2010ze}. In addition, we vary the initial value of the chemical potential as $x_q^0\equiv \mu_q(\tau_0, r_0)/T(\tau_0,r_0) = 0, 0.5, 1.0$ and 1.5 in this study. As noted before, we keep the evolution till the freeze-out criteria $\varepsilon_f=1$ GeV/fm$^{3}$ (Eq.~\eqref{Eq:tot-eps}) is satisfied. 
\par
In Figures~\ref{fig:evo-rho-1} and \ref{fig:evo-rho-2}, we present the evolution of $\That, \hat{\mu}_q, \lambdahat_q$, and $\lambdahat_g$ as a function of Gubser coordinate $\rho$, by varying the initial quark chemical potential $x_q^0$. In all these figures, we have also shown the evolution corresponding to vanishing chemical potential for comparison. We note that the region with $\rho <<0$ corresponds to larger $r$ while $\rho >>0$ probes the later proper time region~\cite{Marrochio:2013wla}. 
In Fig.~\ref{fig:T-rho}, we plot the temperature profile of chemically equilibrating QGP by varying $x_q^0$. We observe that the temperature increases as $\rho$ approaches zero and later decreases with increment in $\rho$. As $\rho$ becomes more negative/positive, $\That$ decreases, indicating that the system cools at larger radial distance/late times. Unlike the equilibrated case Eq.~\eqref{Eq:T-eq} (shown by the dashed curve), the peak of $\That$ profile under chemical non-equilibrium deviates from $\rho = 0$ and temperature decreases rapidly with increase in $\rho$. Further, we find that the $\That$ has a lower maximum value in case of chemical non-equilibrium compared to equilibrium scenario ($\lambda_i =1$). We also observe that the effect of chemical potential on the temperature profiles is more visible around $\rho =0$. It can be seen that the presence of chemical potential increases the temperature and its impact is high for large value of $x_q^0$ chosen. Further, we note that increasing $x_q^0$ slows down the cooling compared to the system with zero chemical potential. Next, in Fig.~\ref{fig:mu-rho}, we depict the evolution of $\hat{\mu}_q$ for different $x_q^0$. We find that with increase in $\rho$, the chemical potential first increases from the initial value, reaches a maximum and then decreases, for any $x_q^0$. 
We have also plotted the evolution corresponding to the equilibrium case (Eq.~\eqref{Eq:mu-eq}). We see that the chemical potential remains larger throughout the $\rho$ range for any value of $x_q^0$ in an equilibrated QGP. Also, the maximum value of $\hat{\mu}_q$ occurs around $\rho =0$ for the equilibrated QGP; while for the non-equilibrium case the peak shifts to $\rho <0$. Moreover, the chemical potential decreases slowly with increasing $\rho$ in an equilibrated QGP, especially for higher value of $x_q^0$. We note that the peak value of $\hat{\mu}_q$ increases with increase in $x_q^0$. 

\begin{figure*} 
  \centering
  \subfigure
  []{\includegraphics[width=8.5cm,height=5.8cm]{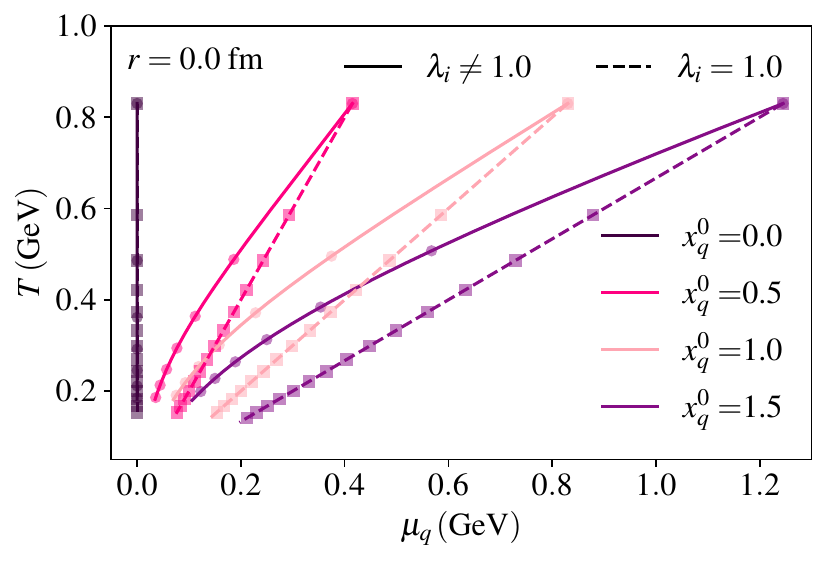}\label{fig:gub-traj-r0}} \quad
  \subfigure[]{\includegraphics[width=8.5cm,height=5.8cm]{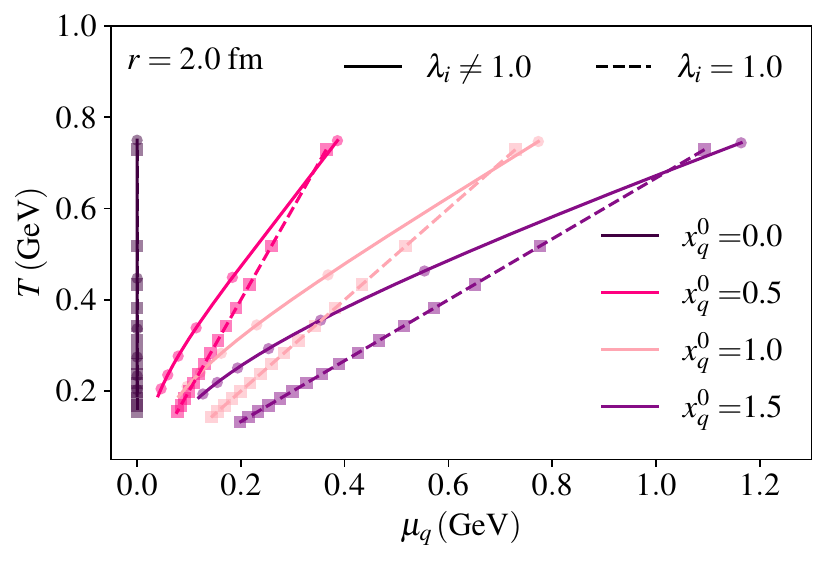}\label{fig:gub-traj-r2} 
  } 
  \caption{Flow trajectories of chemically equilibrating ($\lambda_i\neq 1$) QGP under Gubser flow in the $\mu_q$ - $T$ plane for \textbf{(a)} $r=0$ fm and \textbf{(b)} $r=2$ fm, for different initial quark chemical potential values (initial temperature is fixed as $T_0=0.83$ GeV). Corresponding equilibrium cases ($\lambda_i=1$) are also shown using dashed lines. The markers correspond to proper time with time interval $0.4$ fm. 
  }
  \label{fig:gub-traj}
  \end{figure*}

  \begin{figure}
      \centering
      \includegraphics[width=8.5cm,height=5.8cm]{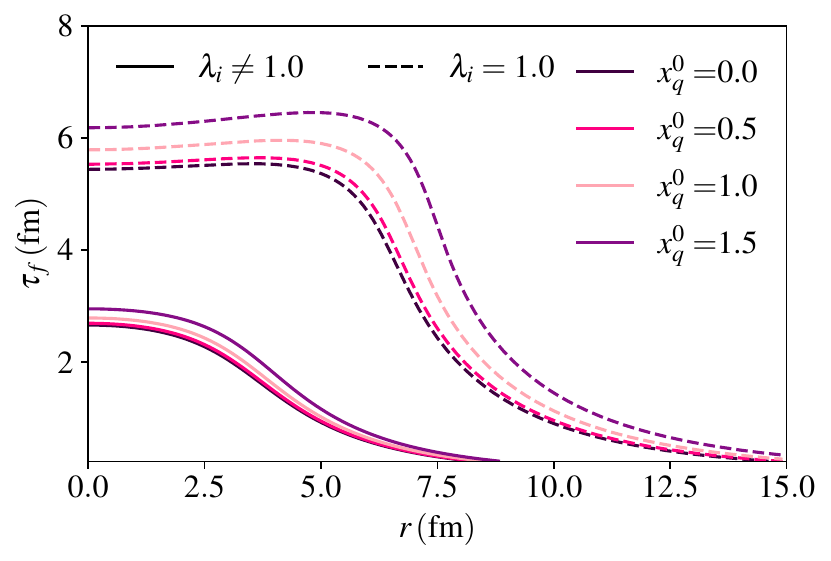}
      \caption{Variation of freeze-out proper time $\tau_f$ of chemically equilibrating QGP as a function of radial co-ordinate $r$, for different values of $x_q^0$. Freeze-out hypersurface is defined by $\varepsilon_f=1$ GeV/fm$^3$. The dashed curves denote the $\tau_f$ values for equilibrated system. }
      \label{fig:tau-r}
  \end{figure}

 \begin{figure*} 
  \centering
  \subfigure
  []{\includegraphics[width=8.5cm,height=5.8cm]{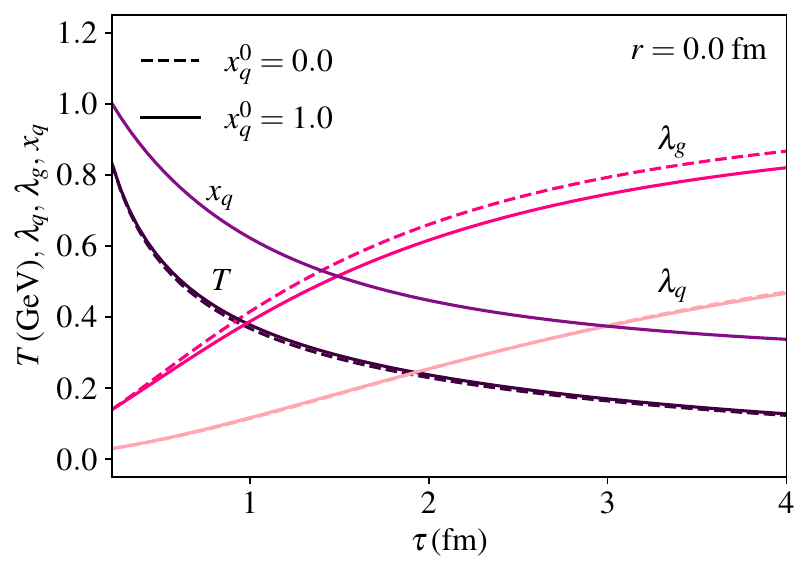}\label{fig:evo-compare-r0}} \quad
  \subfigure[]{\includegraphics[width=8.5cm,height=5.8cm]{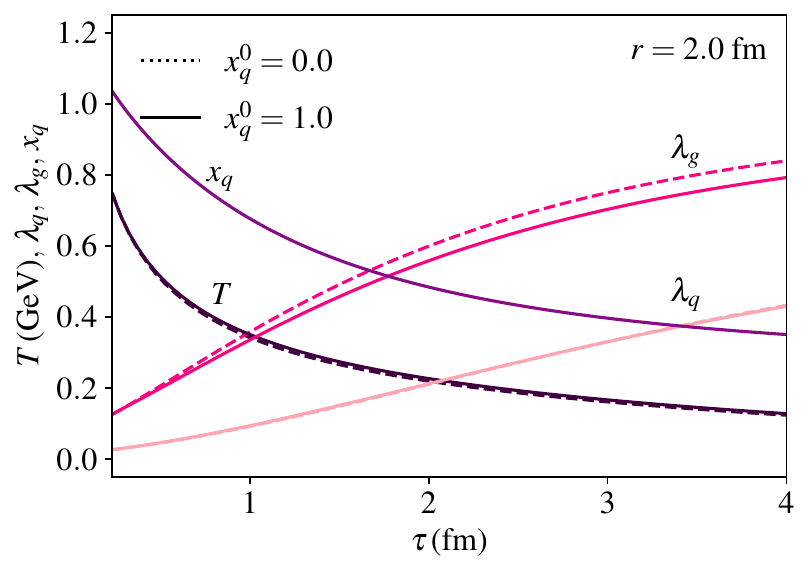}\label{fig:evo-compare-r2} 
  } 
  \caption{Evolution of temperature, quark and gluon fugacities, and quark chemical potential of chemically equilibrating QGP at finite density ($x_q^0 = 1$) as a function of proper time, plotted at \textbf{(a)} $r=0$ fm and \textbf{(b)} $r=2$ fm. Evolution of the quantities at vanishing density ($x_q^0 = 0$) is also plotted for comparison. The initial conditions are taken to be $T_0 = 0.83$ GeV, $\lambda_q^0 = 0.03$ and $\lambda_g^0 = 0.14$ at $\tau_0=0.23$ fm, $r_0 = 0$ fm. }
\label{fig:evo-compare}
  \end{figure*}

   \begin{figure*} 
  \centering
  \subfigure
  []{\includegraphics[width=0.49\textwidth]{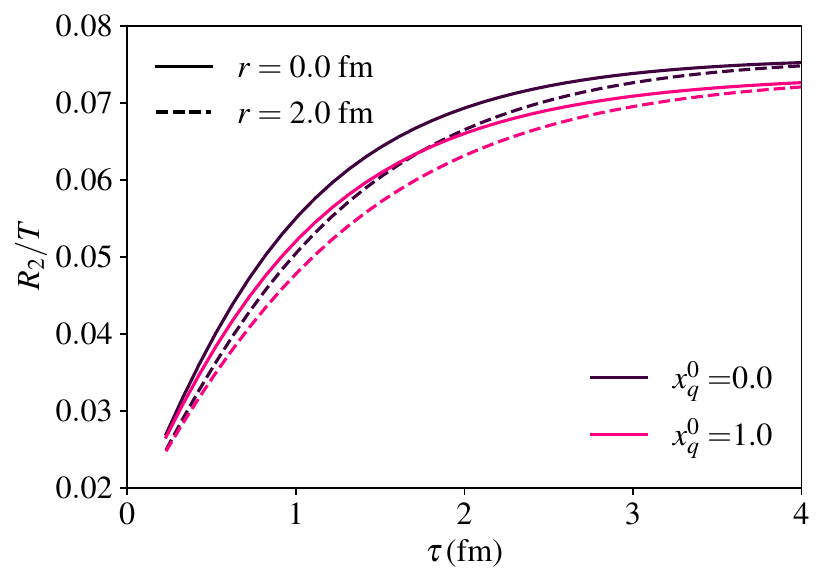}\label{fig:R2byT-tau} } 
  \subfigure[]{\includegraphics[width=0.49\textwidth]{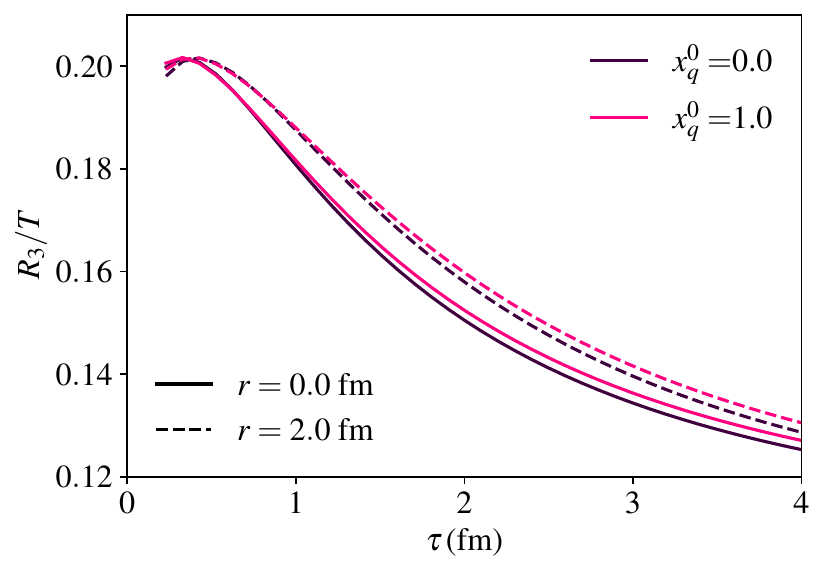}\label{fig:R3byT-tau}}
  \caption{Quark and gluon equilibration rates $R_2$ and $R_3$ scaled to temperature $T$ as a function of proper time $\tau$, for $x_q^0 = 0$ and 1.0. The rates are plotted by fixing $r=0$ and 2 fm.  }
\label{fig:R3-R2-tau}
  \end{figure*}
  
\begin{figure*} 
  \centering
  \subfigure
  []{\includegraphics[width=8.5cm,height=5.8cm]{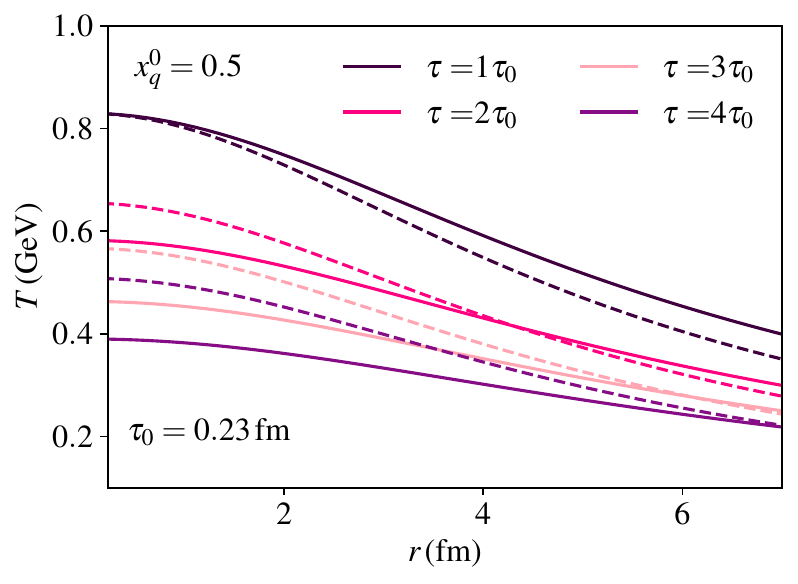}\label{fig:T-r-mu05}} \quad
  \subfigure[]{\includegraphics[width=8.5cm,height=5.8cm]{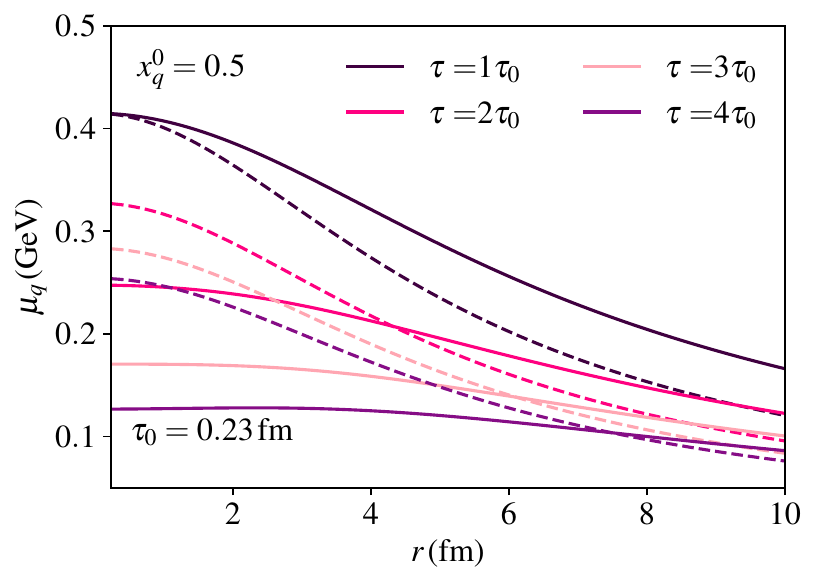}\label{fig:mu-r-mu05} 
  } 
  \caption{\textbf{(a)} Temperature and \textbf{(b)} quark chemical potential from a chemically equilibrating QGP medium as a function of $r$, plotted at different snapshots of proper time. The solid lines represent the evolution in presence of chemical non-equilibrium (with $\lambda_q^0 = 0.03$ and $\lambda_g^0 = 0.14$), and dashed lines correspond to the fully equilibrated case ($\lambda_i=1$). }
  \label{fig:T-mu-r}
  \end{figure*}

 \begin{figure*} 
  \centering
  \subfigure
  []{\includegraphics[width=8.5cm,height=5.8cm]{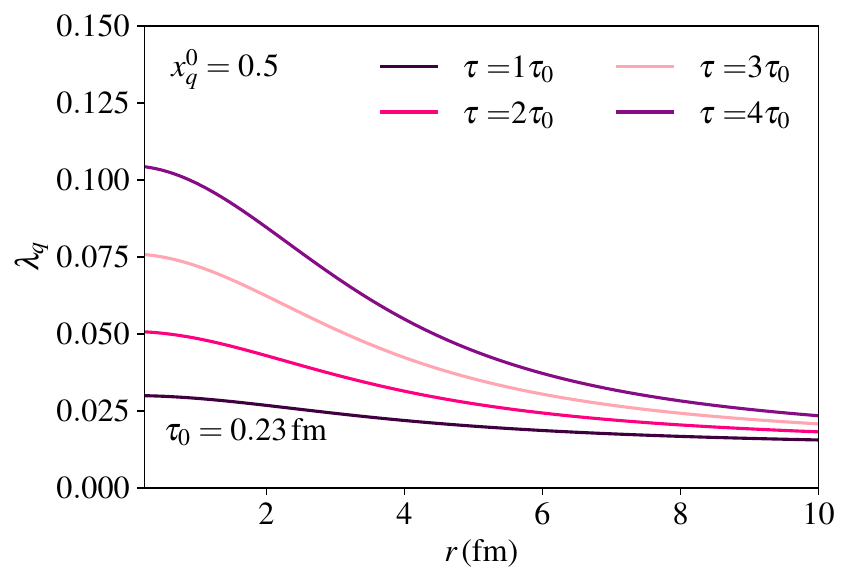}\label{fig:lq-r-mu05}} 
  \subfigure[]{\includegraphics[width=8.5cm,height=5.8cm]{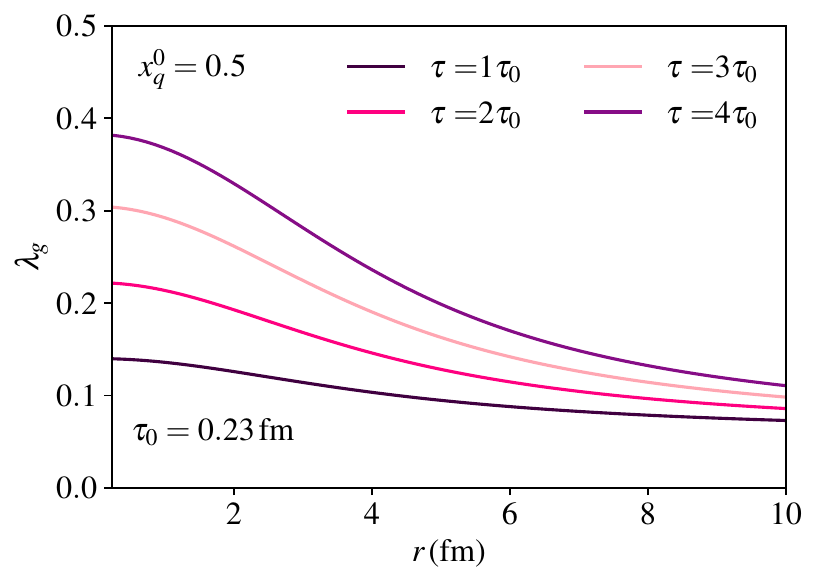}\label{fig:lg-r-mu05} 
  } 
  \caption{\textbf{(a)} Quark and \textbf{(b)} gluon fugacities from a chemically equilibrating QGP medium as a function of $r$, plotted at different snapshots of proper time. Initial value of temperature is fixed as $T_0 = 0.83$ GeV.}
  \label{fig:fugacity-r}
  \end{figure*}

\begin{figure*} 
  \centering
  \subfigure
  []{\includegraphics[width=0.45\linewidth]{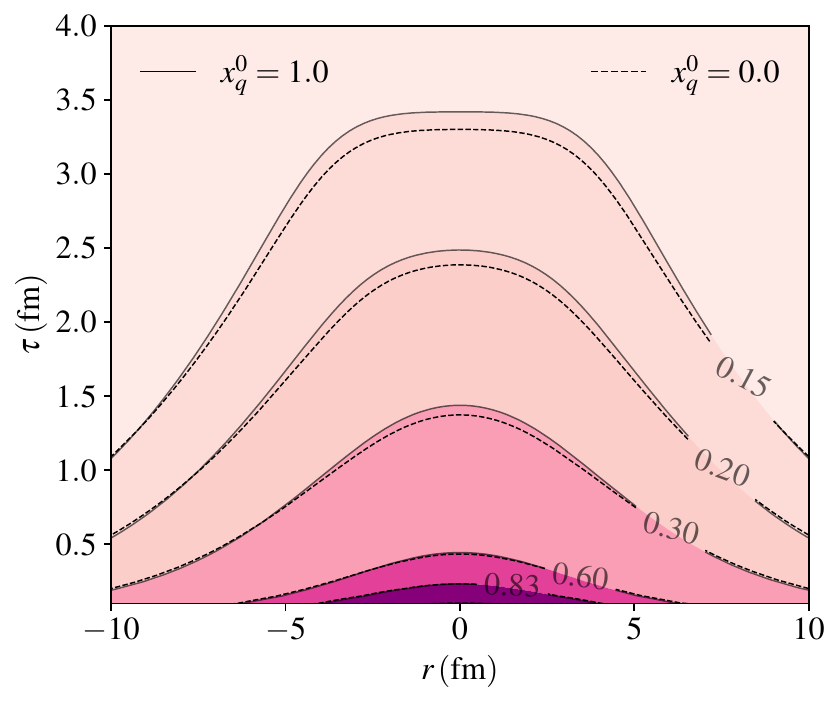}\label{fig:T-contour}} \quad
  \subfigure[]{\includegraphics[width=0.45\linewidth]{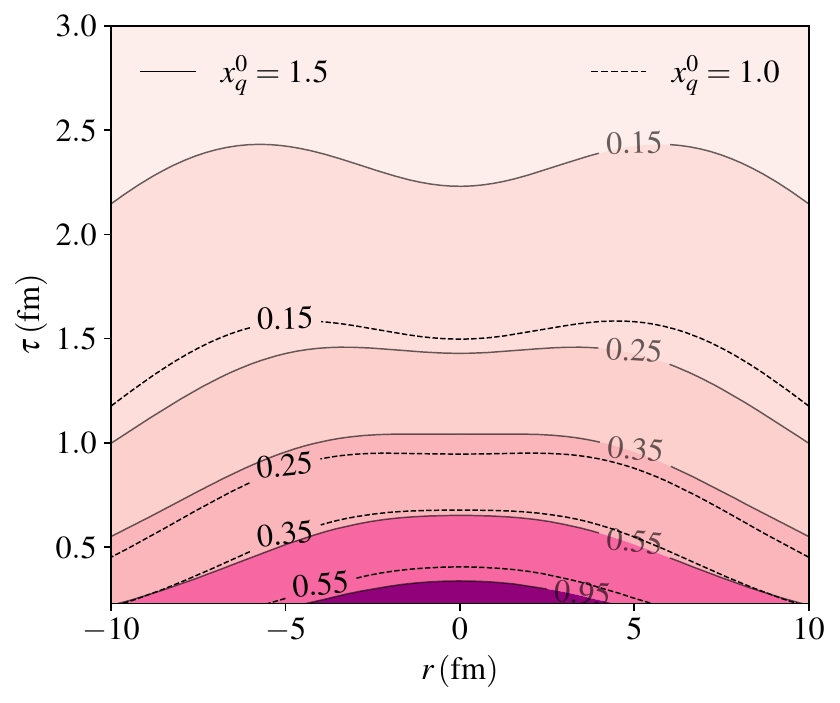}\label{fig:mu-contour} 
  } 
  \caption{\textbf{(a)} Temperature ($T$ in GeV) and \textbf{(b)} chemical potential ($\mu_q$ in GeV) evolution profiles for the chemically equilibrating QGP within the Gubser flow, for different values of $x_q^0$. The initial fugacity values are $\lambda_g^0 = 0.14$ and $\lambda_q^0=0.03$. } 
  \label{fig:evolution-T-xq}
  \end{figure*}

  \begin{figure*} 
  \centering
  \subfigure
  []{\includegraphics[width=0.45\linewidth]{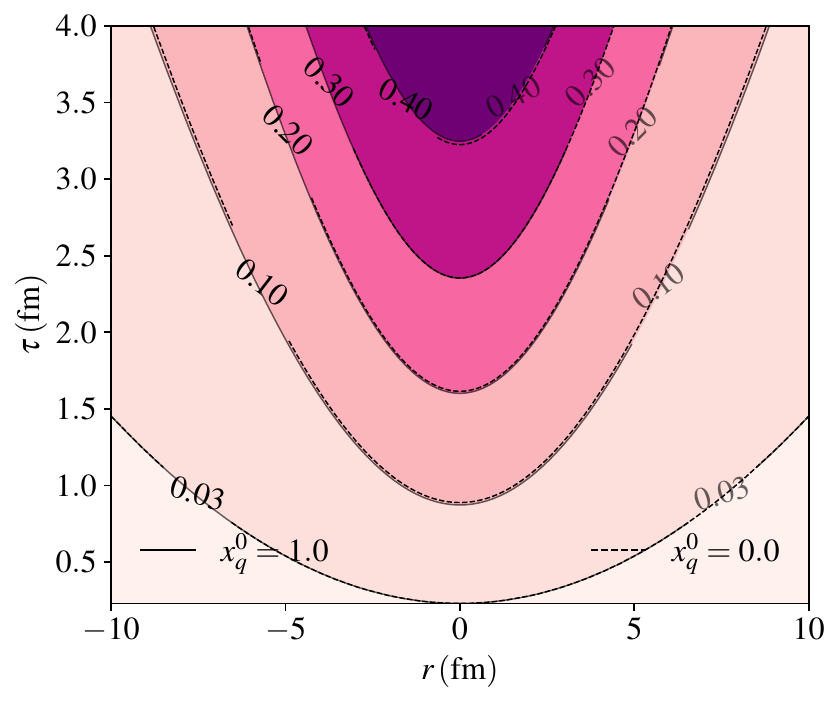}\label{fig:lq-contour}} \quad
  \subfigure[]{\includegraphics[width=0.45\linewidth]{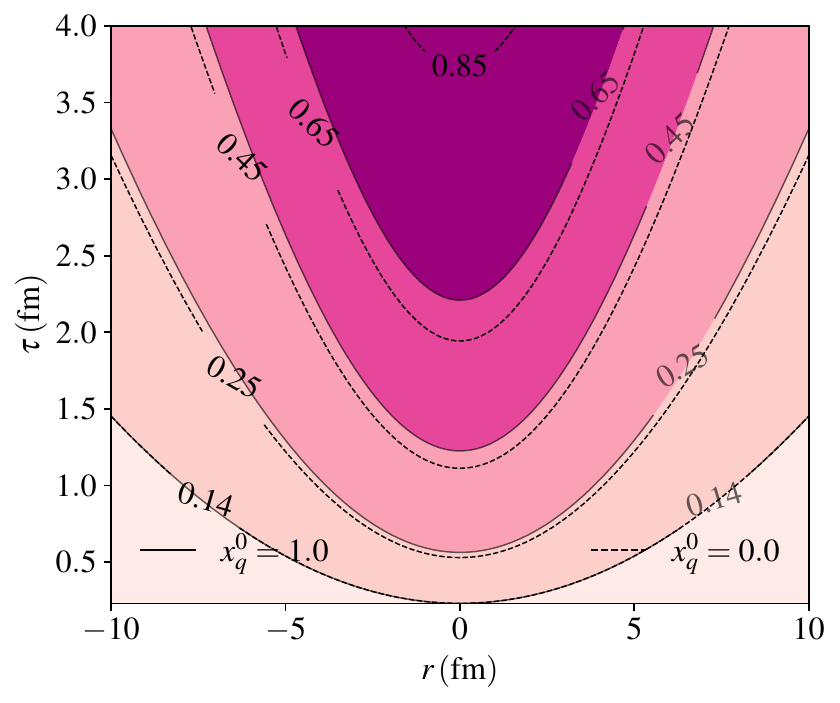}\label{fig:lg-contour} 
  } 
  \caption{\textbf{(a)} Quark and \textbf{(b)} gluon fugacity evolution contours for the chemically equilibrating QGP in the Gubser flow. Dashed lines indicate contours corresponding to $x_q^0=0$ and solid lines denote $x_q^0=1$. Initial value of temperature is fixed as $T_0 = 0.83$ GeV.}
  \label{fig:evolution-lg-lq}
  \end{figure*}

  \begin{figure}
      \centering      \includegraphics[width=0.5\textwidth]{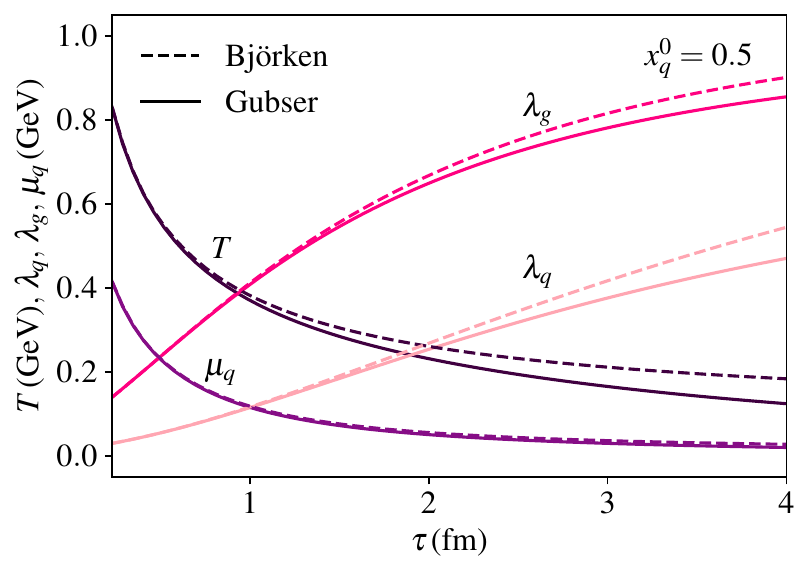}
      \caption{Evolution of temperature, chemical potential and parton fugacities in Gubser ($r=0$) and Bj\"orken geometry, with $x_q^0 = 0.5$.}
      \label{fig:T-gubser-bjorken-compare}
\end{figure}
   \begin{figure*} 
  \centering
  \subfigure
  []{\includegraphics[width=0.49\textwidth]{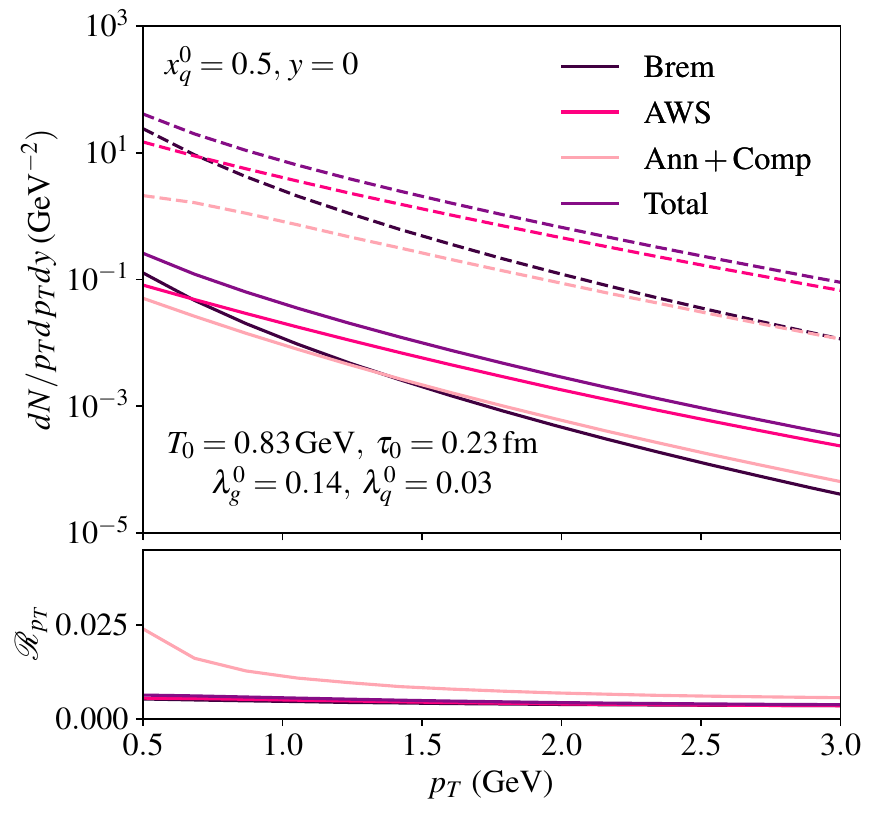}\label{fig:ph-xq05}} 
  \subfigure[]{\includegraphics[width=0.49\textwidth]{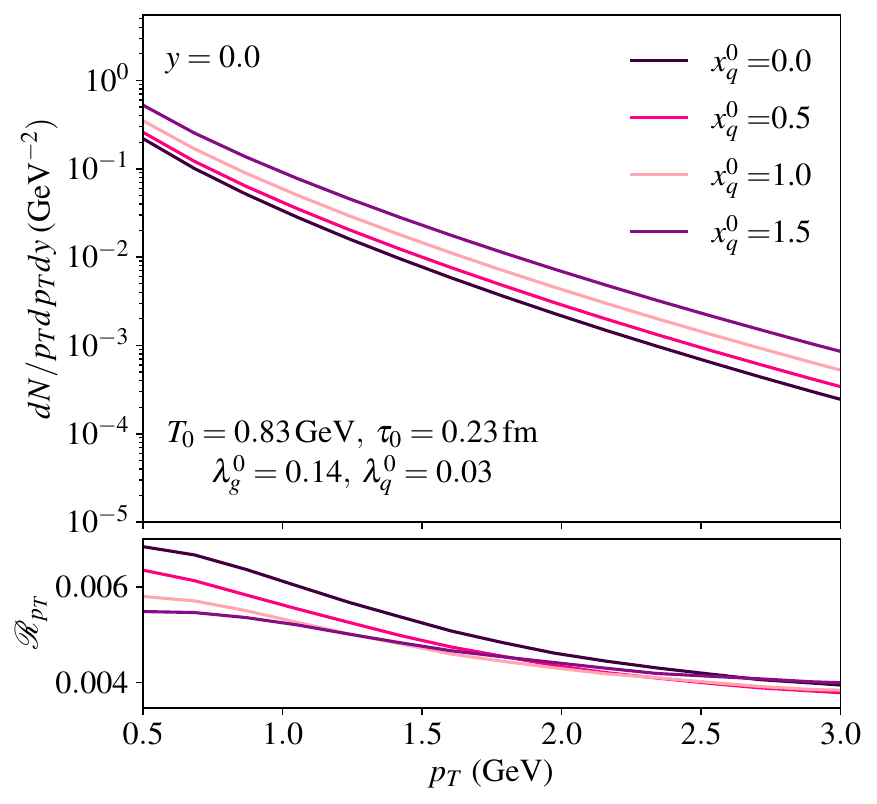}\label{fig:ph-total} 
  } 
  \caption{(\textbf{a}) Photon spectra from chemically equilibrating (solid lines) and fully saturated (dashed lines) QGP with initial finite baryon density, $x_q^0=\mu_q^0/T_0=0.5$. Individual contributions from the processes and total yield are shown. (\textbf{b}) Total hard photon spectra from chemically equilibrating QGP for different values of $x_q^0$. In both the figures, initial values of parton fugacities are $\lambda_g^0=0.14$ and $\lambda_q^0 = 0.03$. The lower panels depict the ratio $\mathscr{R}_{p_T}$ of the spectra obtained within chemically equilibrating medium to that of a fully equilibrated one.}
  \end{figure*}

  \begin{figure}
      \centering
      \includegraphics[width=0.5\textwidth]{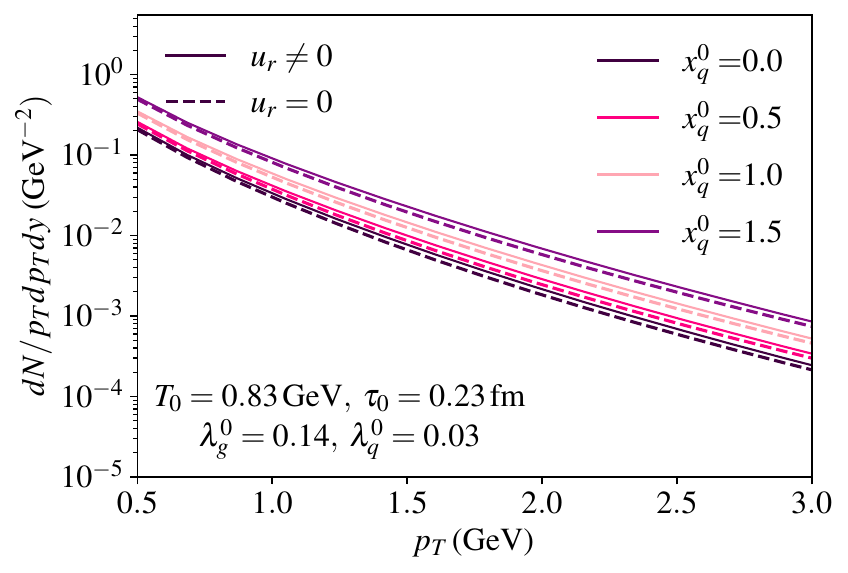}
      \caption{Total thermal photon spectra from chemically equilibrating QGP by varying $x_q^0$, with (solid lines) and without (dashed lines) the Doppler shift. 
      }
      \label{fig:ph-ur}
  \end{figure}

\par
Now, we study the evolution of parton fugacities under Gubser flow in presence of finite chemical potential (Fig.~\ref{fig:evo-rho-2}). We see that both the quark and gluon fugacities increase with $\rho$ and become constant ($\lambdahat_i \approx 1$) for higher $\rho$ (or larger proper time), indicating that the matter eventually reaches chemical equilibrium, given sufficient time to evolve. From both Figs.~\ref{fig:lq-rho} and \ref{fig:lg-rho}, we find that the presence of chemical potential slows down the equilibration of quarks and gluons. Further, we see that the effect of chemical potential is more prominent for gluon fugacity. This is because $x_q$ directly affects the gluon production rate $R_3/T$ (Eq.~\eqref{Eq:rate-R3}); whereas the effect of $x_q$ can only be seen at larger $\lambda_q$ for the quark production rate $R_2/T$~\cite{Dutta:1999hj}.
\par
Next, in order to understand the behaviour of entropy, we plot the quantity $\hat{s}(\rho)\cosh^2\rho$ scaled to its initial value at $\rho = \rho_0$ in Fig.~\ref{fig:s-rho}. As expected, we observe that the entropy increases initially from unity implying that the system is undergoing chemical equilibration. As $\rho$ becomes more positive, entropy becomes constant because the matter attains chemical equilibrium. Further, we note that the entropy of the system decreases with increase in $x_q^0$.
\par
Now, we proceed to study the QGP evolution in Milne coordinates - $i.e.$; proper time $\tau$ and radial coordinate $r$, by using the transformations in Eq.~\eqref{Eq:transformation}. In Fig.~\ref{fig:gub-traj}, we depict the flow trajectories of chemically equilibrating QGP in the $\mu_q$-$T$ plane for different values of $x_q^0$, by fixing the radial coordinate $r$. We also compare the evolution with the equilibrium case $\lambda_i =1$. The markers denote proper time $\tau$ with an equal interval of 0.4 fm. Also, in all the cases, the evolution ends at the freeze-out condition given by $\varepsilon_f=1$ GeV/fm$^{3}$. At the fireball center $r=0$, the evolution begins at the initial values $T_0 = 0.83$ GeV, $\lambda_g^0 = 0.14$, $\lambda_q^0 = 0.03$ and $\tau_0 = 0.23$ fm, for all $x_q^0$. We can see from Fig.~\ref{fig:gub-traj-r0} that in presence of chemical non-equilibrium, the system evolves rapidly compared to the equilibrium scenario and approaches the freeze-out surface very fast. The QGP phase is observed to be shorter in a chemically equilibrating system. It is expected since energy is consumed also for equilibration other than the expansion. Further, while increasing the initial quark chemical potential $x_q^0$, the system remains in the QGP phase for longer time. The freeze-out proper time ($\tau_f$) values in a chemically equilibrating (equilibrated) QGP for $x_q^0 = 0.5$ and 1 are 2.69 (5.53) fm and 2.79 (5.79) fm, respectively at $r=0$. It can be seen that at freeze-out, the chemical potential reduces to a lower value and temperature approaches a higher value compared to the $\lambda_i =1$ case. From Fig.~\ref{fig:gub-traj-r2}, we observe that increase in $r$ affects the flow trajectories of the QGP. At $r=2$ fm, unlike the fireball center, the evolution of the non-equilibrium and equilibrated QGP matter begins at lower values of $T$ and $\mu_q$  
and approaches the freeze-out curve faster. We note that for a chemically equilibrating system, regions away from the center of the fireball freeze out faster. The values of $\tau_f$ at $r=2$ fm are 2.48 fm and 2.59 fm for $x_q^0 = 0.5$ and 1, respectively. Moreover, in the non-equilibrium scenario, the evolution of temperature and chemical potential are coupled and the matter evolves through a curved path; while in the equilibrated system, they decouple from each other resulting in straight-line trajectories.
\par
In Fig.~\ref{fig:tau-r}, we illustrate the variation of freeze-out proper time $\tau_f$ of a chemically equilibrating QGP as a function of $r$, by varying $x_q^0$. We note that the initial system size corresponding to the Gubser parameter $q$ is 4.3 fm. We plot the $\tau_f$ corresponding to equilibrium scenarios as well. For the non-equilibrium case, $\tau_f$ has a maximum value at the center of fireball and they decrease with increase in $r$. 
This decrease in $\tau_f$ with $r$ is slow up to $r\sim 2$ fm, and later it falls rapidly as we move away from the center. Further, we see that $\tau_f$ values increase in presence of finite chemical potential and are higher for larger $x_q^0$ values implying longer QGP phase. It can also be seen that the $\tau_f$ values are almost equal at high $r$. More importantly, we find the freeze-out timescales of equilibrium QGP to be almost double of the chemical non-equilibrium case, for all $x_q^0$. This is because, in presence of chemical non-equilibrium, the system utilizes more energy (compared to the equilibrium case) for the parton equilibration dynamics, resulting in faster expansion and hence shorter QGP phase. We also find that, for the equilibrated case ($\lambda_i =1$), $\tau_f$ remains almost constant up to $r \sim$ 4 fm for small $x_q^0$; while for larger $x_q^0$, $\tau_f$ first increases and then decreases with $r$, as seen in Ref.~\cite{Ingles:2025yrv}. We find that the chemical equilibration process substantially alter the shape of the $\tau_f$-$r$ curve compared to the equilibrium scenario. 
\par
In Fig.~\ref{fig:evo-compare}, we plot the evolution of temperature ($T$), quark and gluon fugacities ($\lambda_q,\lambda_g$), and quark chemical potential $x_q=\mu_q/T$ for $r=0$ fm and $r=2$ fm, as a function of proper time. We consider two initial values of chemical potential $x_q^0:$ 0 and 1. We observe that, though the effect of $x_q^0$ is small on temperature evolution, the cooling of $T$ slows down in presence of finite density. 
Also, the evolution of $x_q$ is similar to that of temperature; $x_q$ decreases with increase in $\tau$. 
It can be seen that quark chemical potential affects the evolution of gluon fugacity considerably. We find that including chemical potential results in slower equilibration of gluon fugacity; while the equilibration rate of quarks increases slightly. The values of fugacities are obtained as $\lambda_q = 0.34\,(0.35)$ and $\lambda_g = 0.76\,(0.72)$ at the freeze-out times $\tau_f \sim $ 2.66 (2.79) fm for $x_q^0 = 0\,(1.0)$. This behavior of $\lambda_q$ and $\lambda_g$ in presence of finite density is similar to the trend found for the analyses within one-dimensional model~\cite{Dutta:1999hj,Dutta:1999dy}. This opposite behavior of quark and gluon fugacities arises because of the dynamical expansion of QGP at finite density. Further, we note that in regions away from the center, the evolution begins at smaller values of $T$ and $\lambda_i$; while $x_q$ starts from a higher value than the initial value at the center. Therefore, in those regions, system takes lower time to reach freeze-out and remains in a more non-equilibrated state. At $r=2$ fm, we obtain $\tau_f \sim 2.44 \,(2.59)$ fm for $x_q^0 = 0\,(1.0)$; and the corresponding values of fugacities are $\lambda_q =0.26\,(0.28)$ and $\lambda_g=0.68\, (0.65)$.
\par
The behavior of fugacities can be better understood from the evolution of quark and gluon equilibration rates ($R_2$ and $R_3$). In Fig.~\ref{fig:R3-R2-tau}, we show the evolution of $R_2$ and $R_3$ scaled to the QGP temperature $T$, for different values of $x_q^0$. We plot the rates for two values of radial coordinate, $r=0$ fm and $r=2$ fm. We find that the rate of gluon equilibration ($R_3$) starts from a higher value compared to that of quark ($R_2$), and initially increases slightly with $\tau$. This is because, at early times, in the gluon rich medium, the $gg \leftrightarrow ggg$ processes may occur abundantly. With increase in $\tau$, $R_3$ then decreases implying the production of quarks via the process $gg \rightarrow q\bar{q}$, albeit slowly, resulting in the increment of the corresponding rate $R_2$. As $\lambda_g (\lambda_q) \rightarrow 1$, $R_3 (R_2)$ saturates, as expected. Further we find that at finite density, $R_2$ increases and $R_3$ decreases slowly compared to the vanishing chemical potential case. 
\par
In Figs.~\ref{fig:T-mu-r} and \ref{fig:fugacity-r}, we depict the evolution at different snapshots of proper time as a function of $r$, by fixing $x_q^0 = 0.5$, for both chemically equilibrating (solid lines) and fully equilibrated (dashed lines) QGP. We observe that both temperature (Fig.~\ref{fig:T-r-mu05}) and chemical potential (Fig.~\ref{fig:mu-r-mu05}) decrease with increase in $r$ and $\tau$. Moreover, we see that the curves flatten over $r$ with increase in proper time. At each $r$, both $T$ and $\mu_q$ decrease slowly for matter in chemical equilibrium. 
Now, from Fig.~\ref{fig:fugacity-r}, we find that the fugacities decrease with $r$ and a decrease in $\tau$ results in flattening of the curves over $r$. Both $\lambda_q$ and $\lambda_g$ increase with $\tau$ for all $r$. At $r=0$ fm, $\lambda_i$ increases rapidly implying that the center of the fireball attains maximum chemical equilibration owing to the larger $\tau_f$ values. At larger $r$, plasma is dilute and cold and therefore, chemical equilibration is difficult to attain in this region before freeze-out. 
\par
Next, we study the evolution of $T$, $\mu_q$ and $\lambda_i$ in Milne coordinates ($r$ and $\tau$) by varying the initial density $x_q^0$ in Figs.~\ref{fig:evolution-T-xq} and \ref{fig:evolution-lg-lq}. 
The temperature contours are plotted for $x_q^0 =0$ and 1.0. From Fig.~\ref{fig:T-contour}, we find that presence of chemical potential alters the $T$ contours, especially for large $\tau$ and small $r$ values. As discussed before, we see that the temperature of QGP cools down slowly at finite chemical potential. At large $r$, the effect of finite density is found to be minimal; whereas, at central region it is more apparent. At $r=0$, the temperature values are reached at larger values of $\tau$ for $x_q^0 = 1$, compared to zero-density case. Similarly, in Fig.~\ref{fig:mu-contour}, we show the $\mu_q$ contours for $x_q^0 = 1.0$ and 1.5. We find that variation in the initial value $x_q^0$ affects the evolution profiles of $\mu_q$ considerably. The fall of $\mu_q$ profile is found to be slower for larger initial baryon density. 
Now, coming to Fig.~\ref{fig:evolution-lg-lq}, we depict the quark and gluon fugacity contours for $x_q^0 = 0$ and $x_q^0 =1.0$. As seen earlier, the effect of finite density is minimal on $\lambda_q$ while it has appreciable effect on $\lambda_g$. The rate of chemical equilibration is found to be faster at $r=0$ fm, while equilibration at outer regions of fireball lags behind. Also, we observe that for any $r$, the gluon equilibration becomes slow at finite density. 
\par
Further, we compare the evolution of the quantities in 1-D Bj\"orken flow with that obtained in our analysis using Gubser geometry (at $r=0$) in Fig.~\ref{fig:T-gubser-bjorken-compare}. For the one-dimensional description of the chemically equilibrating QGP at finite density, we use the evolution equations described by Eqs.~\eqref{Eq:evo-bjorken}. With the initial conditions remaining the same, one can see that the presence of radial velocity, the temperature and chemical potential profiles fall faster as time evolves, compared to pure longitudinal expansion of the Bj\"orken scenario. For $x_q^0 = 0.5$, the freeze-out time taken by the system under Gubser flow ($r=0$) is $\tau_f \sim 2.69$ fm, while within Bj\"orken expansion, the system takes $\sim 5.2$ fm. On the other hand, chemical equilibration rate is higher for Bj\"orken case, since system posses more energy to drive equilibration in the absence of transverse flow. The overestimation of the chemical equilibration in Bj\"orken flow is higher for the quarks $\sim 15.7\%$ at $\tau=4.0$ fm; whereas it is $5.4\%$ for gluons. Now, with increment in $x_q^0$, we observe that the overestimation of gluon equilibration increases, while that of quark equilibration decreases marginally. With $x_q^0 = 1.5$, the overestimation due to Bj\"orken flow becomes $\sim 6.5\%$ for $\lambda_g$; while it becomes $\sim 15.3\%$ for $\lambda_q$, for $\tau = 4$ fm. 
\par
Next, we look into the hard thermal photon spectra from chemically equilibrating QGP under Gubser geometry, at finite chemical potential. 
Initial conditions remain the same: $\tau_0=0.23$ fm, $T_0=0.83$ GeV and $\lambda_q=0.03$ and $\lambda_g=0.14$ with varied values of the chemical potential $x_q^0$. 
We evolve the temperature, chemical potential and fugacities of the matter till the freeze-out condition (Eq.~\eqref{Eq:freeze_out}) is satisfied. Employing the obtained $T$, $x_q$, $\lambda_i$ profiles in the rate terms $\left(2E \frac{dR}{d^3 {\bf p}}\right)_i$, we numerically integrate the yield expression (Eq.~\eqref{Eq:ph-yield}), over the space-time region constrained by the freeze-out condition, to evaluate the individual and total photon yields.
Further, we evaluate all the spectra at midrapidity region $y=0$ alone. 
We note that the expanding QGP at different radial positions reaches the freeze-out surface at different proper times, due to the radial dependence of the freeze-out time, $\tau_f(r)$ (see Fig.~\ref{fig:tau-r}). This is similar to the scenario considered in Ref.~\cite{Naik:2025pjt}, 
where QGP evolution in Gubser flow was restricted using a freeze-out temperature.
\par 
In Fig.~\ref{fig:ph-xq05}, we illustrate the different contributions to the total photon spectra at $x_q^0 = 0.5$. The spectra obtained from a fully equilibrated QGP is also plotted for comparison. 
We find that the presence of chemical non-equilibrium leads to the suppression of the total photon spectra over the entire $p_T$ range considered. The suppression in the total spectra is almost $100\%$ throughout $p_T$ implying a reduction by two orders of magnitude compared to the equilibrium case. The lower panel shows the ratio $\mathscr{R}_{p_T}$ of the photon spectra from chemically undersaturated medium to that of a fully equilibrated one undergoing Gubser expansion. The decrease of ratio with $p_T$ indicates that high $p_T$ photons are more suppressed due to strong initial chemical non-equilibrium. We observe that inclusion of the 2-loop processes (Brem and AWS) has a significant impact on the yield. For QGP under chemical non-equilibrium, the Bremsstrahlung process contributes more to the low $p_T$ region; whereas, the AWS process dominates otherwise. 
Further, we find that 1-loop contribution (Ann $+$ Comp) becomes more than that of Brem for $p_T > 1.5$ GeV, although AWS contribution in this regime is an order of magnitude higher. While, for the fully saturated medium, the 1-loop processes has the lowest contribution throughout the $p_T$ range. We also find that when we increase $x_q^0$, the suppression in the 1-loop contribution due to chemical non-equilibrium decreases. The overall behavior of the spectra observed is qualitatively same for all values of $x_q^0$ considered. 
We note that our results are different from that of Ref.~\cite{Mustafa:2000sg}, in which photon spectra from a chemically equilibrating medium at vanishing chemical potential has been found. However, we note that the authors had neglected the quantum effects while calculating the photon rates. Further, the transverse flow profile superimposed to the hydrodynamical evolution was different along with the initial temperature profiles used. We note that, we follow a self-consistent analysis of hydrodynamics with transverse flow while obtaining the evolution profiles and subsequent spectra.  
\par
In Fig.~\ref{fig:ph-total}, we plot the total photon yield from chemically equilibrating system arising from both 1-loop and 2-loop processes considered, for different values of $x_q^0$. We find that increasing $x_q^0$ results in the enhancement of the particle spectra over the entire $p_T$ range. This is in line with the behaviour seen from Fig.~\ref{fig:rate-total}. We find that increasing $x_q^0$ results in the enhancement of the particle spectra over the entire $p_T$ range. A higher $x_q^0$ implies more quarks than antiquarks in the system, giving more number of photons through relevant processes. An increase in $x_q^0$ results in larger values of freeze-out proper time $\tau_f$ as well, contributing to the enhancement of particle spectra. Also, we find that the impact of chemical potential is more at high values of $p_T$. The enhancement in the spectra due to chemical potential is $\sim 18\%$ at $p_T = 0.5$ GeV, whereas the increment is $\sim 39\%$ at $p_T =3$ GeV for $x_q^0 = 0.5$. We note that this trend is same as that observed for a chemically equilibrating QGP within 1-D Bj\"orken flow~\cite{Long:2005cn}. From both Figs.~\ref{fig:ph-xq05} and \ref{fig:ph-total}, we find that presence of chemical non-equilibrium dominantly suppresses the spectra, compared to the well known enhancement of the equilibrium photon spectra due to the quark chemical potential. The lower panels of the figures reveal that the spectral slope decreases with $p_T$ indicating that the suppression due to chemical non-equilibrium is large for high $p_T$ photons emitted during the initial hot stages of expansion.
\par
In Fig.~\ref{fig:ph-ur}, we study the effect of transverse flow velocity on thermal photon spectra from chemically equilibrating QGP medium at finite chemical potential. To compute the spectra without the Doppler shift, we neglect the term $u_r$ in Eq.~\eqref{Eq:udotp}. We find that the effect of transverse expansion is very minimal on the photon spectra, especially at low $p_T$, for any value of quark chemical potential considered. The enhancement in the spectra due to Doppler shift is less than $\sim$19\% throughout $p_T$ range, for all values of $x_q^0$ considered. Also, we note that the effect of Doppler shift is less than that obtained in Ref.~\cite{Paquet:2023bdx}, where thermal photon spectra from equilibrated QGP at vanishing density has been studied within Gubser flow. We note that though the Doppler shift is minimal over the entire evolution, its effect can be largely seen while considering the spectra only from later times of expansion~\cite{Paquet:2023bdx}. We also find that the effect of Doppler shift is more on thermal photon spectra from equilibrated QGP. The enhancement in the spectra is $\sim 19 \,(27)$\% for $x_q^0 = 0$ and $\sim 23 \,(30)$\% for $x_q^0 = 1.5$ at $p_T = 0.5\,(3)$ GeV. 
\par
\begin{figure}
    \centering
\includegraphics[width=0.5\textwidth]{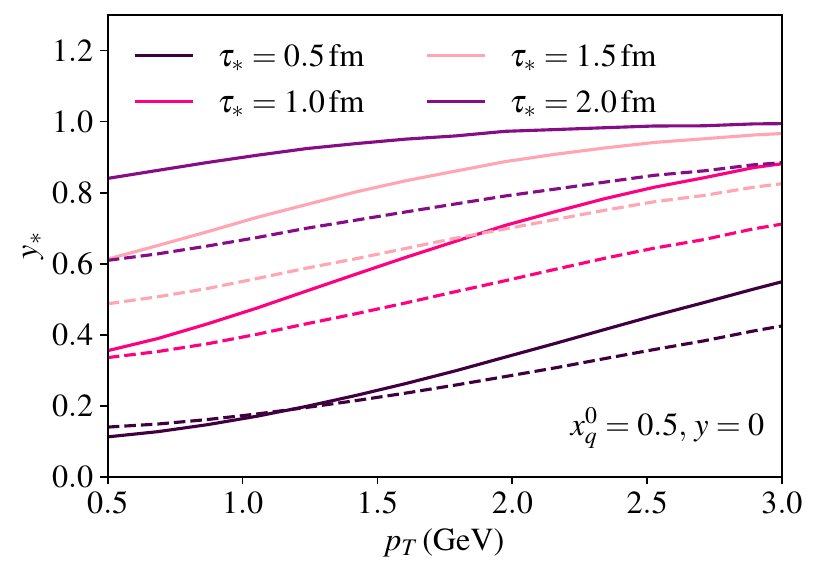}
    \caption{Cumulative early-time fraction of the hard thermal photon yield $(\mathcal{Y}_*)$ as a function of $p_T$ for different values of $\tau_*$, with $x_q^0 = 0.5$. The contribution from the non-equilibrium (equilibrium) QGP is denoted by solid (dashed) curves. The initial conditions are taken same as in Fig.~\ref{fig:ph-ur}.}
    \label{fig:tomo}
\end{figure}

\begin{figure}
    \centering
    \includegraphics[width=0.5\textwidth]{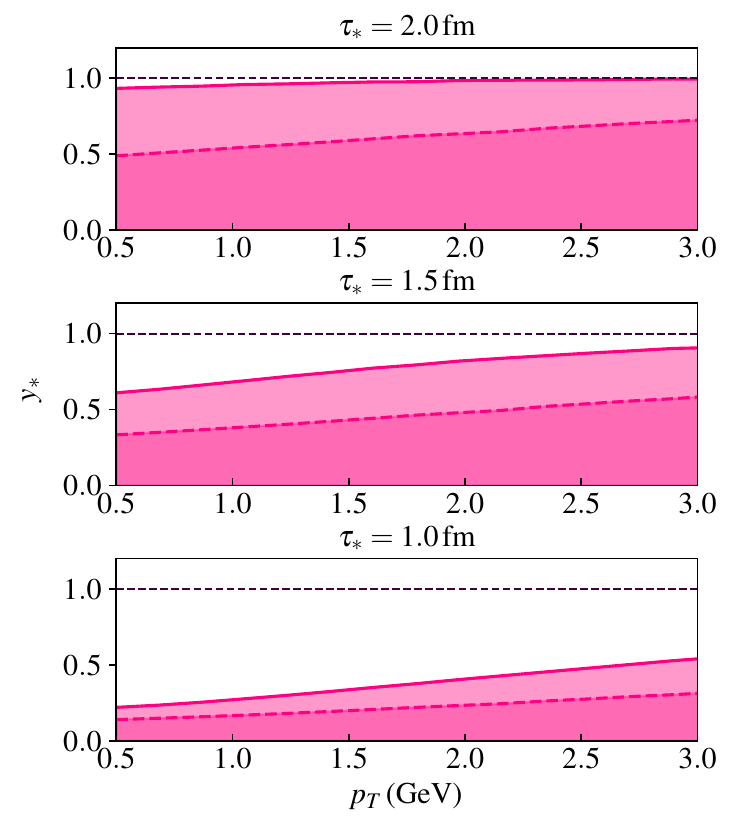}
    \caption{Cumulative early-time fraction of the hard thermal photon yield $(\mathcal{Y}_*)$ for $x_q^0 = 0.5$ as a function of $p_T$ with the initial conditions $T_0 = 0.57$ GeV, $\lambda_g^0=0.08$ and $\lambda_q^0 = 0.02$ at $\tau_0=0.7$ fm. The solid lines denote the contribution from non-equilibrium QGP and dashed lines represent the equilibrium cases.}
    \label{fig:yi-05-fill2}
\end{figure}

 
In order to quantify the temporal distribution of hard thermal photon emission from the finite density system undergoing chemical equilibration, we define the cumulative early–time fraction of the yield
\begin{equation}
\mathcal{Y}_*(p_T;\tau_*) = 
\Bigg(\frac{dN}{p_T dp_T dy}\Bigg)_{\tau<\tau_*} \Bigg/
{\Bigg(\frac{dN}{p_T dp_T dy}\Bigg)}\,;
\end{equation}
which measures the fraction of photons of transverse momentum $p_T$ emitted before a given proper time~$\tau_*$. 
The dependence of $\mathcal{Y}_*$ on $\tau_*$ acts as a tomography of the QGP evolution. In Fig.~\ref{fig:tomo}, we depict the early-time fraction of the photon yield $\mathcal{Y}_*$ from chemically equilibrating QGP (solid lines) at $y=0$ for different values of the proper time $\tau_*$, with the initial conditions $T_0=0.83$ GeV, $\tau_0=0.23$ fm for $x_q^0=0.5$. For reference, we plot the equilibrium yield scenario as well (dotted lines). One can see that the high energy photons are produced predominantly in the initial time period up to $\tau_*=1.0$ fm itself in the non-equilibrium scenario; whereas it takes $\sim 2.0$ fm
to reach same production in the absence of chemical non-equilibrium. One should note that, even though the total yield is suppressed in presence of non-equilibrium, the fractional production of high energy photons are more in this scenario. We report that the behaviour of $\mathcal{Y}_\ast$ remains qualitatively same for $x_q^0 = 0$. We perform the same analysis
for the initial conditions used by the authors of Ref. [96] as well.
With $T_0 = 0.57$ GeV, $\lambda_g^0 = 0.08$, $\lambda_q^0 = 0.02$ at $\tau_0 = 0.7$ fm, the cumulative early time fraction is calculated and displayed in Fig.~\ref{fig:yi-05-fill2}. One can see that, $\mathcal{Y}_\ast$ at large $p_T$ remains distinctly high for non-equilibrium case for all the $\tau_\ast$ values, as seen before. This exceptionally large early time fraction of the high $p_T$ photons, we believe, is the characteristic indicator of the chemically equilibrating system distinguishing it from the equilibrated one. Moreover, the shape of $\mathcal{Y}_*$ curve becomes distinctly different for the non-equilibrium case. This is because of the equilibrating dynamics and consequent faster cooling of the chemically equilibrating plasma. With $x_q^0 = 0.5$ and the first (second) initial condition considered, the system takes only $\tau_f = 2.69$ (2.26) fm to approach the freeze-out surface; whereas, it takes 5.53 (5.58) fm for the equilibrated system, at $r=0$ fm. 

\begin{figure}
    \centering
    \includegraphics[width=0.5\textwidth]{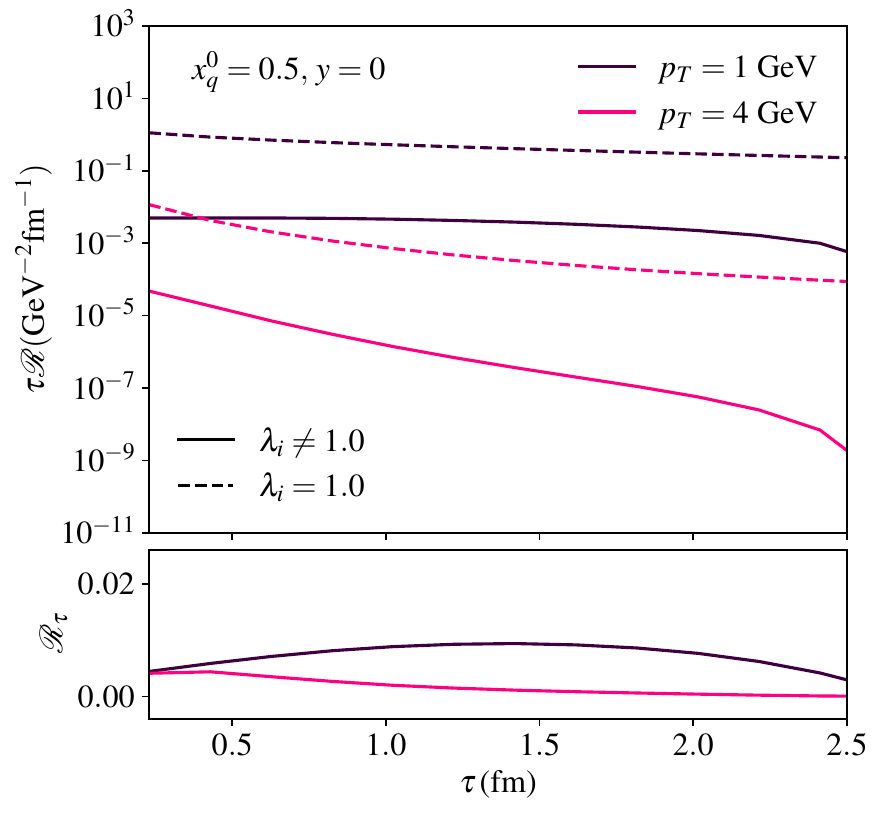}
    \caption{Instantaneous photon rates from a chemically equilibrating QGP to that of a fully equilibrated one within Gubser flow, for $x_q^0 = 0.5$ and by varying $p_T$. The ratio of the rates is depicted in the lower panel.}
    \label{fig:ph-xq05-tau}
\end{figure}

Finally, we plot the instantaneous photon emission rates $\tau \mathscr{R}(p_T,\tau)$, using Eq.~\eqref{Eq:ph-yield}, from a chemically equilibrating system, by varying the value of $p_T$, for $x_q^0 = 0.5$. In Fig.~\ref{fig:ph-xq05-tau}, we depict the instantaneous photon emission profiles within Gubser expansion for $p_T =1$ and $4$ GeV. We observe that the photon emission at each proper time is suppressed in presence of chemical non-equilibrium. The intermediate emission profiles from a chemically equilibrating system show a qualitatively different trend compared to the fully equilibrated case. This trend is more apparent from the lower panel, in which we show the ratio $\mathscr{R}_\tau$ of rates from a chemically undersaturated medium to that of a fully equilibrated one. We observe that the ratio begins from a small value at early time for $p_T = 1$ GeV, implying a strong suppression in the photon production during the hot initial stage, due to the undersaturated quark and gluon densities. As the system evolves, matter begins to approach chemical equilibrium ($\lambda_{g,q}$ increases), resulting in an increase in the production of photons leading to a large $\mathscr{R}_\tau$. However, when $\tau>1.5$ fm, the ratio decreases again because the rapid cooling of the medium dominates. We observe a qualitatively different behaviour for high $p_T$ photons. For $p_T=4$ GeV, the ratio begins from an even smaller value and has a small peak around $\tau =1.5$ fm, and then decreases monotonously with $\tau$. This implies that the high $p_T$ photons are predominantly produced during the early hot stages of evolution. Since the initial quark density is strongly undersaturated, the production of high $p_T$ photons is strongly suppressed. It must be noted that, the ratio $\mathscr{R}_\tau$ reflects the interplay between chemical equilibration and cooling of the plasma. This reveals that the chemical non-equilibrium does not merely suppress the photon spectra, but rather modifies the temporal structure of photon emission. 

\section{Conclusion} \label{Sec:conclusion}
In conclusion, we have studied the chemical equilibration of the hot and dense quark-gluon plasma created in heavy-ion collisions in presence of transverse flow, consistently modeled using conformal Gubser solutions. The chemical non-equilibrium is introduced to the system  by modifying the constituent particle distribution functions via fugacity parameters. By writing down the master rate equations in presence of finite chemical potential, we numerically solved the hydrodynamical equations with relevant initial conditions to obtain the temperature, chemical potential and fugacity evolution profiles of the system in both Gubser and Milne coordinates. We performed a detailed analysis of the interplay of chemical equilibration and finite density dynamics in the system with different initial quark chemical potential values. 
\par 
By tracing the evolution of the system in $\mu_q$-$T$ plane till freeze-out, we find that the chemically undersaturated QGP evolve swiftly through a curved trajectory, compared to the slower straight-line evolution of the equilibrated system. We also observed that in regions far away from the fireball center ($r=0$), the system evolves and approaches the freeze-out surface faster. We found that, for an equilibrating system at the freeze-out, proper time has a distinct dependency on the radial coordinate, different from a chemically saturated system. 
We find that system remains chemically non-equilibrated ($\lambda_i<1$) with gluons reaching more saturation compared to quarks, especially at large $r$. The center of the fireball is found to attain maximum equilibration; while regions far away from the center remain in a more non-equilibrium state. We observed that inclusion of finite density slows down the evolution of the plasma. It delays the gluon equilibration and slightly enhances the quark equilibration of the system. We find that the transverse flow plays an important role in the chemical equilibration and together with finite density dynamics, it drives the system further away from saturation. 
\par 
Further, we calculated the hard thermal photon production rates with the inclusion of chemical non-equilibrium for all the relevant processes in hot and dense QGP. We observe that the effect of non-equilibrium distribution functions is to suppress the rates for all the processes, with two-loop processes dominating the contribution. We also found a power law scaling for the total rate with gluon fugacity, for different baryon densities.  
Then, by convoluting the rates in the space-time history of the collision within Gubser model, we obtained the hard thermal photon yields from the system. It was found that the 2-loop processes have a significant impact on the yield, with 
annihilation with scattering (AWS) process dominating overall, except for the small $p_T$ regime where Bremsstrahlung becomes prominent. The contribution of 1-loop processes increases with chemical potential and at high $p_T$ it overtakes the Bremsstrahlung contribution, but remains significantly low compared to the other 2-loop process of AWS. We observe that with increase in initial quark chemical potential, the total photon yield increases, while the overall behavior of the individual contributions remains the same. Further, by analysing the interplay of chemical non-equilibrium and finite density dynamics on the spectra and it was found that the effect of chemical equilibration is more prominent, resulting in the reduction of the spectra by two orders of magnitude compared to the equilibrium case. We also looked into the impact of Doppler shift on the thermal photon spectra and observed that its effect is found to be minimal. 
\par
Furthermore, we quantified the early-time fraction of photon yield from chemically equilibrating medium at finite density. The instantaneous photon rates in presence of chemical non-equilibrium was also analysed. Our analysis demonstrates that, despite an overall suppression of the total photon yield, the relative contribution of early-time emission to high $p_T$ photon production becomes more pronounced. The instantaneous photon emission rates reveal that the presence of chemical non-equilibrium reshapes the temporal structure of the photon emission. These features may serve as distinctive signatures of a chemically equilibrating quark-gluon plasma.

\section*{Acknowledgements}
\par
We would like to thank the anonymous reviewers, whose suggestions have improved the quality of this manuscript. L. J. N. acknowledges the Department of Science and Technology, Government of India for the
INSPIRE Fellowship. 

\bibliography{ChemEq-Gb}

\begin{thebibliography}{101}%
\makeatletter
\providecommand \@ifxundefined [1]{%
 \@ifx{#1\undefined}
}%
\providecommand \@ifnum [1]{%
 \ifnum #1\expandafter \@firstoftwo
 \else \expandafter \@secondoftwo
 \fi
}%
\providecommand \@ifx [1]{%
 \ifx #1\expandafter \@firstoftwo
 \else \expandafter \@secondoftwo
 \fi
}%
\providecommand \natexlab [1]{#1}%
\providecommand \enquote  [1]{``#1''}%
\providecommand \bibnamefont  [1]{#1}%
\providecommand \bibfnamefont [1]{#1}%
\providecommand \citenamefont [1]{#1}%
\providecommand \href@noop [0]{\@secondoftwo}%
\providecommand \href [0]{\begingroup \@sanitize@url \@href}%
\providecommand \@href[1]{\@@startlink{#1}\@@href}%
\providecommand \@@href[1]{\endgroup#1\@@endlink}%
\providecommand \@sanitize@url [0]{\catcode `\\12\catcode `\$12\catcode
  `\&12\catcode `\#12\catcode `\^12\catcode `\_12\catcode `\%12\relax}%
\providecommand \@@startlink[1]{}%
\providecommand \@@endlink[0]{}%
\providecommand \url  [0]{\begingroup\@sanitize@url \@url }%
\providecommand \@url [1]{\endgroup\@href {#1}{\urlprefix }}%
\providecommand \urlprefix  [0]{URL }%
\providecommand \Eprint [0]{\href }%
\providecommand \doibase [0]{https://doi.org/}%
\providecommand \selectlanguage [0]{\@gobble}%
\providecommand \bibinfo  [0]{\@secondoftwo}%
\providecommand \bibfield  [0]{\@secondoftwo}%
\providecommand \translation [1]{[#1]}%
\providecommand \BibitemOpen [0]{}%
\providecommand \bibitemStop [0]{}%
\providecommand \bibitemNoStop [0]{.\EOS\space}%
\providecommand \EOS [0]{\spacefactor3000\relax}%
\providecommand \BibitemShut  [1]{\csname bibitem#1\endcsname}%
\let\auto@bib@innerbib\@empty
\bibitem [{\citenamefont {Hirano}\ and\ \citenamefont
  {Gyulassy}(2006)}]{Hirano:2005wx}%
  \BibitemOpen
  \bibfield  {author} {\bibinfo {author} {\bibfnamefont {T.}~\bibnamefont
  {Hirano}}\ and\ \bibinfo {author} {\bibfnamefont {M.}~\bibnamefont
  {Gyulassy}},\ }\bibfield  {title} {\bibinfo {title} {{Perfect fluidity of the
  quark gluon plasma core as seen through its dissipative hadronic corona}},\
  }\href {https://doi.org/10.1016/j.nuclphysa.2006.02.005} {\bibfield
  {journal} {\bibinfo  {journal} {Nucl. Phys. A}\ }\textbf {\bibinfo {volume}
  {769}},\ \bibinfo {pages} {71} (\bibinfo {year} {2006})},\ \Eprint
  {https://arxiv.org/abs/nucl-th/0506049} {arXiv:nucl-th/0506049} \BibitemShut
  {NoStop}%
\bibitem [{\citenamefont {Voloshin}\ \emph {et~al.}(2010)\citenamefont
  {Voloshin}, \citenamefont {Poskanzer},\ and\ \citenamefont
  {Snellings}}]{Voloshin:2008dg}%
  \BibitemOpen
  \bibfield  {author} {\bibinfo {author} {\bibfnamefont {S.~A.}\ \bibnamefont
  {Voloshin}}, \bibinfo {author} {\bibfnamefont {A.~M.}\ \bibnamefont
  {Poskanzer}},\ and\ \bibinfo {author} {\bibfnamefont {R.}~\bibnamefont
  {Snellings}},\ }\bibfield  {title} {\bibinfo {title} {{Collective phenomena
  in non-central nuclear collisions}},\ }\href
  {https://doi.org/10.1007/978-3-642-01539-7_10} {\bibfield  {journal}
  {\bibinfo  {journal} {Landolt-Bornstein}\ }\textbf {\bibinfo {volume} {23}},\
  \bibinfo {pages} {293} (\bibinfo {year} {2010})},\ \Eprint
  {https://arxiv.org/abs/0809.2949} {arXiv:0809.2949 [nucl-ex]} \BibitemShut
  {NoStop}%
\bibitem [{\citenamefont {Ablyazimov}\ \emph {et~al.}(2017)\citenamefont
  {Ablyazimov} \emph {et~al.}}]{CBM:2016kpk}%
  \BibitemOpen
  \bibfield  {author} {\bibinfo {author} {\bibfnamefont {T.}~\bibnamefont
  {Ablyazimov}} \emph {et~al.} (\bibinfo {collaboration} {CBM}),\ }\bibfield
  {title} {\bibinfo {title} {{Challenges in QCD matter physics --The scientific
  programme of the Compressed Baryonic Matter experiment at FAIR}},\ }\href
  {https://doi.org/10.1140/epja/i2017-12248-y} {\bibfield  {journal} {\bibinfo
  {journal} {Eur. Phys. J. A}\ }\textbf {\bibinfo {volume} {53}},\ \bibinfo
  {pages} {60} (\bibinfo {year} {2017})},\ \Eprint
  {https://arxiv.org/abs/1607.01487} {arXiv:1607.01487 [nucl-ex]} \BibitemShut
  {NoStop}%
\bibitem [{\citenamefont {Adamczyk}\ \emph {et~al.}(2017)\citenamefont
  {Adamczyk} \emph {et~al.}}]{STAR:2017sal}%
  \BibitemOpen
  \bibfield  {author} {\bibinfo {author} {\bibfnamefont {L.}~\bibnamefont
  {Adamczyk}} \emph {et~al.} (\bibinfo {collaboration} {STAR}),\ }\bibfield
  {title} {\bibinfo {title} {{Bulk Properties of the Medium Produced in
  Relativistic Heavy-Ion Collisions from the Beam Energy Scan Program}},\
  }\href {https://doi.org/10.1103/PhysRevC.96.044904} {\bibfield  {journal}
  {\bibinfo  {journal} {Phys. Rev. C}\ }\textbf {\bibinfo {volume} {96}},\
  \bibinfo {pages} {044904} (\bibinfo {year} {2017})},\ \Eprint
  {https://arxiv.org/abs/1701.07065} {arXiv:1701.07065 [nucl-ex]} \BibitemShut
  {NoStop}%
\bibitem [{\citenamefont {Senger}(2021)}]{Senger:2021cfo}%
  \BibitemOpen
  \bibfield  {author} {\bibinfo {author} {\bibfnamefont {P.}~\bibnamefont
  {Senger}},\ }\bibfield  {title} {\bibinfo {title} {{Heavy-Ion Collisions at
  FAIR-NICA Energies}},\ }\href {https://doi.org/10.3390/particles4020020}
  {\bibfield  {journal} {\bibinfo  {journal} {Particles}\ }\textbf {\bibinfo
  {volume} {4}},\ \bibinfo {pages} {214} (\bibinfo {year} {2021})}\BibitemShut
  {NoStop}%
\bibitem [{\citenamefont {Romatschke}\ and\ \citenamefont
  {Romatschke}(2019)}]{Romatschke:2017ejr}%
  \BibitemOpen
  \bibfield  {author} {\bibinfo {author} {\bibfnamefont {P.}~\bibnamefont
  {Romatschke}}\ and\ \bibinfo {author} {\bibfnamefont {U.}~\bibnamefont
  {Romatschke}},\ }\href {https://doi.org/10.1017/9781108651998} {\emph
  {\bibinfo {title} {{Relativistic Fluid Dynamics In and Out of
  Equilibrium}}}},\ Cambridge Monographs on Mathematical Physics\ (\bibinfo
  {publisher} {Cambridge University Press},\ \bibinfo {year} {2019})\ \Eprint
  {https://arxiv.org/abs/1712.05815} {arXiv:1712.05815 [nucl-th]} \BibitemShut
  {NoStop}%
\bibitem [{\citenamefont {Teaney}\ \emph {et~al.}(2001)\citenamefont {Teaney},
  \citenamefont {Lauret},\ and\ \citenamefont {Shuryak}}]{Teaney:2000cw}%
  \BibitemOpen
  \bibfield  {author} {\bibinfo {author} {\bibfnamefont {D.}~\bibnamefont
  {Teaney}}, \bibinfo {author} {\bibfnamefont {J.}~\bibnamefont {Lauret}},\
  and\ \bibinfo {author} {\bibfnamefont {E.~V.}\ \bibnamefont {Shuryak}},\
  }\bibfield  {title} {\bibinfo {title} {{Flow at the SPS and RHIC as a quark
  gluon plasma signature}},\ }\href
  {https://doi.org/10.1103/PhysRevLett.86.4783} {\bibfield  {journal} {\bibinfo
   {journal} {Phys. Rev. Lett.}\ }\textbf {\bibinfo {volume} {86}},\ \bibinfo
  {pages} {4783} (\bibinfo {year} {2001})},\ \Eprint
  {https://arxiv.org/abs/nucl-th/0011058} {arXiv:nucl-th/0011058} \BibitemShut
  {NoStop}%
\bibitem [{\citenamefont {Muronga}(2004)}]{Muronga:2003ta}%
  \BibitemOpen
  \bibfield  {author} {\bibinfo {author} {\bibfnamefont {A.}~\bibnamefont
  {Muronga}},\ }\bibfield  {title} {\bibinfo {title} {{Causal theories of
  dissipative relativistic fluid dynamics for nuclear collisions}},\ }\href
  {https://doi.org/10.1103/PhysRevC.69.034903} {\bibfield  {journal} {\bibinfo
  {journal} {Phys. Rev. C}\ }\textbf {\bibinfo {volume} {69}},\ \bibinfo
  {pages} {034903} (\bibinfo {year} {2004})},\ \Eprint
  {https://arxiv.org/abs/nucl-th/0309055} {arXiv:nucl-th/0309055} \BibitemShut
  {NoStop}%
\bibitem [{\citenamefont {Romatschke}\ and\ \citenamefont
  {Romatschke}(2007)}]{Romatschke:2007mq}%
  \BibitemOpen
  \bibfield  {author} {\bibinfo {author} {\bibfnamefont {P.}~\bibnamefont
  {Romatschke}}\ and\ \bibinfo {author} {\bibfnamefont {U.}~\bibnamefont
  {Romatschke}},\ }\bibfield  {title} {\bibinfo {title} {{Viscosity Information
  from Relativistic Nuclear Collisions: How Perfect is the Fluid Observed at
  RHIC?}},\ }\href {https://doi.org/10.1103/PhysRevLett.99.172301} {\bibfield
  {journal} {\bibinfo  {journal} {Phys. Rev. Lett.}\ }\textbf {\bibinfo
  {volume} {99}},\ \bibinfo {pages} {172301} (\bibinfo {year} {2007})},\
  \Eprint {https://arxiv.org/abs/0706.1522} {arXiv:0706.1522 [nucl-th]}
  \BibitemShut {NoStop}%
\bibitem [{\citenamefont {Song}\ and\ \citenamefont
  {Heinz}(2008)}]{Song:2007ux}%
  \BibitemOpen
  \bibfield  {author} {\bibinfo {author} {\bibfnamefont {H.}~\bibnamefont
  {Song}}\ and\ \bibinfo {author} {\bibfnamefont {U.~W.}\ \bibnamefont
  {Heinz}},\ }\bibfield  {title} {\bibinfo {title} {{Causal viscous
  hydrodynamics in 2+1 dimensions for relativistic heavy-ion collisions}},\
  }\href {https://doi.org/10.1103/PhysRevC.77.064901} {\bibfield  {journal}
  {\bibinfo  {journal} {Phys. Rev. C}\ }\textbf {\bibinfo {volume} {77}},\
  \bibinfo {pages} {064901} (\bibinfo {year} {2008})},\ \Eprint
  {https://arxiv.org/abs/0712.3715} {arXiv:0712.3715 [nucl-th]} \BibitemShut
  {NoStop}%
\bibitem [{\citenamefont {Gale}\ \emph {et~al.}(2013)\citenamefont {Gale},
  \citenamefont {Jeon},\ and\ \citenamefont {Schenke}}]{Gale:2013da}%
  \BibitemOpen
  \bibfield  {author} {\bibinfo {author} {\bibfnamefont {C.}~\bibnamefont
  {Gale}}, \bibinfo {author} {\bibfnamefont {S.}~\bibnamefont {Jeon}},\ and\
  \bibinfo {author} {\bibfnamefont {B.}~\bibnamefont {Schenke}},\ }\bibfield
  {title} {\bibinfo {title} {{Hydrodynamic Modeling of Heavy-Ion Collisions}},\
  }\href {https://doi.org/10.1142/S0217751X13400113} {\bibfield  {journal}
  {\bibinfo  {journal} {Int. J. Mod. Phys. A}\ }\textbf {\bibinfo {volume}
  {28}},\ \bibinfo {pages} {1340011} (\bibinfo {year} {2013})},\ \Eprint
  {https://arxiv.org/abs/1301.5893} {arXiv:1301.5893 [nucl-th]} \BibitemShut
  {NoStop}%
\bibitem [{\citenamefont {Shuryak}(1992)}]{Shuryak:1992wc}%
  \BibitemOpen
  \bibfield  {author} {\bibinfo {author} {\bibfnamefont {E.~V.}\ \bibnamefont
  {Shuryak}},\ }\bibfield  {title} {\bibinfo {title} {{Two stage equilibration
  in high-energy heavy ion collisions}},\ }\href
  {https://doi.org/10.1103/PhysRevLett.68.3270} {\bibfield  {journal} {\bibinfo
   {journal} {Phys. Rev. Lett.}\ }\textbf {\bibinfo {volume} {68}},\ \bibinfo
  {pages} {3270} (\bibinfo {year} {1992})}\BibitemShut {NoStop}%
\bibitem [{\citenamefont {Geiger}(1992)}]{Geiger:1992si}%
  \BibitemOpen
  \bibfield  {author} {\bibinfo {author} {\bibfnamefont {K.}~\bibnamefont
  {Geiger}},\ }\bibfield  {title} {\bibinfo {title} {{Thermalization in
  ultrarelativistic nuclear collisions. 1. Parton kinetics and quark gluon
  plasma formation}},\ }\href {https://doi.org/10.1103/PhysRevD.46.4965}
  {\bibfield  {journal} {\bibinfo  {journal} {Phys. Rev. D}\ }\textbf {\bibinfo
  {volume} {46}},\ \bibinfo {pages} {4965} (\bibinfo {year}
  {1992})}\BibitemShut {NoStop}%
\bibitem [{\citenamefont {Biro}\ \emph {et~al.}(1993)\citenamefont {Biro},
  \citenamefont {van Doorn}, \citenamefont {Muller}, \citenamefont {Thoma},\
  and\ \citenamefont {Wang}}]{Biro:1993qt}%
  \BibitemOpen
  \bibfield  {author} {\bibinfo {author} {\bibfnamefont {T.~S.}\ \bibnamefont
  {Biro}}, \bibinfo {author} {\bibfnamefont {E.}~\bibnamefont {van Doorn}},
  \bibinfo {author} {\bibfnamefont {B.}~\bibnamefont {Muller}}, \bibinfo
  {author} {\bibfnamefont {M.~H.}\ \bibnamefont {Thoma}},\ and\ \bibinfo
  {author} {\bibfnamefont {X.~N.}\ \bibnamefont {Wang}},\ }\bibfield  {title}
  {\bibinfo {title} {{Parton equilibration in relativistic heavy ion
  collisions}},\ }\href {https://doi.org/10.1103/PhysRevC.48.1275} {\bibfield
  {journal} {\bibinfo  {journal} {Phys. Rev. C}\ }\textbf {\bibinfo {volume}
  {48}},\ \bibinfo {pages} {1275} (\bibinfo {year} {1993})},\ \Eprint
  {https://arxiv.org/abs/nucl-th/9303004} {arXiv:nucl-th/9303004} \BibitemShut
  {NoStop}%
\bibitem [{\citenamefont {Levai}\ \emph {et~al.}(1995)\citenamefont {Levai},
  \citenamefont {Muller},\ and\ \citenamefont {Wang}}]{Levai:1994dx}%
  \BibitemOpen
  \bibfield  {author} {\bibinfo {author} {\bibfnamefont {P.}~\bibnamefont
  {Levai}}, \bibinfo {author} {\bibfnamefont {B.}~\bibnamefont {Muller}},\ and\
  \bibinfo {author} {\bibfnamefont {X.-N.}\ \bibnamefont {Wang}},\ }\bibfield
  {title} {\bibinfo {title} {{Open charm production in an equilibrating parton
  plasma}},\ }\href {https://doi.org/10.1103/PhysRevC.51.3326} {\bibfield
  {journal} {\bibinfo  {journal} {Phys. Rev. C}\ }\textbf {\bibinfo {volume}
  {51}},\ \bibinfo {pages} {3326} (\bibinfo {year} {1995})},\ \Eprint
  {https://arxiv.org/abs/hep-ph/9412352} {arXiv:hep-ph/9412352} \BibitemShut
  {NoStop}%
\bibitem [{\citenamefont {Wong}(1996)}]{Wong:1996va}%
  \BibitemOpen
  \bibfield  {author} {\bibinfo {author} {\bibfnamefont {S.~M.~H.}\
  \bibnamefont {Wong}},\ }\bibfield  {title} {\bibinfo {title} {{Thermal and
  chemical equilibration in relativistic heavy ion collisions}},\ }\href
  {https://doi.org/10.1103/PhysRevC.54.2588} {\bibfield  {journal} {\bibinfo
  {journal} {Phys. Rev. C}\ }\textbf {\bibinfo {volume} {54}},\ \bibinfo
  {pages} {2588} (\bibinfo {year} {1996})},\ \Eprint
  {https://arxiv.org/abs/hep-ph/9609287} {arXiv:hep-ph/9609287} \BibitemShut
  {NoStop}%
\bibitem [{\citenamefont {Baier}\ \emph {et~al.}(2001)\citenamefont {Baier},
  \citenamefont {Mueller}, \citenamefont {Schiff},\ and\ \citenamefont
  {Son}}]{Baier:2000sb}%
  \BibitemOpen
  \bibfield  {author} {\bibinfo {author} {\bibfnamefont {R.}~\bibnamefont
  {Baier}}, \bibinfo {author} {\bibfnamefont {A.~H.}\ \bibnamefont {Mueller}},
  \bibinfo {author} {\bibfnamefont {D.}~\bibnamefont {Schiff}},\ and\ \bibinfo
  {author} {\bibfnamefont {D.~T.}\ \bibnamefont {Son}},\ }\bibfield  {title}
  {\bibinfo {title} {{'Bottom up' thermalization in heavy ion collisions}},\
  }\href {https://doi.org/10.1016/S0370-2693(01)00191-5} {\bibfield  {journal}
  {\bibinfo  {journal} {Phys. Lett. B}\ }\textbf {\bibinfo {volume} {502}},\
  \bibinfo {pages} {51} (\bibinfo {year} {2001})},\ \Eprint
  {https://arxiv.org/abs/hep-ph/0009237} {arXiv:hep-ph/0009237} \BibitemShut
  {NoStop}%
\bibitem [{\citenamefont {Berges}\ \emph {et~al.}(2014)\citenamefont {Berges},
  \citenamefont {Boguslavski}, \citenamefont {Schlichting},\ and\ \citenamefont
  {Venugopalan}}]{Berges:2013eia}%
  \BibitemOpen
  \bibfield  {author} {\bibinfo {author} {\bibfnamefont {J.}~\bibnamefont
  {Berges}}, \bibinfo {author} {\bibfnamefont {K.}~\bibnamefont {Boguslavski}},
  \bibinfo {author} {\bibfnamefont {S.}~\bibnamefont {Schlichting}},\ and\
  \bibinfo {author} {\bibfnamefont {R.}~\bibnamefont {Venugopalan}},\
  }\bibfield  {title} {\bibinfo {title} {{Turbulent thermalization process in
  heavy-ion collisions at ultrarelativistic energies}},\ }\href
  {https://doi.org/10.1103/PhysRevD.89.074011} {\bibfield  {journal} {\bibinfo
  {journal} {Phys. Rev. D}\ }\textbf {\bibinfo {volume} {89}},\ \bibinfo
  {pages} {074011} (\bibinfo {year} {2014})},\ \Eprint
  {https://arxiv.org/abs/1303.5650} {arXiv:1303.5650 [hep-ph]} \BibitemShut
  {NoStop}%
\bibitem [{\citenamefont {Epelbaum}\ and\ \citenamefont
  {Gelis}(2013)}]{Epelbaum:2013ekf}%
  \BibitemOpen
  \bibfield  {author} {\bibinfo {author} {\bibfnamefont {T.}~\bibnamefont
  {Epelbaum}}\ and\ \bibinfo {author} {\bibfnamefont {F.}~\bibnamefont
  {Gelis}},\ }\bibfield  {title} {\bibinfo {title} {{Pressure isotropization in
  high energy heavy ion collisions}},\ }\href
  {https://doi.org/10.1103/PhysRevLett.111.232301} {\bibfield  {journal}
  {\bibinfo  {journal} {Phys. Rev. Lett.}\ }\textbf {\bibinfo {volume} {111}},\
  \bibinfo {pages} {232301} (\bibinfo {year} {2013})},\ \Eprint
  {https://arxiv.org/abs/1307.2214} {arXiv:1307.2214 [hep-ph]} \BibitemShut
  {NoStop}%
\bibitem [{\citenamefont {Kurkela}\ and\ \citenamefont
  {Zhu}(2015)}]{Kurkela:2015qoa}%
  \BibitemOpen
  \bibfield  {author} {\bibinfo {author} {\bibfnamefont {A.}~\bibnamefont
  {Kurkela}}\ and\ \bibinfo {author} {\bibfnamefont {Y.}~\bibnamefont {Zhu}},\
  }\bibfield  {title} {\bibinfo {title} {{Isotropization and hydrodynamization
  in weakly coupled heavy-ion collisions}},\ }\href
  {https://doi.org/10.1103/PhysRevLett.115.182301} {\bibfield  {journal}
  {\bibinfo  {journal} {Phys. Rev. Lett.}\ }\textbf {\bibinfo {volume} {115}},\
  \bibinfo {pages} {182301} (\bibinfo {year} {2015})},\ \Eprint
  {https://arxiv.org/abs/1506.06647} {arXiv:1506.06647 [hep-ph]} \BibitemShut
  {NoStop}%
\bibitem [{\citenamefont {Keegan}\ \emph {et~al.}(2016)\citenamefont {Keegan},
  \citenamefont {Kurkela}, \citenamefont {Mazeliauskas},\ and\ \citenamefont
  {Teaney}}]{Keegan:2016cpi}%
  \BibitemOpen
  \bibfield  {author} {\bibinfo {author} {\bibfnamefont {L.}~\bibnamefont
  {Keegan}}, \bibinfo {author} {\bibfnamefont {A.}~\bibnamefont {Kurkela}},
  \bibinfo {author} {\bibfnamefont {A.}~\bibnamefont {Mazeliauskas}},\ and\
  \bibinfo {author} {\bibfnamefont {D.}~\bibnamefont {Teaney}},\ }\bibfield
  {title} {\bibinfo {title} {{Initial conditions for hydrodynamics from weakly
  coupled pre-equilibrium evolution}},\ }\href
  {https://doi.org/10.1007/JHEP08(2016)171} {\bibfield  {journal} {\bibinfo
  {journal} {JHEP}\ }\textbf {\bibinfo {volume} {08}},\ \bibinfo {pages}
  {171}},\ \Eprint {https://arxiv.org/abs/1605.04287} {arXiv:1605.04287
  [hep-ph]} \BibitemShut {NoStop}%
\bibitem [{\citenamefont {Kurkela}\ \emph {et~al.}(2019)\citenamefont
  {Kurkela}, \citenamefont {Mazeliauskas}, \citenamefont {Paquet},
  \citenamefont {Schlichting},\ and\ \citenamefont {Teaney}}]{Kurkela:2018vqr}%
  \BibitemOpen
  \bibfield  {author} {\bibinfo {author} {\bibfnamefont {A.}~\bibnamefont
  {Kurkela}}, \bibinfo {author} {\bibfnamefont {A.}~\bibnamefont
  {Mazeliauskas}}, \bibinfo {author} {\bibfnamefont {J.-F.}\ \bibnamefont
  {Paquet}}, \bibinfo {author} {\bibfnamefont {S.}~\bibnamefont
  {Schlichting}},\ and\ \bibinfo {author} {\bibfnamefont {D.}~\bibnamefont
  {Teaney}},\ }\bibfield  {title} {\bibinfo {title} {{Effective kinetic
  description of event-by-event pre-equilibrium dynamics in high-energy
  heavy-ion collisions}},\ }\href {https://doi.org/10.1103/PhysRevC.99.034910}
  {\bibfield  {journal} {\bibinfo  {journal} {Phys. Rev. C}\ }\textbf {\bibinfo
  {volume} {99}},\ \bibinfo {pages} {034910} (\bibinfo {year} {2019})},\
  \Eprint {https://arxiv.org/abs/1805.00961} {arXiv:1805.00961 [hep-ph]}
  \BibitemShut {NoStop}%
\bibitem [{\citenamefont {Kurkela}\ and\ \citenamefont
  {Mazeliauskas}(2019{\natexlab{a}})}]{Kurkela:2018oqw}%
  \BibitemOpen
  \bibfield  {author} {\bibinfo {author} {\bibfnamefont {A.}~\bibnamefont
  {Kurkela}}\ and\ \bibinfo {author} {\bibfnamefont {A.}~\bibnamefont
  {Mazeliauskas}},\ }\bibfield  {title} {\bibinfo {title} {{Chemical
  equilibration in weakly coupled QCD}},\ }\href
  {https://doi.org/10.1103/PhysRevD.99.054018} {\bibfield  {journal} {\bibinfo
  {journal} {Phys. Rev. D}\ }\textbf {\bibinfo {volume} {99}},\ \bibinfo
  {pages} {054018} (\bibinfo {year} {2019}{\natexlab{a}})},\ \Eprint
  {https://arxiv.org/abs/1811.03068} {arXiv:1811.03068 [hep-ph]} \BibitemShut
  {NoStop}%
\bibitem [{\citenamefont {Kurkela}\ and\ \citenamefont
  {Mazeliauskas}(2019{\natexlab{b}})}]{Kurkela:2018xxd}%
  \BibitemOpen
  \bibfield  {author} {\bibinfo {author} {\bibfnamefont {A.}~\bibnamefont
  {Kurkela}}\ and\ \bibinfo {author} {\bibfnamefont {A.}~\bibnamefont
  {Mazeliauskas}},\ }\bibfield  {title} {\bibinfo {title} {{Chemical
  Equilibration in Hadronic Collisions}},\ }\href
  {https://doi.org/10.1103/PhysRevLett.122.142301} {\bibfield  {journal}
  {\bibinfo  {journal} {Phys. Rev. Lett.}\ }\textbf {\bibinfo {volume} {122}},\
  \bibinfo {pages} {142301} (\bibinfo {year} {2019}{\natexlab{b}})},\ \Eprint
  {https://arxiv.org/abs/1811.03040} {arXiv:1811.03040 [hep-ph]} \BibitemShut
  {NoStop}%
\bibitem [{\citenamefont {Du}\ and\ \citenamefont
  {Schlichting}(2021)}]{Du:2020zqg}%
  \BibitemOpen
  \bibfield  {author} {\bibinfo {author} {\bibfnamefont {X.}~\bibnamefont
  {Du}}\ and\ \bibinfo {author} {\bibfnamefont {S.}~\bibnamefont
  {Schlichting}},\ }\bibfield  {title} {\bibinfo {title} {{Equilibration of the
  Quark-Gluon Plasma at Finite Net-Baryon Density in QCD Kinetic Theory}},\
  }\href {https://doi.org/10.1103/PhysRevLett.127.122301} {\bibfield  {journal}
  {\bibinfo  {journal} {Phys. Rev. Lett.}\ }\textbf {\bibinfo {volume} {127}},\
  \bibinfo {pages} {122301} (\bibinfo {year} {2021})},\ \Eprint
  {https://arxiv.org/abs/2012.09068} {arXiv:2012.09068 [hep-ph]} \BibitemShut
  {NoStop}%
\bibitem [{\citenamefont {Cabodevila}\ \emph {et~al.}(2024)\citenamefont
  {Cabodevila}, \citenamefont {Salgado},\ and\ \citenamefont
  {Wu}}]{Cabodevila:2023htm}%
  \BibitemOpen
  \bibfield  {author} {\bibinfo {author} {\bibfnamefont {S.~B.}\ \bibnamefont
  {Cabodevila}}, \bibinfo {author} {\bibfnamefont {C.~A.}\ \bibnamefont
  {Salgado}},\ and\ \bibinfo {author} {\bibfnamefont {B.}~\bibnamefont {Wu}},\
  }\bibfield  {title} {\bibinfo {title} {{Quark production and thermalization
  of the quark-gluon plasma}},\ }\href
  {https://doi.org/10.1007/JHEP06(2024)145} {\bibfield  {journal} {\bibinfo
  {journal} {JHEP}\ }\textbf {\bibinfo {volume} {06}},\ \bibinfo {pages}
  {145}},\ \Eprint {https://arxiv.org/abs/2311.07450} {arXiv:2311.07450
  [hep-ph]} \BibitemShut {NoStop}%
\bibitem [{\citenamefont {Bhatt}\ and\ \citenamefont
  {Sreekanth}(2010)}]{Bhatt:2009zg}%
  \BibitemOpen
  \bibfield  {author} {\bibinfo {author} {\bibfnamefont {J.~R.}\ \bibnamefont
  {Bhatt}}\ and\ \bibinfo {author} {\bibfnamefont {V.}~\bibnamefont
  {Sreekanth}},\ }\bibfield  {title} {\bibinfo {title} {{Photon emission from
  out of equilibrium dissipative parton plasma}},\ }\href
  {https://doi.org/10.1142/S0218301310014765} {\bibfield  {journal} {\bibinfo
  {journal} {Int. J. Mod. Phys. E}\ }\textbf {\bibinfo {volume} {19}},\
  \bibinfo {pages} {299} (\bibinfo {year} {2010})},\ \Eprint
  {https://arxiv.org/abs/0901.1363} {arXiv:0901.1363 [hep-ph]} \BibitemShut
  {NoStop}%
\bibitem [{\citenamefont {Elliott}\ and\ \citenamefont
  {Rischke}(2000)}]{Elliott:1999uz}%
  \BibitemOpen
  \bibfield  {author} {\bibinfo {author} {\bibfnamefont {D.~M.}\ \bibnamefont
  {Elliott}}\ and\ \bibinfo {author} {\bibfnamefont {D.~H.}\ \bibnamefont
  {Rischke}},\ }\bibfield  {title} {\bibinfo {title} {{Chemical equilibration
  of quarks and gluons at RHIC and LHC energies}},\ }\href
  {https://doi.org/10.1016/S0375-9474(99)00840-4} {\bibfield  {journal}
  {\bibinfo  {journal} {Nucl. Phys. A}\ }\textbf {\bibinfo {volume} {671}},\
  \bibinfo {pages} {583} (\bibinfo {year} {2000})},\ \Eprint
  {https://arxiv.org/abs/nucl-th/9908004} {arXiv:nucl-th/9908004} \BibitemShut
  {NoStop}%
\bibitem [{\citenamefont {El}\ \emph {et~al.}(2010)\citenamefont {El},
  \citenamefont {Muronga}, \citenamefont {Xu},\ and\ \citenamefont
  {Greiner}}]{El:2010mt}%
  \BibitemOpen
  \bibfield  {author} {\bibinfo {author} {\bibfnamefont {A.}~\bibnamefont
  {El}}, \bibinfo {author} {\bibfnamefont {A.}~\bibnamefont {Muronga}},
  \bibinfo {author} {\bibfnamefont {Z.}~\bibnamefont {Xu}},\ and\ \bibinfo
  {author} {\bibfnamefont {C.}~\bibnamefont {Greiner}},\ }\bibfield  {title}
  {\bibinfo {title} {{A Relativistic dissipative hydrodynamic description for
  systems including particle number changing processes}},\ }\href
  {https://doi.org/10.1016/j.nuclphysa.2010.09.011} {\bibfield  {journal}
  {\bibinfo  {journal} {Nucl. Phys. A}\ }\textbf {\bibinfo {volume} {848}},\
  \bibinfo {pages} {428} (\bibinfo {year} {2010})},\ \Eprint
  {https://arxiv.org/abs/1007.0705} {arXiv:1007.0705 [nucl-th]} \BibitemShut
  {NoStop}%
\bibitem [{\citenamefont {Vovchenko}\ \emph
  {et~al.}(2016{\natexlab{a}})\citenamefont {Vovchenko}, \citenamefont
  {Gorenstein}, \citenamefont {Satarov}, \citenamefont {Mishustin},
  \citenamefont {Csernai}, \citenamefont {Kisel},\ and\ \citenamefont
  {St{\"o}cker}}]{Vovchenko:2015yia}%
  \BibitemOpen
  \bibfield  {author} {\bibinfo {author} {\bibfnamefont {V.}~\bibnamefont
  {Vovchenko}}, \bibinfo {author} {\bibfnamefont {M.~I.}\ \bibnamefont
  {Gorenstein}}, \bibinfo {author} {\bibfnamefont {L.~M.}\ \bibnamefont
  {Satarov}}, \bibinfo {author} {\bibfnamefont {I.~N.}\ \bibnamefont
  {Mishustin}}, \bibinfo {author} {\bibfnamefont {L.~P.}\ \bibnamefont
  {Csernai}}, \bibinfo {author} {\bibfnamefont {I.}~\bibnamefont {Kisel}},\
  and\ \bibinfo {author} {\bibfnamefont {H.}~\bibnamefont {St{\"o}cker}},\
  }\bibfield  {title} {\bibinfo {title} {{Entropy production in chemically
  nonequilibrium quark-gluon plasma created in central Pb+Pb collisions at
  energies available at the CERN Large Hadron Collider}},\ }\href
  {https://doi.org/10.1103/PhysRevC.93.014906} {\bibfield  {journal} {\bibinfo
  {journal} {Phys. Rev. C}\ }\textbf {\bibinfo {volume} {93}},\ \bibinfo
  {pages} {014906} (\bibinfo {year} {2016}{\natexlab{a}})},\ \Eprint
  {https://arxiv.org/abs/1510.01235} {arXiv:1510.01235 [hep-ph]} \BibitemShut
  {NoStop}%
\bibitem [{\citenamefont {Dutta}\ \emph {et~al.}(1999)\citenamefont {Dutta},
  \citenamefont {Kumar}, \citenamefont {Mohanty},\ and\ \citenamefont
  {Choudhury}}]{Dutta:1999hj}%
  \BibitemOpen
  \bibfield  {author} {\bibinfo {author} {\bibfnamefont {D.}~\bibnamefont
  {Dutta}}, \bibinfo {author} {\bibfnamefont {K.}~\bibnamefont {Kumar}},
  \bibinfo {author} {\bibfnamefont {A.~K.}\ \bibnamefont {Mohanty}},\ and\
  \bibinfo {author} {\bibfnamefont {R.~K.}\ \bibnamefont {Choudhury}},\
  }\bibfield  {title} {\bibinfo {title} {{Chemical equilibration and thermal
  dilepton production from the quark gluon plasma at finite baryon density}},\
  }\href {https://doi.org/10.1103/PhysRevC.60.014905} {\bibfield  {journal}
  {\bibinfo  {journal} {Phys. Rev. C}\ }\textbf {\bibinfo {volume} {60}},\
  \bibinfo {pages} {014905} (\bibinfo {year} {1999})},\ \Eprint
  {https://arxiv.org/abs/hep-ph/9905245} {arXiv:hep-ph/9905245} \BibitemShut
  {NoStop}%
\bibitem [{\citenamefont {Kampfer}\ \emph {et~al.}(1995)\citenamefont
  {Kampfer}, \citenamefont {Pavlenko}, \citenamefont {Peshier},\ and\
  \citenamefont {Soff}}]{Kampfer:1995mg}%
  \BibitemOpen
  \bibfield  {author} {\bibinfo {author} {\bibfnamefont {B.}~\bibnamefont
  {Kampfer}}, \bibinfo {author} {\bibfnamefont {O.~P.}\ \bibnamefont
  {Pavlenko}}, \bibinfo {author} {\bibfnamefont {A.}~\bibnamefont {Peshier}},\
  and\ \bibinfo {author} {\bibfnamefont {G.}~\bibnamefont {Soff}},\ }\bibfield
  {title} {\bibinfo {title} {{Dilepton production in a chemically
  equilibrating, expanding, and hadronizing quark - gluon plasma}},\ }\href
  {https://doi.org/10.1103/PhysRevC.52.2704} {\bibfield  {journal} {\bibinfo
  {journal} {Phys. Rev. C}\ }\textbf {\bibinfo {volume} {52}},\ \bibinfo
  {pages} {2704} (\bibinfo {year} {1995})}\BibitemShut {NoStop}%
\bibitem [{\citenamefont {Srivastava}\ \emph {et~al.}(1997)\citenamefont
  {Srivastava}, \citenamefont {Mustafa},\ and\ \citenamefont
  {Muller}}]{Srivastava:1996qd}%
  \BibitemOpen
  \bibfield  {author} {\bibinfo {author} {\bibfnamefont {D.~K.}\ \bibnamefont
  {Srivastava}}, \bibinfo {author} {\bibfnamefont {M.~G.}\ \bibnamefont
  {Mustafa}},\ and\ \bibinfo {author} {\bibfnamefont {B.}~\bibnamefont
  {Muller}},\ }\bibfield  {title} {\bibinfo {title} {{Expanding quark - gluon
  plasmas: Transverse flow, chemical equilibration and electromagnetic
  radiation}},\ }\href {https://doi.org/10.1103/PhysRevC.56.1064} {\bibfield
  {journal} {\bibinfo  {journal} {Phys. Rev. C}\ }\textbf {\bibinfo {volume}
  {56}},\ \bibinfo {pages} {1064} (\bibinfo {year} {1997})},\ \Eprint
  {https://arxiv.org/abs/nucl-th/9611041} {arXiv:nucl-th/9611041} \BibitemShut
  {NoStop}%
\bibitem [{\citenamefont {Gelis}\ \emph {et~al.}(2004)\citenamefont {Gelis},
  \citenamefont {Niemi}, \citenamefont {Ruuskanen},\ and\ \citenamefont
  {Rasanen}}]{Gelis:2004ep}%
  \BibitemOpen
  \bibfield  {author} {\bibinfo {author} {\bibfnamefont {F.}~\bibnamefont
  {Gelis}}, \bibinfo {author} {\bibfnamefont {H.}~\bibnamefont {Niemi}},
  \bibinfo {author} {\bibfnamefont {P.~V.}\ \bibnamefont {Ruuskanen}},\ and\
  \bibinfo {author} {\bibfnamefont {S.~S.}\ \bibnamefont {Rasanen}},\
  }\bibfield  {title} {\bibinfo {title} {{Photon production from nonequilibrium
  QGP in heavy ion collisions}},\ }\href
  {https://doi.org/10.1088/0954-3899/30/8/053} {\bibfield  {journal} {\bibinfo
  {journal} {J. Phys. G}\ }\textbf {\bibinfo {volume} {30}},\ \bibinfo {pages}
  {S1031} (\bibinfo {year} {2004})},\ \Eprint
  {https://arxiv.org/abs/nucl-th/0403040} {arXiv:nucl-th/0403040} \BibitemShut
  {NoStop}%
\bibitem [{\citenamefont {Vovchenko}\ \emph
  {et~al.}(2016{\natexlab{b}})\citenamefont {Vovchenko}, \citenamefont
  {Karpenko}, \citenamefont {Gorenstein}, \citenamefont {Satarov},
  \citenamefont {Mishustin}, \citenamefont {K{\"a}mpfer},\ and\ \citenamefont
  {Stoecker}}]{Vovchenko:2016ijt}%
  \BibitemOpen
  \bibfield  {author} {\bibinfo {author} {\bibfnamefont {V.}~\bibnamefont
  {Vovchenko}}, \bibinfo {author} {\bibfnamefont {I.~A.}\ \bibnamefont
  {Karpenko}}, \bibinfo {author} {\bibfnamefont {M.~I.}\ \bibnamefont
  {Gorenstein}}, \bibinfo {author} {\bibfnamefont {L.~M.}\ \bibnamefont
  {Satarov}}, \bibinfo {author} {\bibfnamefont {I.~N.}\ \bibnamefont
  {Mishustin}}, \bibinfo {author} {\bibfnamefont {B.}~\bibnamefont
  {K{\"a}mpfer}},\ and\ \bibinfo {author} {\bibfnamefont {H.}~\bibnamefont
  {Stoecker}},\ }\bibfield  {title} {\bibinfo {title} {{Electromagnetic probes
  of a pure-glue initial state in nucleus-nucleus collisions at energies
  available at the CERN Large Hadron Collider}},\ }\href
  {https://doi.org/10.1103/PhysRevC.94.024906} {\bibfield  {journal} {\bibinfo
  {journal} {Phys. Rev. C}\ }\textbf {\bibinfo {volume} {94}},\ \bibinfo
  {pages} {024906} (\bibinfo {year} {2016}{\natexlab{b}})},\ \Eprint
  {https://arxiv.org/abs/1604.06346} {arXiv:1604.06346 [nucl-th]} \BibitemShut
  {NoStop}%
\bibitem [{\citenamefont {Landau}(1953)}]{Landau:1953gs}%
  \BibitemOpen
  \bibfield  {author} {\bibinfo {author} {\bibfnamefont {L.~D.}\ \bibnamefont
  {Landau}},\ }\bibfield  {title} {\bibinfo {title} {{On the multiparticle
  production in high-energy collisions}},\ }\href@noop {} {\bibfield  {journal}
  {\bibinfo  {journal} {Izv. Akad. Nauk Ser. Fiz.}\ }\textbf {\bibinfo {volume}
  {17}},\ \bibinfo {pages} {51} (\bibinfo {year} {1953})}\BibitemShut {NoStop}%
\bibitem [{\citenamefont {Bjorken}(1983)}]{Bjorken:1982qr}%
  \BibitemOpen
  \bibfield  {author} {\bibinfo {author} {\bibfnamefont {J.~D.}\ \bibnamefont
  {Bjorken}},\ }\bibfield  {title} {\bibinfo {title} {{Highly Relativistic
  Nucleus-Nucleus Collisions: The Central Rapidity Region}},\ }\href
  {https://doi.org/10.1103/PhysRevD.27.140} {\bibfield  {journal} {\bibinfo
  {journal} {Phys. Rev. D}\ }\textbf {\bibinfo {volume} {27}},\ \bibinfo
  {pages} {140} (\bibinfo {year} {1983})}\BibitemShut {NoStop}%
\bibitem [{\citenamefont {Csorgo}\ \emph {et~al.}(2008)\citenamefont {Csorgo},
  \citenamefont {Nagy},\ and\ \citenamefont {Csanad}}]{Csorgo:2006ax}%
  \BibitemOpen
  \bibfield  {author} {\bibinfo {author} {\bibfnamefont {T.}~\bibnamefont
  {Csorgo}}, \bibinfo {author} {\bibfnamefont {M.~I.}\ \bibnamefont {Nagy}},\
  and\ \bibinfo {author} {\bibfnamefont {M.}~\bibnamefont {Csanad}},\
  }\bibfield  {title} {\bibinfo {title} {{A New family of simple solutions of
  perfect fluid hydrodynamics}},\ }\href
  {https://doi.org/10.1016/j.physletb.2008.04.038} {\bibfield  {journal}
  {\bibinfo  {journal} {Phys. Lett. B}\ }\textbf {\bibinfo {volume} {663}},\
  \bibinfo {pages} {306} (\bibinfo {year} {2008})},\ \Eprint
  {https://arxiv.org/abs/nucl-th/0605070} {arXiv:nucl-th/0605070} \BibitemShut
  {NoStop}%
\bibitem [{\citenamefont {Bialas}\ \emph {et~al.}(2007)\citenamefont {Bialas},
  \citenamefont {Janik},\ and\ \citenamefont {Peschanski}}]{Bialas:2007iu}%
  \BibitemOpen
  \bibfield  {author} {\bibinfo {author} {\bibfnamefont {A.}~\bibnamefont
  {Bialas}}, \bibinfo {author} {\bibfnamefont {R.~A.}\ \bibnamefont {Janik}},\
  and\ \bibinfo {author} {\bibfnamefont {R.~B.}\ \bibnamefont {Peschanski}},\
  }\bibfield  {title} {\bibinfo {title} {{Unified description of Bjorken and
  Landau 1+1 hydrodynamics}},\ }\href
  {https://doi.org/10.1103/PhysRevC.76.054901} {\bibfield  {journal} {\bibinfo
  {journal} {Phys. Rev. C}\ }\textbf {\bibinfo {volume} {76}},\ \bibinfo
  {pages} {054901} (\bibinfo {year} {2007})},\ \Eprint
  {https://arxiv.org/abs/0706.2108} {arXiv:0706.2108 [nucl-th]} \BibitemShut
  {NoStop}%
\bibitem [{\citenamefont {Gubser}\ and\ \citenamefont
  {Yarom}(2011)}]{Gubser:2010ui}%
  \BibitemOpen
  \bibfield  {author} {\bibinfo {author} {\bibfnamefont {S.~S.}\ \bibnamefont
  {Gubser}}\ and\ \bibinfo {author} {\bibfnamefont {A.}~\bibnamefont {Yarom}},\
  }\bibfield  {title} {\bibinfo {title} {{Conformal hydrodynamics in Minkowski
  and de Sitter spacetimes}},\ }\href
  {https://doi.org/10.1016/j.nuclphysb.2011.01.012} {\bibfield  {journal}
  {\bibinfo  {journal} {Nucl. Phys. B}\ }\textbf {\bibinfo {volume} {846}},\
  \bibinfo {pages} {469} (\bibinfo {year} {2011})},\ \Eprint
  {https://arxiv.org/abs/1012.1314} {arXiv:1012.1314 [hep-th]} \BibitemShut
  {NoStop}%
\bibitem [{\citenamefont {Gubser}(2010)}]{Gubser:2010ze}%
  \BibitemOpen
  \bibfield  {author} {\bibinfo {author} {\bibfnamefont {S.~S.}\ \bibnamefont
  {Gubser}},\ }\bibfield  {title} {\bibinfo {title} {{Symmetry constraints on
  generalizations of Bjorken flow}},\ }\href
  {https://doi.org/10.1103/PhysRevD.82.085027} {\bibfield  {journal} {\bibinfo
  {journal} {Phys. Rev. D}\ }\textbf {\bibinfo {volume} {82}},\ \bibinfo
  {pages} {085027} (\bibinfo {year} {2010})},\ \Eprint
  {https://arxiv.org/abs/1006.0006} {arXiv:1006.0006 [hep-th]} \BibitemShut
  {NoStop}%
\bibitem [{\citenamefont {Shi}\ \emph {et~al.}(2022)\citenamefont {Shi},
  \citenamefont {Jeon},\ and\ \citenamefont {Gale}}]{Shi:2022iyb}%
  \BibitemOpen
  \bibfield  {author} {\bibinfo {author} {\bibfnamefont {S.}~\bibnamefont
  {Shi}}, \bibinfo {author} {\bibfnamefont {S.}~\bibnamefont {Jeon}},\ and\
  \bibinfo {author} {\bibfnamefont {C.}~\bibnamefont {Gale}},\ }\bibfield
  {title} {\bibinfo {title} {{Family of new exact solutions for longitudinally
  expanding ideal fluids}},\ }\href
  {https://doi.org/10.1103/PhysRevC.105.L021902} {\bibfield  {journal}
  {\bibinfo  {journal} {Phys. Rev. C}\ }\textbf {\bibinfo {volume} {105}},\
  \bibinfo {pages} {L021902} (\bibinfo {year} {2022})},\ \Eprint
  {https://arxiv.org/abs/2201.06670} {arXiv:2201.06670 [hep-ph]} \BibitemShut
  {NoStop}%
\bibitem [{\citenamefont {Marrochio}\ \emph {et~al.}(2015)\citenamefont
  {Marrochio}, \citenamefont {Noronha}, \citenamefont {Denicol}, \citenamefont
  {Luzum}, \citenamefont {Jeon},\ and\ \citenamefont
  {Gale}}]{Marrochio:2013wla}%
  \BibitemOpen
  \bibfield  {author} {\bibinfo {author} {\bibfnamefont {H.}~\bibnamefont
  {Marrochio}}, \bibinfo {author} {\bibfnamefont {J.}~\bibnamefont {Noronha}},
  \bibinfo {author} {\bibfnamefont {G.~S.}\ \bibnamefont {Denicol}}, \bibinfo
  {author} {\bibfnamefont {M.}~\bibnamefont {Luzum}}, \bibinfo {author}
  {\bibfnamefont {S.}~\bibnamefont {Jeon}},\ and\ \bibinfo {author}
  {\bibfnamefont {C.}~\bibnamefont {Gale}},\ }\bibfield  {title} {\bibinfo
  {title} {{Solutions of Conformal Israel-Stewart Relativistic Viscous Fluid
  Dynamics}},\ }\href {https://doi.org/10.1103/PhysRevC.91.014903} {\bibfield
  {journal} {\bibinfo  {journal} {Phys. Rev. C}\ }\textbf {\bibinfo {volume}
  {91}},\ \bibinfo {pages} {014903} (\bibinfo {year} {2015})},\ \Eprint
  {https://arxiv.org/abs/1307.6130} {arXiv:1307.6130 [nucl-th]} \BibitemShut
  {NoStop}%
\bibitem [{\citenamefont {Denicol}\ \emph
  {et~al.}(2014{\natexlab{a}})\citenamefont {Denicol}, \citenamefont {Heinz},
  \citenamefont {Martinez}, \citenamefont {Noronha},\ and\ \citenamefont
  {Strickland}}]{Denicol:2014tha}%
  \BibitemOpen
  \bibfield  {author} {\bibinfo {author} {\bibfnamefont {G.~S.}\ \bibnamefont
  {Denicol}}, \bibinfo {author} {\bibfnamefont {U.~W.}\ \bibnamefont {Heinz}},
  \bibinfo {author} {\bibfnamefont {M.}~\bibnamefont {Martinez}}, \bibinfo
  {author} {\bibfnamefont {J.}~\bibnamefont {Noronha}},\ and\ \bibinfo {author}
  {\bibfnamefont {M.}~\bibnamefont {Strickland}},\ }\bibfield  {title}
  {\bibinfo {title} {{Studying the validity of relativistic hydrodynamics with
  a new exact solution of the Boltzmann equation}},\ }\href
  {https://doi.org/10.1103/PhysRevD.90.125026} {\bibfield  {journal} {\bibinfo
  {journal} {Phys. Rev. D}\ }\textbf {\bibinfo {volume} {90}},\ \bibinfo
  {pages} {125026} (\bibinfo {year} {2014}{\natexlab{a}})},\ \Eprint
  {https://arxiv.org/abs/1408.7048} {arXiv:1408.7048 [hep-ph]} \BibitemShut
  {NoStop}%
\bibitem [{\citenamefont {Denicol}\ \emph
  {et~al.}(2014{\natexlab{b}})\citenamefont {Denicol}, \citenamefont {Heinz},
  \citenamefont {Martinez}, \citenamefont {Noronha},\ and\ \citenamefont
  {Strickland}}]{Denicol:2014xca}%
  \BibitemOpen
  \bibfield  {author} {\bibinfo {author} {\bibfnamefont {G.~S.}\ \bibnamefont
  {Denicol}}, \bibinfo {author} {\bibfnamefont {U.~W.}\ \bibnamefont {Heinz}},
  \bibinfo {author} {\bibfnamefont {M.}~\bibnamefont {Martinez}}, \bibinfo
  {author} {\bibfnamefont {J.}~\bibnamefont {Noronha}},\ and\ \bibinfo {author}
  {\bibfnamefont {M.}~\bibnamefont {Strickland}},\ }\bibfield  {title}
  {\bibinfo {title} {{New Exact Solution of the Relativistic Boltzmann Equation
  and its Hydrodynamic Limit}},\ }\href
  {https://doi.org/10.1103/PhysRevLett.113.202301} {\bibfield  {journal}
  {\bibinfo  {journal} {Phys. Rev. Lett.}\ }\textbf {\bibinfo {volume} {113}},\
  \bibinfo {pages} {202301} (\bibinfo {year} {2014}{\natexlab{b}})},\ \Eprint
  {https://arxiv.org/abs/1408.5646} {arXiv:1408.5646 [hep-ph]} \BibitemShut
  {NoStop}%
\bibitem [{\citenamefont {Nopoush}\ \emph {et~al.}(2015)\citenamefont
  {Nopoush}, \citenamefont {Ryblewski},\ and\ \citenamefont
  {Strickland}}]{Nopoush:2014qba}%
  \BibitemOpen
  \bibfield  {author} {\bibinfo {author} {\bibfnamefont {M.}~\bibnamefont
  {Nopoush}}, \bibinfo {author} {\bibfnamefont {R.}~\bibnamefont {Ryblewski}},\
  and\ \bibinfo {author} {\bibfnamefont {M.}~\bibnamefont {Strickland}},\
  }\bibfield  {title} {\bibinfo {title} {{Anisotropic hydrodynamics for
  conformal Gubser flow}},\ }\href {https://doi.org/10.1103/PhysRevD.91.045007}
  {\bibfield  {journal} {\bibinfo  {journal} {Phys. Rev. D}\ }\textbf {\bibinfo
  {volume} {91}},\ \bibinfo {pages} {045007} (\bibinfo {year} {2015})},\
  \Eprint {https://arxiv.org/abs/1410.6790} {arXiv:1410.6790 [nucl-th]}
  \BibitemShut {NoStop}%
\bibitem [{\citenamefont {Martinez}\ \emph {et~al.}(2017)\citenamefont
  {Martinez}, \citenamefont {McNelis},\ and\ \citenamefont
  {Heinz}}]{Martinez:2017ibh}%
  \BibitemOpen
  \bibfield  {author} {\bibinfo {author} {\bibfnamefont {M.}~\bibnamefont
  {Martinez}}, \bibinfo {author} {\bibfnamefont {M.}~\bibnamefont {McNelis}},\
  and\ \bibinfo {author} {\bibfnamefont {U.}~\bibnamefont {Heinz}},\ }\bibfield
   {title} {\bibinfo {title} {{Anisotropic fluid dynamics for Gubser flow}},\
  }\href {https://doi.org/10.1103/PhysRevC.95.054907} {\bibfield  {journal}
  {\bibinfo  {journal} {Phys. Rev. C}\ }\textbf {\bibinfo {volume} {95}},\
  \bibinfo {pages} {054907} (\bibinfo {year} {2017})},\ \Eprint
  {https://arxiv.org/abs/1703.10955} {arXiv:1703.10955 [nucl-th]} \BibitemShut
  {NoStop}%
\bibitem [{\citenamefont {Chattopadhyay}\ \emph {et~al.}(2018)\citenamefont
  {Chattopadhyay}, \citenamefont {Heinz}, \citenamefont {Pal},\ and\
  \citenamefont {Vujanovic}}]{Chattopadhyay:2018apf}%
  \BibitemOpen
  \bibfield  {author} {\bibinfo {author} {\bibfnamefont {C.}~\bibnamefont
  {Chattopadhyay}}, \bibinfo {author} {\bibfnamefont {U.}~\bibnamefont
  {Heinz}}, \bibinfo {author} {\bibfnamefont {S.}~\bibnamefont {Pal}},\ and\
  \bibinfo {author} {\bibfnamefont {G.}~\bibnamefont {Vujanovic}},\ }\bibfield
  {title} {\bibinfo {title} {{Higher order and anisotropic hydrodynamics for
  Bjorken and Gubser flows}},\ }\href
  {https://doi.org/10.1103/PhysRevC.97.064909} {\bibfield  {journal} {\bibinfo
  {journal} {Phys. Rev. C}\ }\textbf {\bibinfo {volume} {97}},\ \bibinfo
  {pages} {064909} (\bibinfo {year} {2018})},\ \Eprint
  {https://arxiv.org/abs/1801.07755} {arXiv:1801.07755 [nucl-th]} \BibitemShut
  {NoStop}%
\bibitem [{\citenamefont {Shokri}\ and\ \citenamefont
  {Sadooghi}(2018)}]{Shokri:2018qcu}%
  \BibitemOpen
  \bibfield  {author} {\bibinfo {author} {\bibfnamefont {M.}~\bibnamefont
  {Shokri}}\ and\ \bibinfo {author} {\bibfnamefont {N.}~\bibnamefont
  {Sadooghi}},\ }\bibfield  {title} {\bibinfo {title} {{Evolution of magnetic
  fields from the 3 + 1 dimensional self-similar and Gubser flows in ideal
  relativistic magnetohydrodynamics}},\ }\href
  {https://doi.org/10.1007/JHEP11(2018)181} {\bibfield  {journal} {\bibinfo
  {journal} {JHEP}\ }\textbf {\bibinfo {volume} {11}},\ \bibinfo {pages}
  {181}},\ \Eprint {https://arxiv.org/abs/1807.09487} {arXiv:1807.09487
  [nucl-th]} \BibitemShut {NoStop}%
\bibitem [{\citenamefont {Wang}\ \emph {et~al.}(2022)\citenamefont {Wang},
  \citenamefont {Xie}, \citenamefont {Fang},\ and\ \citenamefont
  {Pu}}]{Wang:2021wqq}%
  \BibitemOpen
  \bibfield  {author} {\bibinfo {author} {\bibfnamefont {D.-L.}\ \bibnamefont
  {Wang}}, \bibinfo {author} {\bibfnamefont {X.-Q.}\ \bibnamefont {Xie}},
  \bibinfo {author} {\bibfnamefont {S.}~\bibnamefont {Fang}},\ and\ \bibinfo
  {author} {\bibfnamefont {S.}~\bibnamefont {Pu}},\ }\bibfield  {title}
  {\bibinfo {title} {{Analytic solutions of relativistic dissipative spin
  hydrodynamics with radial expansion in Gubser flow}},\ }\href
  {https://doi.org/10.1103/PhysRevD.105.114050} {\bibfield  {journal} {\bibinfo
   {journal} {Phys. Rev. D}\ }\textbf {\bibinfo {volume} {105}},\ \bibinfo
  {pages} {114050} (\bibinfo {year} {2022})},\ \Eprint
  {https://arxiv.org/abs/2112.15535} {arXiv:2112.15535 [hep-ph]} \BibitemShut
  {NoStop}%
\bibitem [{\citenamefont {Behtash}\ \emph {et~al.}(2018)\citenamefont
  {Behtash}, \citenamefont {Cruz-Camacho},\ and\ \citenamefont
  {Martinez}}]{Behtash:2017wqg}%
  \BibitemOpen
  \bibfield  {author} {\bibinfo {author} {\bibfnamefont {A.}~\bibnamefont
  {Behtash}}, \bibinfo {author} {\bibfnamefont {C.~N.}\ \bibnamefont
  {Cruz-Camacho}},\ and\ \bibinfo {author} {\bibfnamefont {M.}~\bibnamefont
  {Martinez}},\ }\bibfield  {title} {\bibinfo {title} {{Far-from-equilibrium
  attractors and nonlinear dynamical systems approach to the Gubser flow}},\
  }\href {https://doi.org/10.1103/PhysRevD.97.044041} {\bibfield  {journal}
  {\bibinfo  {journal} {Phys. Rev. D}\ }\textbf {\bibinfo {volume} {97}},\
  \bibinfo {pages} {044041} (\bibinfo {year} {2018})},\ \Eprint
  {https://arxiv.org/abs/1711.01745} {arXiv:1711.01745 [hep-th]} \BibitemShut
  {NoStop}%
\bibitem [{\citenamefont {Denicol}\ and\ \citenamefont
  {Noronha}(2019)}]{Denicol:2018pak}%
  \BibitemOpen
  \bibfield  {author} {\bibinfo {author} {\bibfnamefont {G.~S.}\ \bibnamefont
  {Denicol}}\ and\ \bibinfo {author} {\bibfnamefont {J.}~\bibnamefont
  {Noronha}},\ }\bibfield  {title} {\bibinfo {title} {{Hydrodynamic attractor
  and the fate of perturbative expansions in Gubser flow}},\ }\href
  {https://doi.org/10.1103/PhysRevD.99.116004} {\bibfield  {journal} {\bibinfo
  {journal} {Phys. Rev. D}\ }\textbf {\bibinfo {volume} {99}},\ \bibinfo
  {pages} {116004} (\bibinfo {year} {2019})},\ \Eprint
  {https://arxiv.org/abs/1804.04771} {arXiv:1804.04771 [nucl-th]} \BibitemShut
  {NoStop}%
\bibitem [{\citenamefont {Ingles}\ \emph {et~al.}(2025)\citenamefont {Ingles},
  \citenamefont {Salinas San~Mart{\'\i}n}, \citenamefont {Serenone},\ and\
  \citenamefont {Noronha-Hostler}}]{Ingles:2025yrv}%
  \BibitemOpen
  \bibfield  {author} {\bibinfo {author} {\bibfnamefont {K.}~\bibnamefont
  {Ingles}}, \bibinfo {author} {\bibfnamefont {J.}~\bibnamefont {Salinas
  San~Mart{\'\i}n}}, \bibinfo {author} {\bibfnamefont {W.}~\bibnamefont
  {Serenone}},\ and\ \bibinfo {author} {\bibfnamefont {J.}~\bibnamefont
  {Noronha-Hostler}},\ }\bibfield  {title} {\bibinfo {title} {{Viscous Gubser
  flow with conserved charges to benchmark fluid simulations}},\ }\href@noop {}
  {\  (\bibinfo {year} {2025})},\ \Eprint {https://arxiv.org/abs/2503.20021}
  {arXiv:2503.20021 [nucl-th]} \BibitemShut {NoStop}%
\bibitem [{\citenamefont {Singh}\ \emph {et~al.}(2024)\citenamefont {Singh},
  \citenamefont {Dey},\ and\ \citenamefont {Sahoo}}]{Singh:2024emy}%
  \BibitemOpen
  \bibfield  {author} {\bibinfo {author} {\bibfnamefont {K.}~\bibnamefont
  {Singh}}, \bibinfo {author} {\bibfnamefont {J.}~\bibnamefont {Dey}},\ and\
  \bibinfo {author} {\bibfnamefont {R.}~\bibnamefont {Sahoo}},\ }\bibfield
  {title} {\bibinfo {title} {{Electric field induction in quark-gluon plasma
  due to thermoelectric effects}},\ }\href
  {https://doi.org/10.1103/PhysRevD.110.114051} {\bibfield  {journal} {\bibinfo
   {journal} {Phys. Rev. D}\ }\textbf {\bibinfo {volume} {110}},\ \bibinfo
  {pages} {114051} (\bibinfo {year} {2024})},\ \Eprint
  {https://arxiv.org/abs/2405.12510} {arXiv:2405.12510 [hep-ph]} \BibitemShut
  {NoStop}%
\bibitem [{\citenamefont {Hatta}\ and\ \citenamefont
  {Xiao}(2014)}]{Hatta:2014upa}%
  \BibitemOpen
  \bibfield  {author} {\bibinfo {author} {\bibfnamefont {Y.}~\bibnamefont
  {Hatta}}\ and\ \bibinfo {author} {\bibfnamefont {B.-W.}\ \bibnamefont
  {Xiao}},\ }\bibfield  {title} {\bibinfo {title} {{Building up the elliptic
  flow: analytical insights}},\ }\href
  {https://doi.org/10.1016/j.physletb.2014.07.017} {\bibfield  {journal}
  {\bibinfo  {journal} {Phys. Lett. B}\ }\textbf {\bibinfo {volume} {736}},\
  \bibinfo {pages} {180} (\bibinfo {year} {2014})},\ \Eprint
  {https://arxiv.org/abs/1405.1984} {arXiv:1405.1984 [nucl-th]} \BibitemShut
  {NoStop}%
\bibitem [{\citenamefont {Gursoy}\ \emph {et~al.}(2014)\citenamefont {Gursoy},
  \citenamefont {Kharzeev},\ and\ \citenamefont {Rajagopal}}]{Gursoy:2014aka}%
  \BibitemOpen
  \bibfield  {author} {\bibinfo {author} {\bibfnamefont {U.}~\bibnamefont
  {Gursoy}}, \bibinfo {author} {\bibfnamefont {D.}~\bibnamefont {Kharzeev}},\
  and\ \bibinfo {author} {\bibfnamefont {K.}~\bibnamefont {Rajagopal}},\
  }\bibfield  {title} {\bibinfo {title} {{Magnetohydrodynamics, charged
  currents and directed flow in heavy ion collisions}},\ }\href
  {https://doi.org/10.1103/PhysRevC.89.054905} {\bibfield  {journal} {\bibinfo
  {journal} {Phys. Rev. C}\ }\textbf {\bibinfo {volume} {89}},\ \bibinfo
  {pages} {054905} (\bibinfo {year} {2014})},\ \Eprint
  {https://arxiv.org/abs/1401.3805} {arXiv:1401.3805 [hep-ph]} \BibitemShut
  {NoStop}%
\bibitem [{\citenamefont {Hatta}\ \emph {et~al.}(2015)\citenamefont {Hatta},
  \citenamefont {Monnai},\ and\ \citenamefont {Xiao}}]{Hatta:2015era}%
  \BibitemOpen
  \bibfield  {author} {\bibinfo {author} {\bibfnamefont {Y.}~\bibnamefont
  {Hatta}}, \bibinfo {author} {\bibfnamefont {A.}~\bibnamefont {Monnai}},\ and\
  \bibinfo {author} {\bibfnamefont {B.-W.}\ \bibnamefont {Xiao}},\ }\bibfield
  {title} {\bibinfo {title} {{Flow harmonics $v_n$ at finite density}},\ }\href
  {https://doi.org/10.1103/PhysRevD.92.114010} {\bibfield  {journal} {\bibinfo
  {journal} {Phys. Rev. D}\ }\textbf {\bibinfo {volume} {92}},\ \bibinfo
  {pages} {114010} (\bibinfo {year} {2015})},\ \Eprint
  {https://arxiv.org/abs/1505.04226} {arXiv:1505.04226 [hep-ph]} \BibitemShut
  {NoStop}%
\bibitem [{\citenamefont {Bagchi}\ \emph {et~al.}(2024)\citenamefont {Bagchi},
  \citenamefont {Das},\ and\ \citenamefont {Mishra}}]{Bagchi:2023vfv}%
  \BibitemOpen
  \bibfield  {author} {\bibinfo {author} {\bibfnamefont {P.}~\bibnamefont
  {Bagchi}}, \bibinfo {author} {\bibfnamefont {A.}~\bibnamefont {Das}},\ and\
  \bibinfo {author} {\bibfnamefont {A.~P.}\ \bibnamefont {Mishra}},\ }\bibfield
   {title} {\bibinfo {title} {{Does Quarkonia Suppression serve as a probe for
  the deconfinement in small systems?}},\ }\href
  {https://doi.org/10.1103/PhysRevD.110.014017} {\bibfield  {journal} {\bibinfo
   {journal} {Phys. Rev. D}\ }\textbf {\bibinfo {volume} {110}},\ \bibinfo
  {pages} {014017} (\bibinfo {year} {2024})},\ \Eprint
  {https://arxiv.org/abs/2310.12267} {arXiv:2310.12267 [nucl-th]} \BibitemShut
  {NoStop}%
\bibitem [{\citenamefont {Singh}\ \emph {et~al.}(2025)\citenamefont {Singh},
  \citenamefont {Bagchi}, \citenamefont {Sahoo},\ and\ \citenamefont
  {Alam}}]{Singh:2025xrd}%
  \BibitemOpen
  \bibfield  {author} {\bibinfo {author} {\bibfnamefont {C.~R.}\ \bibnamefont
  {Singh}}, \bibinfo {author} {\bibfnamefont {P.}~\bibnamefont {Bagchi}},
  \bibinfo {author} {\bibfnamefont {R.}~\bibnamefont {Sahoo}},\ and\ \bibinfo
  {author} {\bibfnamefont {J.-e.}\ \bibnamefont {Alam}},\ }\bibfield  {title}
  {\bibinfo {title} {{Exploring QGP-like phenomena with charmonia in p+p
  collisions at s=13{\,}{\,}TeV}},\ }\href {https://doi.org/10.1103/b3zr-8yg9}
  {\bibfield  {journal} {\bibinfo  {journal} {Phys. Rev. D}\ }\textbf {\bibinfo
  {volume} {112}},\ \bibinfo {pages} {014017} (\bibinfo {year} {2025})},\
  \Eprint {https://arxiv.org/abs/2501.00753} {arXiv:2501.00753 [hep-ph]}
  \BibitemShut {NoStop}%
\bibitem [{\citenamefont {Paquet}(2023)}]{Paquet:2023bdx}%
  \BibitemOpen
  \bibfield  {author} {\bibinfo {author} {\bibfnamefont {J.-F.}\ \bibnamefont
  {Paquet}},\ }\bibfield  {title} {\bibinfo {title} {{Thermal photon production
  in Gubser inviscid relativistic fluid dynamics}},\ }\href
  {https://doi.org/10.1103/PhysRevC.108.064912} {\bibfield  {journal} {\bibinfo
   {journal} {Phys. Rev. C}\ }\textbf {\bibinfo {volume} {108}},\ \bibinfo
  {pages} {064912} (\bibinfo {year} {2023})},\ \Eprint
  {https://arxiv.org/abs/2305.10669} {arXiv:2305.10669 [nucl-th]} \BibitemShut
  {NoStop}%
\bibitem [{\citenamefont {Naik}\ and\ \citenamefont
  {Sreekanth}(2025)}]{Naik:2025pjt}%
  \BibitemOpen
  \bibfield  {author} {\bibinfo {author} {\bibfnamefont {L.~J.}\ \bibnamefont
  {Naik}}\ and\ \bibinfo {author} {\bibfnamefont {V.}~\bibnamefont
  {Sreekanth}},\ }\bibfield  {title} {\bibinfo {title} {{Thermal dilepton
  production within conformal viscous Gubser flow}},\ }\href
  {https://doi.org/10.1140/epjc/s10052-025-14393-6} {\bibfield  {journal}
  {\bibinfo  {journal} {Eur. Phys. J. C}\ }\textbf {\bibinfo {volume} {85}},\
  \bibinfo {pages} {664} (\bibinfo {year} {2025})},\ \Eprint
  {https://arxiv.org/abs/2506.01500} {arXiv:2506.01500 [hep-ph]} \BibitemShut
  {NoStop}%
\bibitem [{\citenamefont {Dwibedi}\ \emph {et~al.}(2025)\citenamefont
  {Dwibedi}, \citenamefont {Panda}, \citenamefont {Ghosh},\ and\ \citenamefont
  {Roy}}]{Dwibedi:2025xho}%
  \BibitemOpen
  \bibfield  {author} {\bibinfo {author} {\bibfnamefont {A.}~\bibnamefont
  {Dwibedi}}, \bibinfo {author} {\bibfnamefont {A.~K.}\ \bibnamefont {Panda}},
  \bibinfo {author} {\bibfnamefont {S.}~\bibnamefont {Ghosh}},\ and\ \bibinfo
  {author} {\bibfnamefont {V.}~\bibnamefont {Roy}},\ }\bibfield  {title}
  {\bibinfo {title} {{Probing Dynamical Electrical Conductivity via Dilepton
  Emission: A Kinetic theory approach}},\ }\href@noop {} {\  (\bibinfo {year}
  {2025})},\ \Eprint {https://arxiv.org/abs/2508.16988} {arXiv:2508.16988
  [nucl-th]} \BibitemShut {NoStop}%
\bibitem [{\citenamefont {Feinberg}(1976)}]{Feinberg:1976ua}%
  \BibitemOpen
  \bibfield  {author} {\bibinfo {author} {\bibfnamefont {E.~L.}\ \bibnamefont
  {Feinberg}},\ }\bibfield  {title} {\bibinfo {title} {{Direct Production of
  Photons and Dileptons in Thermodynamical Models of Multiple Hadron
  Production}},\ }\href@noop {} {\bibfield  {journal} {\bibinfo  {journal}
  {Nuovo Cim. A}\ }\textbf {\bibinfo {volume} {34}},\ \bibinfo {pages} {391}
  (\bibinfo {year} {1976})}\BibitemShut {NoStop}%
\bibitem [{\citenamefont {Shuryak}(1978)}]{Shuryak:1978ij}%
  \BibitemOpen
  \bibfield  {author} {\bibinfo {author} {\bibfnamefont {E.~V.}\ \bibnamefont
  {Shuryak}},\ }\bibfield  {title} {\bibinfo {title} {{Quark-Gluon Plasma and
  Hadronic Production of Leptons, Photons and Psions}},\ }\href
  {https://doi.org/10.1016/0370-2693(78)90370-2} {\bibfield  {journal}
  {\bibinfo  {journal} {Phys. Lett. B}\ }\textbf {\bibinfo {volume} {78}},\
  \bibinfo {pages} {150} (\bibinfo {year} {1978})}\BibitemShut {NoStop}%
\bibitem [{\citenamefont {Kapusta}\ \emph {et~al.}(1991)\citenamefont
  {Kapusta}, \citenamefont {Lichard},\ and\ \citenamefont
  {Seibert}}]{Kapusta:1991qp}%
  \BibitemOpen
  \bibfield  {author} {\bibinfo {author} {\bibfnamefont {J.~I.}\ \bibnamefont
  {Kapusta}}, \bibinfo {author} {\bibfnamefont {P.}~\bibnamefont {Lichard}},\
  and\ \bibinfo {author} {\bibfnamefont {D.}~\bibnamefont {Seibert}},\
  }\bibfield  {title} {\bibinfo {title} {{High-energy photons from quark -
  gluon plasma versus hot hadronic gas}},\ }\href
  {https://doi.org/10.1103/PhysRevD.47.4171} {\bibfield  {journal} {\bibinfo
  {journal} {Phys. Rev. D}\ }\textbf {\bibinfo {volume} {44}},\ \bibinfo
  {pages} {2774} (\bibinfo {year} {1991})},\ \bibinfo {note} {[Erratum:
  Phys.Rev.D 47, 4171 (1993)]}\BibitemShut {NoStop}%
\bibitem [{\citenamefont {Peitzmann}\ and\ \citenamefont
  {Thoma}(2002)}]{Peitzmann:2001mz}%
  \BibitemOpen
  \bibfield  {author} {\bibinfo {author} {\bibfnamefont {T.}~\bibnamefont
  {Peitzmann}}\ and\ \bibinfo {author} {\bibfnamefont {M.~H.}\ \bibnamefont
  {Thoma}},\ }\bibfield  {title} {\bibinfo {title} {{Direct photons from
  relativistic heavy ion collisions}},\ }\href
  {https://doi.org/10.1016/S0370-1573(02)00012-1} {\bibfield  {journal}
  {\bibinfo  {journal} {Phys. Rept.}\ }\textbf {\bibinfo {volume} {364}},\
  \bibinfo {pages} {175} (\bibinfo {year} {2002})},\ \Eprint
  {https://arxiv.org/abs/hep-ph/0111114} {arXiv:hep-ph/0111114} \BibitemShut
  {NoStop}%
\bibitem [{\citenamefont {Rapp}\ and\ \citenamefont {van
  Hees}(2016)}]{Rapp:2016xzw}%
  \BibitemOpen
  \bibfield  {author} {\bibinfo {author} {\bibfnamefont {R.}~\bibnamefont
  {Rapp}}\ and\ \bibinfo {author} {\bibfnamefont {H.}~\bibnamefont {van
  Hees}},\ }\bibfield  {title} {\bibinfo {title} {{Thermal Electromagnetic
  Radiation in Heavy-Ion Collisions}},\ }\href
  {https://doi.org/10.1140/epja/i2016-16257-0} {\bibfield  {journal} {\bibinfo
  {journal} {Eur. Phys. J. A}\ }\textbf {\bibinfo {volume} {52}},\ \bibinfo
  {pages} {257} (\bibinfo {year} {2016})},\ \Eprint
  {https://arxiv.org/abs/1608.05279} {arXiv:1608.05279 [hep-ph]} \BibitemShut
  {NoStop}%
\bibitem [{\citenamefont {David}(2020)}]{David:2019wpt}%
  \BibitemOpen
  \bibfield  {author} {\bibinfo {author} {\bibfnamefont {G.}~\bibnamefont
  {David}},\ }\bibfield  {title} {\bibinfo {title} {{Direct real photons in
  relativistic heavy ion collisions}},\ }\href
  {https://doi.org/10.1088/1361-6633/ab6f57} {\bibfield  {journal} {\bibinfo
  {journal} {Rept. Prog. Phys.}\ }\textbf {\bibinfo {volume} {83}},\ \bibinfo
  {pages} {046301} (\bibinfo {year} {2020})},\ \Eprint
  {https://arxiv.org/abs/1907.08893} {arXiv:1907.08893 [nucl-ex]} \BibitemShut
  {NoStop}%
\bibitem [{\citenamefont {Geurts}\ and\ \citenamefont
  {Tripolt}(2023)}]{Geurts:2022xmk}%
  \BibitemOpen
  \bibfield  {author} {\bibinfo {author} {\bibfnamefont {F.}~\bibnamefont
  {Geurts}}\ and\ \bibinfo {author} {\bibfnamefont {R.-A.}\ \bibnamefont
  {Tripolt}},\ }\bibfield  {title} {\bibinfo {title} {{Electromagnetic probes:
  Theory and experiment}},\ }\href {https://doi.org/10.1016/j.ppnp.2022.104004}
  {\bibfield  {journal} {\bibinfo  {journal} {Prog. Part. Nucl. Phys.}\
  }\textbf {\bibinfo {volume} {128}},\ \bibinfo {pages} {104004} (\bibinfo
  {year} {2023})},\ \Eprint {https://arxiv.org/abs/2210.01622}
  {arXiv:2210.01622 [hep-ph]} \BibitemShut {NoStop}%
\bibitem [{\citenamefont {Bhatt}\ \emph {et~al.}(2012)\citenamefont {Bhatt},
  \citenamefont {Mishra},\ and\ \citenamefont {Sreekanth}}]{Bhatt:2011kx}%
  \BibitemOpen
  \bibfield  {author} {\bibinfo {author} {\bibfnamefont {J.~R.}\ \bibnamefont
  {Bhatt}}, \bibinfo {author} {\bibfnamefont {H.}~\bibnamefont {Mishra}},\ and\
  \bibinfo {author} {\bibfnamefont {V.}~\bibnamefont {Sreekanth}},\ }\bibfield
  {title} {\bibinfo {title} {{Cavitation and thermal dilepton production in
  QGP}},\ }\href {https://doi.org/10.1016/j.nuclphysa.2011.11.012} {\bibfield
  {journal} {\bibinfo  {journal} {Nucl. Phys. A}\ }\textbf {\bibinfo {volume}
  {875}},\ \bibinfo {pages} {181} (\bibinfo {year} {2012})},\ \Eprint
  {https://arxiv.org/abs/1101.5597} {arXiv:1101.5597 [hep-ph]} \BibitemShut
  {NoStop}%
\bibitem [{\citenamefont {Linnyk}\ \emph {et~al.}(2013)\citenamefont {Linnyk},
  \citenamefont {Konchakovski}, \citenamefont {Cassing},\ and\ \citenamefont
  {Bratkovskaya}}]{Linnyk:2013hta}%
  \BibitemOpen
  \bibfield  {author} {\bibinfo {author} {\bibfnamefont {O.}~\bibnamefont
  {Linnyk}}, \bibinfo {author} {\bibfnamefont {V.~P.}\ \bibnamefont
  {Konchakovski}}, \bibinfo {author} {\bibfnamefont {W.}~\bibnamefont
  {Cassing}},\ and\ \bibinfo {author} {\bibfnamefont {E.~L.}\ \bibnamefont
  {Bratkovskaya}},\ }\bibfield  {title} {\bibinfo {title} {{Photon elliptic
  flow in relativistic heavy-ion collisions: hadronic versus partonic
  sources}},\ }\href {https://doi.org/10.1103/PhysRevC.88.034904} {\bibfield
  {journal} {\bibinfo  {journal} {Phys. Rev. C}\ }\textbf {\bibinfo {volume}
  {88}},\ \bibinfo {pages} {034904} (\bibinfo {year} {2013})},\ \Eprint
  {https://arxiv.org/abs/1304.7030} {arXiv:1304.7030 [nucl-th]} \BibitemShut
  {NoStop}%
\bibitem [{\citenamefont {Chandra}\ and\ \citenamefont
  {Sreekanth}(2015)}]{Chandra:2015rdz}%
  \BibitemOpen
  \bibfield  {author} {\bibinfo {author} {\bibfnamefont {V.}~\bibnamefont
  {Chandra}}\ and\ \bibinfo {author} {\bibfnamefont {V.}~\bibnamefont
  {Sreekanth}},\ }\bibfield  {title} {\bibinfo {title} {{Quark and gluon
  distribution functions in a viscous quark-gluon plasma medium and dilepton
  production via $q\bar{q}$ annihilation}},\ }\href
  {https://doi.org/10.1103/PhysRevD.92.094027} {\bibfield  {journal} {\bibinfo
  {journal} {Phys. Rev. D}\ }\textbf {\bibinfo {volume} {92}},\ \bibinfo
  {pages} {094027} (\bibinfo {year} {2015})},\ \Eprint
  {https://arxiv.org/abs/1511.01208} {arXiv:1511.01208 [nucl-th]} \BibitemShut
  {NoStop}%
\bibitem [{\citenamefont {Vujanovic}\ \emph {et~al.}(2016)\citenamefont
  {Vujanovic}, \citenamefont {Paquet}, \citenamefont {Denicol}, \citenamefont
  {Luzum}, \citenamefont {Jeon},\ and\ \citenamefont
  {Gale}}]{Vujanovic:2016anq}%
  \BibitemOpen
  \bibfield  {author} {\bibinfo {author} {\bibfnamefont {G.}~\bibnamefont
  {Vujanovic}}, \bibinfo {author} {\bibfnamefont {J.-F.}\ \bibnamefont
  {Paquet}}, \bibinfo {author} {\bibfnamefont {G.~S.}\ \bibnamefont {Denicol}},
  \bibinfo {author} {\bibfnamefont {M.}~\bibnamefont {Luzum}}, \bibinfo
  {author} {\bibfnamefont {S.}~\bibnamefont {Jeon}},\ and\ \bibinfo {author}
  {\bibfnamefont {C.}~\bibnamefont {Gale}},\ }\bibfield  {title} {\bibinfo
  {title} {{Electromagnetic radiation as a probe of the initial state and of
  viscous dynamics in relativistic nuclear collisions}},\ }\href
  {https://doi.org/10.1103/PhysRevC.94.014904} {\bibfield  {journal} {\bibinfo
  {journal} {Phys. Rev. C}\ }\textbf {\bibinfo {volume} {94}},\ \bibinfo
  {pages} {014904} (\bibinfo {year} {2016})},\ \Eprint
  {https://arxiv.org/abs/1602.01455} {arXiv:1602.01455 [nucl-th]} \BibitemShut
  {NoStop}%
\bibitem [{\citenamefont {Naik}\ \emph {et~al.}(2022)\citenamefont {Naik},
  \citenamefont {Sreekanth}, \citenamefont {Kurian},\ and\ \citenamefont
  {Chandra}}]{Naik:2020jfc}%
  \BibitemOpen
  \bibfield  {author} {\bibinfo {author} {\bibfnamefont {L.~J.}\ \bibnamefont
  {Naik}}, \bibinfo {author} {\bibfnamefont {V.}~\bibnamefont {Sreekanth}},
  \bibinfo {author} {\bibfnamefont {M.}~\bibnamefont {Kurian}},\ and\ \bibinfo
  {author} {\bibfnamefont {V.}~\bibnamefont {Chandra}},\ }\bibfield  {title}
  {\bibinfo {title} {{Thermal dilepton production in collisional hot QCD medium
  in the presence of chromo-turbulent fields}},\ }\href
  {https://doi.org/10.1088/1361-6471/ac65a5} {\bibfield  {journal} {\bibinfo
  {journal} {J. Phys. G}\ }\textbf {\bibinfo {volume} {49}},\ \bibinfo {pages}
  {075103} (\bibinfo {year} {2022})},\ \Eprint
  {https://arxiv.org/abs/2003.13645} {arXiv:2003.13645 [hep-ph]} \BibitemShut
  {NoStop}%
\bibitem [{\citenamefont {G{\"o}tz}\ \emph {et~al.}(2022)\citenamefont
  {G{\"o}tz}, \citenamefont {Sch{\"a}fer}, \citenamefont {Garcia-Montero},
  \citenamefont {Paquet}, \citenamefont {Elfner},\ and\ \citenamefont
  {Gale}}]{Gotz:2021dco}%
  \BibitemOpen
  \bibfield  {author} {\bibinfo {author} {\bibfnamefont {N.}~\bibnamefont
  {G{\"o}tz}}, \bibinfo {author} {\bibfnamefont {A.}~\bibnamefont
  {Sch{\"a}fer}}, \bibinfo {author} {\bibfnamefont {O.}~\bibnamefont
  {Garcia-Montero}}, \bibinfo {author} {\bibfnamefont {J.-F.}\ \bibnamefont
  {Paquet}}, \bibinfo {author} {\bibfnamefont {H.}~\bibnamefont {Elfner}},\
  and\ \bibinfo {author} {\bibfnamefont {C.}~\bibnamefont {Gale}} (\bibinfo
  {collaboration} {SMASH}),\ }\bibfield  {title} {\bibinfo {title}
  {{Out-of-equilibrium photon production in the late stages of relativistic
  heavy-ion collisions}},\ }\href {https://doi.org/10.1103/PhysRevC.105.044910}
  {\bibfield  {journal} {\bibinfo  {journal} {Phys. Rev. C}\ }\textbf {\bibinfo
  {volume} {105}},\ \bibinfo {pages} {044910} (\bibinfo {year} {2022})},\
  \bibinfo {note} {[Erratum: Phys.Rev.C 109, 049901 (2024)]},\ \Eprint
  {https://arxiv.org/abs/2111.13603} {arXiv:2111.13603 [hep-ph]} \BibitemShut
  {NoStop}%
\bibitem [{\citenamefont {Naik}\ and\ \citenamefont
  {Sreekanth}(2023)}]{Naik:2022pyk}%
  \BibitemOpen
  \bibfield  {author} {\bibinfo {author} {\bibfnamefont {L.~J.}\ \bibnamefont
  {Naik}}\ and\ \bibinfo {author} {\bibfnamefont {V.}~\bibnamefont
  {Sreekanth}},\ }\bibfield  {title} {\bibinfo {title} {{Second order
  hydrodynamics based on effective kinetic theory and electromagnetic signals
  from QGP}},\ }\href {https://doi.org/10.1088/1361-6471/aca924} {\bibfield
  {journal} {\bibinfo  {journal} {J. Phys. G}\ }\textbf {\bibinfo {volume}
  {50}},\ \bibinfo {pages} {025102} (\bibinfo {year} {2023})},\ \Eprint
  {https://arxiv.org/abs/2207.05310} {arXiv:2207.05310 [nucl-th]} \BibitemShut
  {NoStop}%
\bibitem [{\citenamefont {Garcia-Montero}\ \emph {et~al.}(2024)\citenamefont
  {Garcia-Montero}, \citenamefont {Mazeliauskas}, \citenamefont {Plaschke},\
  and\ \citenamefont {Schlichting}}]{Garcia-Montero:2023lrd}%
  \BibitemOpen
  \bibfield  {author} {\bibinfo {author} {\bibfnamefont {O.}~\bibnamefont
  {Garcia-Montero}}, \bibinfo {author} {\bibfnamefont {A.}~\bibnamefont
  {Mazeliauskas}}, \bibinfo {author} {\bibfnamefont {P.}~\bibnamefont
  {Plaschke}},\ and\ \bibinfo {author} {\bibfnamefont {S.}~\bibnamefont
  {Schlichting}},\ }\bibfield  {title} {\bibinfo {title} {{Pre-equilibrium
  photons from the early stages of heavy-ion collisions}},\ }\href
  {https://doi.org/10.1007/JHEP03(2024)053} {\bibfield  {journal} {\bibinfo
  {journal} {JHEP}\ }\textbf {\bibinfo {volume} {03}},\ \bibinfo {pages}
  {053}},\ \Eprint {https://arxiv.org/abs/2308.09747} {arXiv:2308.09747
  [hep-ph]} \BibitemShut {NoStop}%
\bibitem [{\citenamefont {Churchill}\ \emph {et~al.}(2024)\citenamefont
  {Churchill}, \citenamefont {Du}, \citenamefont {Gale}, \citenamefont
  {Jackson},\ and\ \citenamefont {Jeon}}]{Churchill:2023vpt}%
  \BibitemOpen
  \bibfield  {author} {\bibinfo {author} {\bibfnamefont {J.}~\bibnamefont
  {Churchill}}, \bibinfo {author} {\bibfnamefont {L.}~\bibnamefont {Du}},
  \bibinfo {author} {\bibfnamefont {C.}~\bibnamefont {Gale}}, \bibinfo {author}
  {\bibfnamefont {G.}~\bibnamefont {Jackson}},\ and\ \bibinfo {author}
  {\bibfnamefont {S.}~\bibnamefont {Jeon}},\ }\bibfield  {title} {\bibinfo
  {title} {{Dilepton production at next-to-leading order and intermediate
  invariant-mass observables}},\ }\href
  {https://doi.org/10.1103/PhysRevC.109.044915} {\bibfield  {journal} {\bibinfo
   {journal} {Phys. Rev. C}\ }\textbf {\bibinfo {volume} {109}},\ \bibinfo
  {pages} {044915} (\bibinfo {year} {2024})},\ \Eprint
  {https://arxiv.org/abs/2311.06675} {arXiv:2311.06675 [nucl-th]} \BibitemShut
  {NoStop}%
\bibitem [{\citenamefont {Aurenche}\ \emph {et~al.}(1998)\citenamefont
  {Aurenche}, \citenamefont {Gelis}, \citenamefont {Kobes},\ and\ \citenamefont
  {Zaraket}}]{Aurenche:1998nw}%
  \BibitemOpen
  \bibfield  {author} {\bibinfo {author} {\bibfnamefont {P.}~\bibnamefont
  {Aurenche}}, \bibinfo {author} {\bibfnamefont {F.}~\bibnamefont {Gelis}},
  \bibinfo {author} {\bibfnamefont {R.}~\bibnamefont {Kobes}},\ and\ \bibinfo
  {author} {\bibfnamefont {H.}~\bibnamefont {Zaraket}},\ }\bibfield  {title}
  {\bibinfo {title} {{Bremsstrahlung and photon production in thermal QCD}},\
  }\href {https://doi.org/10.1103/PhysRevD.58.085003} {\bibfield  {journal}
  {\bibinfo  {journal} {Phys. Rev. D}\ }\textbf {\bibinfo {volume} {58}},\
  \bibinfo {pages} {085003} (\bibinfo {year} {1998})},\ \Eprint
  {https://arxiv.org/abs/hep-ph/9804224} {arXiv:hep-ph/9804224} \BibitemShut
  {NoStop}%
\bibitem [{\citenamefont {Arnold}\ \emph {et~al.}(2001)\citenamefont {Arnold},
  \citenamefont {Moore},\ and\ \citenamefont {Yaffe}}]{Arnold:2001ms}%
  \BibitemOpen
  \bibfield  {author} {\bibinfo {author} {\bibfnamefont {P.~B.}\ \bibnamefont
  {Arnold}}, \bibinfo {author} {\bibfnamefont {G.~D.}\ \bibnamefont {Moore}},\
  and\ \bibinfo {author} {\bibfnamefont {L.~G.}\ \bibnamefont {Yaffe}},\
  }\bibfield  {title} {\bibinfo {title} {{Photon emission from quark gluon
  plasma: Complete leading order results}},\ }\href
  {https://doi.org/10.1088/1126-6708/2001/12/009} {\bibfield  {journal}
  {\bibinfo  {journal} {JHEP}\ }\textbf {\bibinfo {volume} {12}},\ \bibinfo
  {pages} {009}},\ \Eprint {https://arxiv.org/abs/hep-ph/0111107}
  {arXiv:hep-ph/0111107} \BibitemShut {NoStop}%
\bibitem [{\citenamefont {Steffen}\ and\ \citenamefont
  {Thoma}(2001)}]{Steffen:2001pv}%
  \BibitemOpen
  \bibfield  {author} {\bibinfo {author} {\bibfnamefont {F.~D.}\ \bibnamefont
  {Steffen}}\ and\ \bibinfo {author} {\bibfnamefont {M.~H.}\ \bibnamefont
  {Thoma}},\ }\bibfield  {title} {\bibinfo {title} {{Hard thermal photon
  production in relativistic heavy ion collisions}},\ }\href
  {https://doi.org/10.1016/S0370-2693(01)00525-1} {\bibfield  {journal}
  {\bibinfo  {journal} {Phys. Lett. B}\ }\textbf {\bibinfo {volume} {510}},\
  \bibinfo {pages} {98} (\bibinfo {year} {2001})},\ \bibinfo {note} {[Erratum:
  Phys.Lett.B 660, 604--606 (2008)]},\ \Eprint
  {https://arxiv.org/abs/hep-ph/0103044} {arXiv:hep-ph/0103044} \BibitemShut
  {NoStop}%
\bibitem [{\citenamefont {Traxler}\ \emph {et~al.}(1995)\citenamefont
  {Traxler}, \citenamefont {Vija},\ and\ \citenamefont
  {Thoma}}]{Traxler:1994hy}%
  \BibitemOpen
  \bibfield  {author} {\bibinfo {author} {\bibfnamefont {C.~T.}\ \bibnamefont
  {Traxler}}, \bibinfo {author} {\bibfnamefont {H.}~\bibnamefont {Vija}},\ and\
  \bibinfo {author} {\bibfnamefont {M.~H.}\ \bibnamefont {Thoma}},\ }\bibfield
  {title} {\bibinfo {title} {{Hard photon production rate of a quark - gluon
  plasma at finite quark chemical potential}},\ }\href
  {https://doi.org/10.1016/0370-2693(95)00004-5} {\bibfield  {journal}
  {\bibinfo  {journal} {Phys. Lett. B}\ }\textbf {\bibinfo {volume} {346}},\
  \bibinfo {pages} {329} (\bibinfo {year} {1995})},\ \Eprint
  {https://arxiv.org/abs/hep-ph/9410309} {arXiv:hep-ph/9410309} \BibitemShut
  {NoStop}%
\bibitem [{\citenamefont {Schenke}\ and\ \citenamefont
  {Strickland}(2007)}]{Schenke:2006yp}%
  \BibitemOpen
  \bibfield  {author} {\bibinfo {author} {\bibfnamefont {B.}~\bibnamefont
  {Schenke}}\ and\ \bibinfo {author} {\bibfnamefont {M.}~\bibnamefont
  {Strickland}},\ }\bibfield  {title} {\bibinfo {title} {{Photon production
  from an anisotropic quark-gluon plasma}},\ }\href
  {https://doi.org/10.1103/PhysRevD.76.025023} {\bibfield  {journal} {\bibinfo
  {journal} {Phys. Rev. D}\ }\textbf {\bibinfo {volume} {76}},\ \bibinfo
  {pages} {025023} (\bibinfo {year} {2007})},\ \Eprint
  {https://arxiv.org/abs/hep-ph/0611332} {arXiv:hep-ph/0611332} \BibitemShut
  {NoStop}%
\bibitem [{\citenamefont {Bhatt}\ \emph {et~al.}(2010)\citenamefont {Bhatt},
  \citenamefont {Mishra},\ and\ \citenamefont {Sreekanth}}]{Bhatt:2010cy}%
  \BibitemOpen
  \bibfield  {author} {\bibinfo {author} {\bibfnamefont {J.~R.}\ \bibnamefont
  {Bhatt}}, \bibinfo {author} {\bibfnamefont {H.}~\bibnamefont {Mishra}},\ and\
  \bibinfo {author} {\bibfnamefont {V.}~\bibnamefont {Sreekanth}},\ }\bibfield
  {title} {\bibinfo {title} {{Thermal photons in QGP and non-ideal effects}},\
  }\href {https://doi.org/10.1007/JHEP11(2010)106} {\bibfield  {journal}
  {\bibinfo  {journal} {JHEP}\ }\textbf {\bibinfo {volume} {11}},\ \bibinfo
  {pages} {106}},\ \Eprint {https://arxiv.org/abs/1011.1969} {arXiv:1011.1969
  [hep-ph]} \BibitemShut {NoStop}%
\bibitem [{\citenamefont {Shen}\ \emph {et~al.}(2015)\citenamefont {Shen},
  \citenamefont {Heinz}, \citenamefont {Paquet}, \citenamefont {Kozlov},\ and\
  \citenamefont {Gale}}]{Shen:2013cca}%
  \BibitemOpen
  \bibfield  {author} {\bibinfo {author} {\bibfnamefont {C.}~\bibnamefont
  {Shen}}, \bibinfo {author} {\bibfnamefont {U.~W.}\ \bibnamefont {Heinz}},
  \bibinfo {author} {\bibfnamefont {J.-F.}\ \bibnamefont {Paquet}}, \bibinfo
  {author} {\bibfnamefont {I.}~\bibnamefont {Kozlov}},\ and\ \bibinfo {author}
  {\bibfnamefont {C.}~\bibnamefont {Gale}},\ }\bibfield  {title} {\bibinfo
  {title} {{Anisotropic flow of thermal photons as a quark-gluon plasma
  viscometer}},\ }\href {https://doi.org/10.1103/PhysRevC.91.024908} {\bibfield
   {journal} {\bibinfo  {journal} {Phys. Rev. C}\ }\textbf {\bibinfo {volume}
  {91}},\ \bibinfo {pages} {024908} (\bibinfo {year} {2015})},\ \Eprint
  {https://arxiv.org/abs/1308.2111} {arXiv:1308.2111 [nucl-th]} \BibitemShut
  {NoStop}%
\bibitem [{\citenamefont {Bhattacharya}\ \emph {et~al.}(2016)\citenamefont
  {Bhattacharya}, \citenamefont {Ryblewski},\ and\ \citenamefont
  {Strickland}}]{Bhattacharya:2015ada}%
  \BibitemOpen
  \bibfield  {author} {\bibinfo {author} {\bibfnamefont {L.}~\bibnamefont
  {Bhattacharya}}, \bibinfo {author} {\bibfnamefont {R.}~\bibnamefont
  {Ryblewski}},\ and\ \bibinfo {author} {\bibfnamefont {M.}~\bibnamefont
  {Strickland}},\ }\bibfield  {title} {\bibinfo {title} {{Photon production
  from a nonequilibrium quark-gluon plasma}},\ }\href
  {https://doi.org/10.1103/PhysRevD.93.065005} {\bibfield  {journal} {\bibinfo
  {journal} {Phys. Rev. D}\ }\textbf {\bibinfo {volume} {93}},\ \bibinfo
  {pages} {065005} (\bibinfo {year} {2016})},\ \Eprint
  {https://arxiv.org/abs/1507.06605} {arXiv:1507.06605 [hep-ph]} \BibitemShut
  {NoStop}%
\bibitem [{\citenamefont {Hauksson}\ \emph {et~al.}(2018)\citenamefont
  {Hauksson}, \citenamefont {Jeon},\ and\ \citenamefont
  {Gale}}]{Hauksson:2017udm}%
  \BibitemOpen
  \bibfield  {author} {\bibinfo {author} {\bibfnamefont {S.}~\bibnamefont
  {Hauksson}}, \bibinfo {author} {\bibfnamefont {S.}~\bibnamefont {Jeon}},\
  and\ \bibinfo {author} {\bibfnamefont {C.}~\bibnamefont {Gale}},\ }\bibfield
  {title} {\bibinfo {title} {{Photon emission from quark-gluon plasma out of
  equilibrium}},\ }\href {https://doi.org/10.1103/PhysRevC.97.014901}
  {\bibfield  {journal} {\bibinfo  {journal} {Phys. Rev. C}\ }\textbf {\bibinfo
  {volume} {97}},\ \bibinfo {pages} {014901} (\bibinfo {year} {2018})},\
  \Eprint {https://arxiv.org/abs/1709.03598} {arXiv:1709.03598 [nucl-th]}
  \BibitemShut {NoStop}%
\bibitem [{\citenamefont {Naik}\ \emph {et~al.}(2021)\citenamefont {Naik},
  \citenamefont {Jaiswal}, \citenamefont {Sreelakshmi}, \citenamefont
  {Jaiswal},\ and\ \citenamefont {Sreekanth}}]{Naik:2021yph}%
  \BibitemOpen
  \bibfield  {author} {\bibinfo {author} {\bibfnamefont {L.~J.}\ \bibnamefont
  {Naik}}, \bibinfo {author} {\bibfnamefont {S.}~\bibnamefont {Jaiswal}},
  \bibinfo {author} {\bibfnamefont {K.}~\bibnamefont {Sreelakshmi}}, \bibinfo
  {author} {\bibfnamefont {A.}~\bibnamefont {Jaiswal}},\ and\ \bibinfo {author}
  {\bibfnamefont {V.}~\bibnamefont {Sreekanth}},\ }\bibfield  {title} {\bibinfo
  {title} {{Hydrodynamical attractor and thermal particle production in
  heavy-ion collision}},\ }\href@noop {} {\  (\bibinfo {year} {2021})},\
  \Eprint {https://arxiv.org/abs/2107.08791} {arXiv:2107.08791 [hep-ph]}
  \BibitemShut {NoStop}%
\bibitem [{\citenamefont {Xiong}\ \emph {et~al.}(2025)\citenamefont {Xiong},
  \citenamefont {Fan}, \citenamefont {Jing}, \citenamefont {Yang},
  \citenamefont {She},\ and\ \citenamefont {Jiang}}]{Xiong:2025koa}%
  \BibitemOpen
  \bibfield  {author} {\bibinfo {author} {\bibfnamefont {J.}~\bibnamefont
  {Xiong}}, \bibinfo {author} {\bibfnamefont {X.}~\bibnamefont {Fan}}, \bibinfo
  {author} {\bibfnamefont {J.}~\bibnamefont {Jing}}, \bibinfo {author}
  {\bibfnamefont {W.}~\bibnamefont {Yang}}, \bibinfo {author} {\bibfnamefont
  {D.}~\bibnamefont {She}},\ and\ \bibinfo {author} {\bibfnamefont {Z.-F.}\
  \bibnamefont {Jiang}},\ }\bibfield  {title} {\bibinfo {title} {{Thermal
  photon emission from quark-gluon plasma: 1+1D magnetohydrodynamics
  results}},\ }\href@noop {} {\  (\bibinfo {year} {2025})},\ \Eprint
  {https://arxiv.org/abs/2510.06604} {arXiv:2510.06604 [hep-ph]} \BibitemShut
  {NoStop}%
\bibitem [{\citenamefont {Kampfer}\ and\ \citenamefont
  {Pavlenko}(1994)}]{Kampfer:1994rr}%
  \BibitemOpen
  \bibfield  {author} {\bibinfo {author} {\bibfnamefont {B.}~\bibnamefont
  {Kampfer}}\ and\ \bibinfo {author} {\bibfnamefont {O.~P.}\ \bibnamefont
  {Pavlenko}},\ }\bibfield  {title} {\bibinfo {title} {{Photon production in an
  expanding and chemically equilibrating gluon enriched plasma}},\ }\href
  {https://doi.org/10.1007/BF01555909} {\bibfield  {journal} {\bibinfo
  {journal} {Z. Phys. C}\ }\textbf {\bibinfo {volume} {62}},\ \bibinfo {pages}
  {491} (\bibinfo {year} {1994})}\BibitemShut {NoStop}%
\bibitem [{\citenamefont {Strickland}(1994)}]{Strickland:1994rf}%
  \BibitemOpen
  \bibfield  {author} {\bibinfo {author} {\bibfnamefont {M.}~\bibnamefont
  {Strickland}},\ }\bibfield  {title} {\bibinfo {title} {{Thermal photons and
  dileptons from nonequilibrium quark - gluon plasma}},\ }\href
  {https://doi.org/10.1016/0370-2693(94)91045-6} {\bibfield  {journal}
  {\bibinfo  {journal} {Phys. Lett. B}\ }\textbf {\bibinfo {volume} {331}},\
  \bibinfo {pages} {245} (\bibinfo {year} {1994})}\BibitemShut {NoStop}%
\bibitem [{\citenamefont {Traxler}\ and\ \citenamefont
  {Thoma}(1996)}]{Traxler:1995kx}%
  \BibitemOpen
  \bibfield  {author} {\bibinfo {author} {\bibfnamefont {C.~T.}\ \bibnamefont
  {Traxler}}\ and\ \bibinfo {author} {\bibfnamefont {M.~H.}\ \bibnamefont
  {Thoma}},\ }\bibfield  {title} {\bibinfo {title} {{Photon emission from a
  parton gas at chemical nonequilibrium}},\ }\href
  {https://doi.org/10.1103/PhysRevC.53.1348} {\bibfield  {journal} {\bibinfo
  {journal} {Phys. Rev. C}\ }\textbf {\bibinfo {volume} {53}},\ \bibinfo
  {pages} {1348} (\bibinfo {year} {1996})},\ \Eprint
  {https://arxiv.org/abs/hep-ph/9507444} {arXiv:hep-ph/9507444} \BibitemShut
  {NoStop}%
\bibitem [{\citenamefont {Dutta}\ \emph {et~al.}(2000)\citenamefont {Dutta},
  \citenamefont {Mohanty}, \citenamefont {Kumar},\ and\ \citenamefont
  {Choudhury}}]{Dutta:1999dy}%
  \BibitemOpen
  \bibfield  {author} {\bibinfo {author} {\bibfnamefont {D.}~\bibnamefont
  {Dutta}}, \bibinfo {author} {\bibfnamefont {A.~K.}\ \bibnamefont {Mohanty}},
  \bibinfo {author} {\bibfnamefont {K.}~\bibnamefont {Kumar}},\ and\ \bibinfo
  {author} {\bibfnamefont {R.~K.}\ \bibnamefont {Choudhury}},\ }\bibfield
  {title} {\bibinfo {title} {{Effect of baryon density on parton production,
  chemical equilibration and thermal photon emission from quark gluon
  plasma}},\ }\href {https://doi.org/10.1103/PhysRevC.61.064911} {\bibfield
  {journal} {\bibinfo  {journal} {Phys. Rev. C}\ }\textbf {\bibinfo {volume}
  {61}},\ \bibinfo {pages} {064911} (\bibinfo {year} {2000})},\ \Eprint
  {https://arxiv.org/abs/hep-ph/9912352} {arXiv:hep-ph/9912352} \BibitemShut
  {NoStop}%
\bibitem [{\citenamefont {Mustafa}\ and\ \citenamefont
  {Thoma}(2000)}]{Mustafa:2000sg}%
  \BibitemOpen
  \bibfield  {author} {\bibinfo {author} {\bibfnamefont {M.~G.}\ \bibnamefont
  {Mustafa}}\ and\ \bibinfo {author} {\bibfnamefont {M.~H.}\ \bibnamefont
  {Thoma}},\ }\bibfield  {title} {\bibinfo {title} {{Bremsstrahlung from an
  equilibrating quark - gluon plasma}},\ }\href
  {https://doi.org/10.1103/PhysRevC.62.014902} {\bibfield  {journal} {\bibinfo
  {journal} {Phys. Rev. C}\ }\textbf {\bibinfo {volume} {62}},\ \bibinfo
  {pages} {014902} (\bibinfo {year} {2000})},\ \Eprint
  {https://arxiv.org/abs/hep-ph/0001230} {arXiv:hep-ph/0001230} \BibitemShut
  {NoStop}%
\bibitem [{\citenamefont {Dutta}\ \emph {et~al.}(2002)\citenamefont {Dutta},
  \citenamefont {Suryanarayana}, \citenamefont {Mohanty}, \citenamefont
  {Kumar},\ and\ \citenamefont {Choudhury}}]{Dutta:2001ii}%
  \BibitemOpen
  \bibfield  {author} {\bibinfo {author} {\bibfnamefont {D.}~\bibnamefont
  {Dutta}}, \bibinfo {author} {\bibfnamefont {S.~S.~V.}\ \bibnamefont
  {Suryanarayana}}, \bibinfo {author} {\bibfnamefont {A.~K.}\ \bibnamefont
  {Mohanty}}, \bibinfo {author} {\bibfnamefont {K.}~\bibnamefont {Kumar}},\
  and\ \bibinfo {author} {\bibfnamefont {R.~K.}\ \bibnamefont {Choudhury}},\
  }\bibfield  {title} {\bibinfo {title} {{Hard photon production from
  unsaturated quark gluon plasma at two loop level}},\ }\href
  {https://doi.org/10.1016/S0375-9474(02)01166-1} {\bibfield  {journal}
  {\bibinfo  {journal} {Nucl. Phys. A}\ }\textbf {\bibinfo {volume} {710}},\
  \bibinfo {pages} {415} (\bibinfo {year} {2002})},\ \Eprint
  {https://arxiv.org/abs/hep-ph/0104134} {arXiv:hep-ph/0104134} \BibitemShut
  {NoStop}%
\bibitem [{\citenamefont {Long}\ \emph {et~al.}(2005)\citenamefont {Long},
  \citenamefont {He}, \citenamefont {Ma},\ and\ \citenamefont
  {Liu}}]{Long:2005cn}%
  \BibitemOpen
  \bibfield  {author} {\bibinfo {author} {\bibfnamefont {J.~L.}\ \bibnamefont
  {Long}}, \bibinfo {author} {\bibfnamefont {Z.~J.}\ \bibnamefont {He}},
  \bibinfo {author} {\bibfnamefont {Y.~G.}\ \bibnamefont {Ma}},\ and\ \bibinfo
  {author} {\bibfnamefont {B.}~\bibnamefont {Liu}},\ }\bibfield  {title}
  {\bibinfo {title} {{Hard photon production from a chemically equilibrating
  quark-gluon plasma with finite baryon density at one loop and two loop}},\
  }\href {https://doi.org/10.1103/PhysRevC.72.064907} {\bibfield  {journal}
  {\bibinfo  {journal} {Phys. Rev. C}\ }\textbf {\bibinfo {volume} {72}},\
  \bibinfo {pages} {064907} (\bibinfo {year} {2005})}\BibitemShut {NoStop}%
\bibitem [{\citenamefont {Monnai}(2014)}]{Monnai:2014kqa}%
  \BibitemOpen
  \bibfield  {author} {\bibinfo {author} {\bibfnamefont {A.}~\bibnamefont
  {Monnai}},\ }\bibfield  {title} {\bibinfo {title} {{Thermal photon $v_2$ with
  slow quark chemical equilibration}},\ }\href
  {https://doi.org/10.1103/PhysRevC.90.021901} {\bibfield  {journal} {\bibinfo
  {journal} {Phys. Rev. C}\ }\textbf {\bibinfo {volume} {90}},\ \bibinfo
  {pages} {021901} (\bibinfo {year} {2014})},\ \Eprint
  {https://arxiv.org/abs/1403.4225} {arXiv:1403.4225 [nucl-th]} \BibitemShut
  {NoStop}%
\bibitem [{\citenamefont {Srivastava}\ \emph {et~al.}(2018)\citenamefont
  {Srivastava}, \citenamefont {Chatterjee},\ and\ \citenamefont
  {Mustafa}}]{Srivastava:2016hwr}%
  \BibitemOpen
  \bibfield  {author} {\bibinfo {author} {\bibfnamefont {D.~K.}\ \bibnamefont
  {Srivastava}}, \bibinfo {author} {\bibfnamefont {R.}~\bibnamefont
  {Chatterjee}},\ and\ \bibinfo {author} {\bibfnamefont {M.~G.}\ \bibnamefont
  {Mustafa}},\ }\bibfield  {title} {\bibinfo {title} {{Initial temperature and
  extent of chemical equilibration of partons in relativistic collisions of
  heavy nuclei}},\ }\href {https://doi.org/10.1088/1361-6471/aa9421} {\bibfield
   {journal} {\bibinfo  {journal} {J. Phys. G}\ }\textbf {\bibinfo {volume}
  {45}},\ \bibinfo {pages} {015103} (\bibinfo {year} {2018})},\ \Eprint
  {https://arxiv.org/abs/1609.06496} {arXiv:1609.06496 [nucl-th]} \BibitemShut
  {NoStop}%
\bibitem [{\citenamefont {Gordeev}\ \emph {et~al.}(2025)\citenamefont
  {Gordeev}, \citenamefont {Bass}, \citenamefont {Mueller},\ and\ \citenamefont
  {Paquet}}]{Gordeev:2025vog}%
  \BibitemOpen
  \bibfield  {author} {\bibinfo {author} {\bibfnamefont {A.}~\bibnamefont
  {Gordeev}}, \bibinfo {author} {\bibfnamefont {S.~A.}\ \bibnamefont {Bass}},
  \bibinfo {author} {\bibfnamefont {B.}~\bibnamefont {Mueller}},\ and\ \bibinfo
  {author} {\bibfnamefont {J.-F.}\ \bibnamefont {Paquet}},\ }\bibfield  {title}
  {\bibinfo {title} {{Quark flavor equilibration of the quark-gluon plasma}},\
  }\href@noop {} {\  (\bibinfo {year} {2025})},\ \Eprint
  {https://arxiv.org/abs/2501.06433} {arXiv:2501.06433 [hep-ph]} \BibitemShut
  {NoStop}%
\bibitem [{\citenamefont {Aurenche}\ \emph {et~al.}(2002)\citenamefont
  {Aurenche}, \citenamefont {Gelis},\ and\ \citenamefont
  {Zaraket}}]{Aurenche:2002pd}%
  \BibitemOpen
  \bibfield  {author} {\bibinfo {author} {\bibfnamefont {P.}~\bibnamefont
  {Aurenche}}, \bibinfo {author} {\bibfnamefont {F.}~\bibnamefont {Gelis}},\
  and\ \bibinfo {author} {\bibfnamefont {H.}~\bibnamefont {Zaraket}},\
  }\bibfield  {title} {\bibinfo {title} {{A Simple sum rule for the thermal
  gluon spectral function and applications}},\ }\href
  {https://doi.org/10.1088/1126-6708/2002/05/043} {\bibfield  {journal}
  {\bibinfo  {journal} {JHEP}\ }\textbf {\bibinfo {volume} {05}},\ \bibinfo
  {pages} {043}},\ \Eprint {https://arxiv.org/abs/hep-ph/0204146}
  {arXiv:hep-ph/0204146} \BibitemShut {NoStop}%
\bibitem [{\citenamefont {Wang}\ and\ \citenamefont
  {Gyulassy}(1991)}]{Wang:1991hta}%
  \BibitemOpen
  \bibfield  {author} {\bibinfo {author} {\bibfnamefont {X.-N.}\ \bibnamefont
  {Wang}}\ and\ \bibinfo {author} {\bibfnamefont {M.}~\bibnamefont
  {Gyulassy}},\ }\bibfield  {title} {\bibinfo {title} {{HIJING: A Monte Carlo
  model for multiple jet production in p p, p A and A A collisions}},\ }\href
  {https://doi.org/10.1103/PhysRevD.44.3501} {\bibfield  {journal} {\bibinfo
  {journal} {Phys. Rev. D}\ }\textbf {\bibinfo {volume} {44}},\ \bibinfo
  {pages} {3501} (\bibinfo {year} {1991})}\BibitemShut {NoStop}%
\end{thebibliography}%
\end{document}